\documentclass{emulateapj}

\shorttitle{Abundances in MW Dwarf Satellites}
\shortauthors{Kirby et al.}
\slugcomment{Accepted to ApJS, 2010 October 26}

\begin{document}
\newcommand{\teff}{$T_{\rm{eff}}$}
\newcommand{\mathteff}{T_{\rm eff}}
\newcommand{\logg}{$\log g$}
\newcommand{\mathlogg}{\log g}
\newcommand{\feh}{[Fe/H]}
\newcommand{\mathfeh}{{\rm [Fe/H]}}
\newcommand{\afe}{[$\alpha$/Fe]}
\newcommand{\mathafe}{{\rm [\alpha/Fe]}}
\newcommand{\xfe}{[$\alpha$/Fe]}
\newcommand{\mathxfe}{{\rm [X/Fe]}}
\newcommand{\ah}{[$\alpha$/H]}
\newcommand{\mathah}{{\rm [\alpha/H]}}
\newcommand{\vt}{$v_t$}
\newcommand{\mathvt}{v_t}
\newcommand{\ngcaf}{NGC~1904 (M79)}
\newcommand{\ngca}{M79}
\newcommand{\ngcb}{NGC 2419}
\newcommand{\ngccf}{NGC~6205 (M13)}
\newcommand{\ngcc}{M13}
\newcommand{\ngcdf}{NGC~6838 (M71)}
\newcommand{\ngcd}{M71}
\newcommand{\ngce}{NGC 7006}
\newcommand{\ngcff}{NGC~7078 (M15)}
\newcommand{\ngcf}{M15}
\newcommand{\ngcg}{NGC 7492}
\newcommand{\bd}{BD+28$^{\circ}$~4211}
\newcommand{\mathaa}{\mathrm{\AA}}
\newcommand\smion[2]{#1$\;$\protect \footnotesize{#2}\small}%
\newcommand\scion[2]{#1$\;$\protect \tiny{#2}\scriptsize}%

\newcommand{\ngcstars}{445}
\newcommand{\fehsyserr}{0.113}

\newcommand{\nmgfeerr}{462}
\newcommand{\nsifeerr}{420}
\newcommand{\ncafeerr}{574}
\newcommand{\ntifeerr}{364}
\newcommand{\gcteffdiffmean}{-29}
\newcommand{\gcteffdiffsigma}{79}
\newcommand{\gcloggdiffmean}{+0.06}
\newcommand{\gcloggdiffsigma}{0.18}
\newcommand{\gcvtdiffmean}{+0.1}
\newcommand{\gcvtdiffsigma}{0.3}
\newcommand{\gcfehdiffmean}{-0.09}
\newcommand{\gcfehdiffsigma}{0.13}
\newcommand{\gcmgfediffmean}{-0.07}
\newcommand{\gcmgfediffsigma}{0.24}
\newcommand{\gcsifediffmean}{+0.04}
\newcommand{\gcsifediffsigma}{0.17}
\newcommand{\gccafediffmean}{+0.02}
\newcommand{\gccafediffsigma}{0.16}
\newcommand{\gctifediffmean}{-0.04}
\newcommand{\gctifediffsigma}{0.14}
\newcommand{\gcalphafediffmean}{+0.01}
\newcommand{\gcalphafediffsigma}{0.14}

\newcommand{\haloteffdiffmean}{+11}
\newcommand{\haloteffdiffsigma}{129}
\newcommand{\halologgdiffmean}{+0.03}
\newcommand{\halologgdiffsigma}{0.17}
\newcommand{\halovtdiffmean}{-0.2}
\newcommand{\halovtdiffsigma}{0.2}
\newcommand{\halofehdiffmean}{-0.14}
\newcommand{\halofehdiffsigma}{0.17}
\newcommand{\halomgfediffmean}{+0.05}
\newcommand{\halomgfediffsigma}{0.24}
\newcommand{\halosifediffmean}{+0.19}
\newcommand{\halosifediffsigma}{0.18}
\newcommand{\halocafediffmean}{-0.04}
\newcommand{\halocafediffsigma}{0.17}
\newcommand{\halotifediffmean}{+0.02}
\newcommand{\halotifediffsigma}{0.25}
\newcommand{\haloalphafediffmean}{+0.02}
\newcommand{\haloalphafediffsigma}{0.17}

\newcommand{\ndup}{167}
\newcommand{\nalphadup}{141}
\newcommand{\fehdupsigma}{0.98}
\newcommand{\fehdupkurtosis}{-0.14}
\newcommand{\alphadupsigma}{1.00}
\newcommand{\alphadupkurtosis}{-0.05}
\newcommand{\verr}{3.1}
\newcommand{\verrerr}{0.2}
\newcommand{\medtefferrrand}{53}
\newcommand{\medtefferrsys}{60}
\newcommand{\medtefferrtot}{85}
\newcommand{\medloggerrrand}{0.05}
\newcommand{\medloggerrsys}{0.02}
\newcommand{\medloggerrtot}{0.06}
\newcommand{\ndsphstars}{2961}
\newcommand{\fehbatsigma}{0.17}
\newcommand{\cafebatsigma}{1.38}
\newcommand{\cafebatcorr}{0.00}
\newcommand{\nbat}{57}
\newcommand{\dsphteffdiffmean}{-40}
\newcommand{\dsphteffdiffsigma}{132}
\newcommand{\dsphloggdiffmean}{+0.05}
\newcommand{\dsphloggdiffsigma}{0.41}
\newcommand{\dsphvtdiffmean}{-0.2}
\newcommand{\dsphvtdiffsigma}{0.3}
\newcommand{\dsphfehdiffmean}{-0.04}
\newcommand{\dsphfehdiffsigma}{0.15}
\newcommand{\dsphmgfediffmean}{-0.06}
\newcommand{\dsphmgfediffsigma}{0.18}
\newcommand{\dsphsifediffmean}{-0.16}
\newcommand{\dsphsifediffsigma}{0.28}
\newcommand{\dsphcafediffmean}{-0.03}
\newcommand{\dsphcafediffsigma}{0.23}
\newcommand{\dsphtifediffmean}{-0.05}
\newcommand{\dsphtifediffsigma}{0.24}
\newcommand{\dsphalphafediffmean}{-0.05}
\newcommand{\dsphalphafediffsigma}{0.19}

\newcommand{\nhrs}{132}
\newcommand{\allteffdiffmean}{-27}
\newcommand{\allteffdiffsigma}{115}
\newcommand{\allloggdiffmean}{+0.05}
\newcommand{\allloggdiffsigma}{0.33}
\newcommand{\allvtdiffmean}{-0.1}
\newcommand{\allvtdiffsigma}{0.3}
\newcommand{\allfehdiffmean}{-0.07}
\newcommand{\allfehdiffsigma}{0.15}
\newcommand{\allmgfediffmean}{-0.05}
\newcommand{\allmgfediffsigma}{0.20}
\newcommand{\allsifediffmean}{-0.04}
\newcommand{\allsifediffsigma}{0.26}
\newcommand{\allcafediffmean}{-0.01}
\newcommand{\allcafediffsigma}{0.20}
\newcommand{\alltifediffmean}{-0.04}
\newcommand{\alltifediffsigma}{0.22}
\newcommand{\allalphafediffmean}{-0.02}
\newcommand{\allalphafediffsigma}{0.16}
\newcommand{\tefffehint}{-0.06}
\newcommand{\tefffehinterr}{0.01}
\newcommand{\loggfehint}{-0.08}
\newcommand{\loggfehinterr}{0.01}
\newcommand{\vtfehint}{-0.09}
\newcommand{\vtfehinterr}{0.01}
\newcommand{\fehalphafeint}{-0.02}
\newcommand{\fehalphafeinterr}{0.02}
\newcommand{\gcfecorcoeff}{0.98}
\newcommand{\gcmgcorcoeff}{0.41}
\newcommand{\gcsicorcoeff}{0.49}
\newcommand{\gccacorcoeff}{0.33}
\newcommand{\gcticorcoeff}{0.26}
\newcommand{\vtfehslope}{-0.20}
\newcommand{\fehscatter}{0.13}

\newcommand{\sclnbad}{1}
\newcommand{\sclngood}{411}
\newcommand{\sclndup}{17}
\newcommand{\sclnunique}{394}
\newcommand{\alphasyserr}{0.081}
\newcommand{\mgfesyserr}{0.095}
\newcommand{\sifesyserr}{0.104}
\newcommand{\cafesyserr}{0.118}
\newcommand{\tifesyserr}{0.083}
\newcommand{\scllowv}{85.1}
\newcommand{\sclhighv}{138.0}
\newcommand{\sclmeanv}{111.6}
\newcommand{\sclmeanverr}{0.5}
\newcommand{\sclvdupsigma}{3.0}
\newcommand{\sclsigmav}{8.3}
\newcommand{\sclsigmaverr}{0.5}
\newcommand{\sclmeanvother}{110.4}
\newcommand{\sclmeanverrother}{0.8}
\newcommand{\sclsigmavother}{8.8}
\newcommand{\sclsigmaverrother}{0.6}
\newcommand{\sclmeanvdiff}{+1.2}
\newcommand{\sclsigmavdiff}{-0.7}
\newcommand{\sclmeanvw}{111.3}
\newcommand{\sclmeanverrw}{0.2}
\newcommand{\sclsigmavw}{8.8}
\newcommand{\sclsigmaverrw}{0.2}
\newcommand{\sclmeanvdiffw}{+0.5}
\newcommand{\sclsigmavdiffw}{-0.9}
\newcommand{\sclalphaerfsigma}{2.4}
\newcommand{\sclnnonmember}{18}
\newcommand{\sclnmember}{376}
\newcommand{\sclnvmp}{115}
\newcommand{\sclmedtefferrrand}{94}
\newcommand{\sclmedtefferrsys}{58}
\newcommand{\sclmedtefferrtot}{110}
\newcommand{\sclmedloggerrrand}{0.06}
\newcommand{\sclmedloggerrsys}{0.02}
\newcommand{\sclmedloggerrtot}{0.06}
\newcommand{\sclfehhrssigma}{0.16}
\newcommand{\sclfehbatsigma}{0.17}
\newcommand{\sclcafebatsigma}{1.47}
\newcommand{\sclcafebatcorr}{****}
\newcommand{\sclnbat}{50}
\newcommand{\sclmdfhelmean}{-1.82}
\newcommand{\sclmdfhelsigma}{0.35}
\newcommand{\sclmdfbatmean}{-1.56}
\newcommand{\sclmdfbatsigma}{0.38}
\newcommand{\sclnbathrs}{7}
\newcommand{\sclfehbathrssigma}{0.16}
\newcommand{\sclfehcuthrssigma}{0.19}
\newcommand{\sclfehslope}{-1.856}
\newcommand{\sclfehslopeerr}{0.160}
\newcommand{\sclfehslopekpc}{-1.24}
\newcommand{\sclfehslopekpcerr}{0.11}
\newcommand{\sclfehsloperc}{-0.18}
\newcommand{\sclfehslopercerr}{0.02}
\newcommand{\sclafeslope}{+0.013}
\newcommand{\sclafeslopeerr}{0.003}
\newcommand{\sclafeslopekpc}{+0.54}
\newcommand{\sclafeslopekpcerr}{0.10}
\newcommand{\sclalphafehrssigma}{0.17}
\newcommand{\sclfehrange}{****}
\newcommand{\sclfehinitial}{-4.85}
\newcommand{\sclfehinitialerr}{0.00}
\newcommand{\sclfehfinal}{-4.85}
\newcommand{\sclfehfinalerr}{0.00}
\newcommand{\sclpureyield}{0.030}
\newcommand{\sclsimpleyield}{0.028}
\newcommand{\sclsimpleyielderr}{0.001}
\newcommand{\sclinfallm}{1.28}
\newcommand{\sclinfallyield}{0.030}
\newcommand{\sclprobratio}{1.59}
\newcommand{\sclprobratiopure}{0.72}
\newcommand{\sclprobratioinfallpre}{0.86}
\newcommand{\sclprobratiopreinfall}{-0.86}
\newcommand{\sclfehspread}{0.46}
\newcommand{\sclmgfespread}{0.21}
\newcommand{\sclsifespread}{0.29}
\newcommand{\sclcafespread}{0.17}
\newcommand{\scltifespread}{0.22}
\newcommand{\sclalphafespread}{0.24}
\newcommand{\sclmgfespreadsub}{-0.10}
\newcommand{\sclsifespreadsub}{0.10}
\newcommand{\sclcafespreadsub}{0.05}
\newcommand{\scltifespreadsub}{0.09}
\newcommand{\sclalphafespreadsub}{0.09}
\newcommand{\sclfehmean}{-1.68}
\newcommand{\sclfehsigma}{0.48}
\newcommand{\sclfehmedian}{-1.67}
\newcommand{\sclfehmad}{0.37}
\newcommand{\sclfehiqr}{0.75}
\newcommand{\sclempfehone}{-3.02}
\newcommand{\sclempfeherrone}{0.15}
\newcommand{\sclempvone}{18.06}
\newcommand{\sclempfehtwo}{-3.27}
\newcommand{\sclempfeherrtwo}{0.59}
\newcommand{\sclempvtwo}{20.08}
\newcommand{\sclempfehthree}{-3.87}
\newcommand{\sclempfeherrthree}{0.21}
\newcommand{\sclempvthree}{18.19}
\newcommand{\sclcmdfeh}{-1.53}

\title{Multi-Element Abundance Measurements from Medium-Resolution
Spectra. \\ II. Catalog of Stars in Milky Way Dwarf Satellite
Galaxies\altaffilmark{1}}

\author{Evan~N.~Kirby\altaffilmark{2,3},
  Puragra~Guhathakurta\altaffilmark{4},
  Joshua~D.~Simon\altaffilmark{5},
  Marla~C.~Geha\altaffilmark{6},
  Constance~M.~Rockosi\altaffilmark{4},
  Christopher~Sneden\altaffilmark{7},
  Judith~G.~Cohen\altaffilmark{2},
  Sangmo~Tony~Sohn\altaffilmark{8},
  Steven~R.~Majewski\altaffilmark{9},
  Michael~Siegel\altaffilmark{10}}

\altaffiltext{1}{Data herein were obtained at the W.~M. Keck
  Observatory, which is operated as a scientific partnership among the
  California Institute of Technology, the University of California,
  and NASA.  The Observatory was made possible by the generous
  financial support of the W.~M. Keck Foundation.}
\altaffiltext{2}{California Institute of Technology, Department of
  Astronomy, Mail Stop 249-17, Pasadena, CA 91106, USA}
\altaffiltext{3}{Hubble Fellow}
\altaffiltext{4}{University of California Observatories/Lick
  Observatory, Department of Astronomy \& Astrophysics, University of
  California, Santa Cruz, CA 95064, USA}
\altaffiltext{5}{Observatories of the Carnegie Institution of
  Washington, 813 Santa Barbara Street, Pasadena, CA 91101, USA}
\altaffiltext{6}{Astronomy Department, Yale University, New Haven, CT
  06520, USA}
\altaffiltext{7}{McDonald Observatory, University of Texas, Austin, TX
  78712, USA}
\altaffiltext{8}{Space Telescope Science Institute, 3700 San Martin
  Drive, Baltimore, MD 21218, USA}
\altaffiltext{9}{Department of Astronomy, University of Virginia,
  P.O. Box 400325, Charlottesville, VA 22904-4325, USA}
\altaffiltext{10}{Pennsylvania State University, 525 Davey Lab, State
  College, PA 16801, USA}

\keywords{galaxies: abundances --- galaxies: dwarf --- Galaxy:
  evolution --- Local Group}


\begin{abstract}

We present a catalog of Fe, Mg, Si, Ca, and Ti abundances for
\ndsphstars\ red giant stars that are likely members of eight dwarf
satellite galaxies of the Milky Way (MW): Sculptor, Fornax, Leo~I,
Sextans, Leo~II, Canes Venatici~I, Ursa Minor, and Draco.  For the
purposes of validating our measurements, we also observed
\ngcstars\ red giants in MW globular clusters and 21 field red giants
in the MW halo.  The measurements are based on Keck/DEIMOS
medium-resolution spectroscopy combined with spectral synthesis.  We
estimate uncertainties in \feh\ by quantifying the dispersion of
\feh\ measurements in a sample of stars in monometallic globular
clusters.  We estimate uncertainties in Mg, Si, Ca, and Ti abundances
by comparing our medium-resolution spectroscopic measurements to
high-resolution spectroscopic abundances of the same stars.  For this
purpose, our DEIMOS sample included \nhrs\ red giants with published
high-resolution spectroscopy in globular clusters, the MW halo field,
and dwarf galaxies.  The standard deviations of the differences in
\feh\ and $\langle[\alpha/\rm{Fe}]\rangle$ (the average of [Mg/Fe],
          [Si/Fe], [Ca/Fe], and [Ti/Fe]) between the two samples is
          $\allfehdiffsigma$ and $\allalphafediffsigma$, respectively.
          This catalog represents the largest sample of multi-element
          abundances in dwarf galaxies to date.  The next papers in
          this series draw conclusions on the chemical evolution, gas
          dynamics, and star formation histories from the catalog
          presented here.  The wide range of dwarf galaxy luminosity
          reveals the dependence of dwarf galaxy chemical evolution on
          galaxy stellar mass.

\end{abstract}


\section{Introduction}
\label{sec:intro}

The Milky Way contains $\sim 5 \times 10^{10}~M_{\odot}$ of
stars.  Approximately 90\% of those stars lie in the disk, 9\% in the
bulge, and 1\% in the halo \citep{bin08}.  However, these structures
did not form their stars en masse.  The sites of star formation are
far less massive than any of these components.  For example, star
formation in the disk occurs in open clusters, which typically have
just $300~M_{\odot}$ of stars \citep{pis07}.  Dwarf galaxies, which
may have built the bulge and halo \citep{sea78,whi78}, range in mass
from $10^3~M_{\odot}$ \citep[e.g., Segue~1;][]{mar08} to more than
$10^7~M_{\odot}$ \citep[e.g., Fornax;][]{irw95,mat98}.  The pockets of
star formation within the dwarf galaxies likely contained just as
little mass as the open clusters in the Galactic disk.  Dynamical
processes assemble the many disparate sites of star formation into the
primary Galactic components, such as the disk, bulge, and halo.
Therefore, the key to discovering the origins of stars lies in
studying stellar populations in less massive structures.

The dwarf galaxies that orbit the Milky Way (MW) present the
opportunity to study the formation of stellar populations most similar
to the MW stellar halo.  Unlike most open clusters, dwarf galaxies
contain old, metal-poor stars.  Unlike most globular clusters, dwarf
galaxies enable the study of temporally extended star formation, which
ultimately results in chemical evolution.  Furthermore, galaxies of
this class probably created at least some of the stellar halo of the
MW by gravitational dissolution \citep[e.g.,][]{maj93,maj96,bel08}.
If so, then the stellar population of the present MW halo contains a
mixture of many different galaxies.  The individual stellar
populations are difficult to disentangle, especially in the inner halo
\citep[e.g.,][]{coo10}.  Surviving dwarf galaxies may offer a look at
single, though evolved, counterparts to an accreted component of the
halo.  The stellar populations in surviving and accreted dwarfs differ
because the accretion time is correlated with stellar mass and
therefore star formation history.  Nonetheless, the observed abundance
differences can be compared with models of surviving galaxies and
hierarchically formed halos \citep[e.g.,][]{rob05,fon06}.

Alternatively, the halo may have been created by a combination of
monolithic collapse \citep*{egg62} and hierarchical assembly.
Comparisons of the abundances of dwarf galaxy stars and the
chemodynamics of halo stars \citep[e.g.,][]{gra03,ven04} show that
only some of the stars in the halo are consistent with accretion of
dwarf galaxies similar to the surviving dwarfs.  Although it is clear
that the nearby halo stars have different abundance patterns than
dwarf satellite galaxies, the number of dwarf galaxy stars with
published multi-element abundance measurements is less than for the
halo.  The relative distance of dwarf galaxies compared to the nearest
observable halo stars has prevented the sample of dwarf galaxy
abundances from growing as fast as for halo stars.  Studying the role
of dwarf galaxies in building the halo would benefit from a larger
sample of dwarf galaxy elemental abundances.  (\citeauthor{roe09}
\citeyear{roe09} points out that even the halo abundances have not
been well-sampled because the more distant halo stars have not been
observed nearly as completely as nearby halo stars.)

Stellar elemental abundances reveal many characteristics of the
stellar population of a galaxy.  Its metallicity distribution, usually
represented by the stellar iron content, is a function of the galaxy's
total mass, the amount of gas infall and outflow, the chemical yield
of supernovae, whether the galaxy formed out of gas pre-enriched with
metals, and other variables.  Beyond the one-dimensional metallicity
distribution, the ratios of other elements to iron reflect other
details of the star formation history.  In particular, the ratios of
alpha elements, such as magnesium, to iron depend on the duration and
intensity of star formation.

A large sample of multi-element abundances for individual stars in
dwarf galaxies would be a useful tool in studying star formation in
small systems.  However, obtaining elemental abundances is expensive.
High-quality, high-resolution spectroscopy (HRS, $R > 15000$) allows
the measurements of tens of elements to precisions of $<0.2$~dex for
stars with $V \la 18.5$.  This magnitude limit restricts the
measurement of elemental abundances to the upper 1--2 magnitudes of
the red giant branch for most MW dwarf galaxies.  In addition, most
current high-resolution spectrographs operate on one star at a time.
The exposure time required for adequate signal to derive accurate
abundance measurements has limited the sample of stars in intact MW
dwarf satellite galaxies (excluding Sagittarius) with published
high-resolution spectroscopy to fewer than 200.

With current technology, a statistical analysis of the elemental
abundance distributions in dwarf galaxies must rely on
medium-resolution spectroscopy (MRS, $R \sim 7000$).  MRS relaxes the
magnitude limit to $V \la 21.5$ while permitting multiplexing of more
than one hundred stars in one exposure.  Building a large sample of
multi-element abundances in MW dwarf galaxies requires both of these
capabilities.  The disadvantages of MRS are that fewer elements are
accessible than with HRS, and the measurements are less precise.
\citet{tol01} presented one of the first applications of MRS to dwarf
galaxy metallicity distributions.  They based their metallicity
measurements on a calibration of infrared calcium triplet equivalent
width to \feh\ \citep[e.g.,][]{arm91,rut97}.  However, spectral
synthesis is required to measure the abundances of multiple elements.
Several recent spectral synthesis-based studies have employed MRS to
begin building the sample of abundances in MW dwarf galaxies.
\citet{kir08b} discovered extremely metal-poor ($\mathfeh < -3$) stars
in the ultra-faint dwarf galaxies using Keck/DEIMOS spectroscopy.
\citet{she09} measured Fe, Mg, Ca, and Ti abundances for 27 stars in
the Leo~II galaxy using Keck/LRIS spectroscopy.  Finally,
\citet[][Paper~I]{kir09} measured Fe, Mg, Si, Ca, and Ti abundances
for 388 stars in the Sculptor dwarf galaxy using Keck/DEIMOS
spectroscopy.

In this article, we present a catalog of elemental abundance
measurements of \ndsphstars\ stars in eight MW dwarf galaxies.  The
technique is very similar to that developed by \citet[][KGS08]{kir08a}
and modified as noted in \citeauthor*{kir09}.  We describe the
observations in Section~\ref{sec:observations}.  In
Section~\ref{sec:measurements}, we briefly summarize the technique and
present the catalog.  In Section~\ref{sec:accuracy}, we explore the
accuracy of the measurements using repeat observations and comparison
to HRS measurements of the same stars.  We summarize our work and
discuss our articles that interpret these data in
Section~\ref{sec:summary}.

The next papers in this series apply the data presented here to
deducing the past chemical evolution histories of the eight dwarf
spheroidal galaxies (dSphs).  \citeauthor*{kir10b} \citep{kir10b}
includes galactic chemical evolution models of the metallicity
distributions.  \citeauthor*{kir10a} \citep{kir10a} shows how the
\afe\ ratios change with \feh, a diagnostic of the past intensity of
star formation.  Both papers examine trends with the total luminosity
of the dSph.  This large data set, encompassing a wide luminosity
range of MW satellites, will uniquely enable an exploration of the
dependence of chemical evolution on dwarf galaxy mass.


\section{Observations}
\label{sec:observations}

\begin{deluxetable}{lc@{~~}c@{~~}c@{~~}c}
\tablewidth{0pt}
\tablecolumns{5}
\tablecaption{Spectroscopic Targets\label{tab:targets}}
\tablehead{\colhead{Target} & \colhead{RA} & \colhead{Dec} & \colhead{Distance} & $(m-M)_0$\tablenotemark{a} \\
                            & \colhead{(J2000)} & \colhead{(J2000)} & \colhead{(kpc)} & \colhead{(mag)}} \\
\startdata
\cutinhead{Globular Clusters}
NGC 288           & $00^{\mathrm{h}} 52^{\mathrm{m}} 45^{\mathrm{s}}$ & $-26\arcdeg 34\arcmin 43\arcsec$ &    \phn\phn8.8 & 14.74 \\ 
NGC 1904 (M79)    & $05^{\mathrm{h}} 24^{\mathrm{m}} 11^{\mathrm{s}}$ & $-24\arcdeg 31\arcmin 27\arcsec$ &       \phn12.9 & 15.56 \\ 
NGC 2419          & $07^{\mathrm{h}} 38^{\mathrm{m}} 09^{\mathrm{s}}$ & $+38\arcdeg 52\arcmin 55\arcsec$ &       \phn84.3 & 19.63 \\ 
NGC 5904 (M5)     & $15^{\mathrm{h}} 18^{\mathrm{m}} 34^{\mathrm{s}}$ & $+02\arcdeg 04\arcmin 58\arcsec$ &    \phn\phn7.5 & 14.37 \\ 
NGC 6205 (M13)    & $16^{\mathrm{h}} 41^{\mathrm{m}} 41^{\mathrm{s}}$ & $+36\arcdeg 27\arcmin 37\arcsec$ &    \phn\phn7.6 & 14.42 \\ 
NGC 6341 (M92)    & $17^{\mathrm{h}} 17^{\mathrm{m}} 07^{\mathrm{s}}$ & $+43\arcdeg 08\arcmin 11\arcsec$ &    \phn\phn8.2 & 14.58 \\ 
NGC 6838 (M71)    & $19^{\mathrm{h}} 53^{\mathrm{m}} 46^{\mathrm{s}}$ & $+18\arcdeg 46\arcmin 42\arcsec$ &    \phn\phn4.0 & 13.02 \\ 
NGC 7006          & $21^{\mathrm{h}} 01^{\mathrm{m}} 29^{\mathrm{s}}$ & $+16\arcdeg 11\arcmin 14\arcsec$ &       \phn41.4 & 18.09 \\ 
NGC 7078 (M15)    & $21^{\mathrm{h}} 29^{\mathrm{m}} 58^{\mathrm{s}}$ & $+12\arcdeg 10\arcmin 01\arcsec$ &       \phn10.3 & 15.06 \\ 
NGC 7089 (M2)     & $21^{\mathrm{h}} 33^{\mathrm{m}} 27^{\mathrm{s}}$ & $-00\arcdeg 49\arcmin 24\arcsec$ &       \phn11.5 & 15.30 \\ 
Pal 13            & $23^{\mathrm{h}} 06^{\mathrm{m}} 44^{\mathrm{s}}$ & $+12\arcdeg 46\arcmin 19\arcsec$ &       \phn25.8 & 17.05 \\ 
NGC 7492          & $23^{\mathrm{h}} 08^{\mathrm{m}} 27^{\mathrm{s}}$ & $-15\arcdeg 36\arcmin 41\arcsec$ &       \phn25.8 & 17.06 \\ 
\cutinhead{dSphs}
Sculptor          & $01^{\mathrm{h}} 00^{\mathrm{m}} 09^{\mathrm{s}}$ & $-33\arcdeg 42\arcmin 32\arcsec$ & \phn85\phd\phn & 19.67 \\ 
Fornax            & $02^{\mathrm{h}} 39^{\mathrm{m}} 59^{\mathrm{s}}$ & $-34\arcdeg 26\arcmin 57\arcsec$ &    139\phd\phn & 20.72 \\ 
Leo I             & $10^{\mathrm{h}} 08^{\mathrm{m}} 28^{\mathrm{s}}$ & $+12\arcdeg 18\arcmin 23\arcsec$ &    254\phd\phn & 22.02 \\ 
Sextans           & $10^{\mathrm{h}} 13^{\mathrm{m}} 03^{\mathrm{s}}$ & $-01\arcdeg 36\arcmin 52\arcsec$ & \phn95\phd\phn & 19.90 \\ 
Leo II            & $11^{\mathrm{h}} 13^{\mathrm{m}} 29^{\mathrm{s}}$ & $+22\arcdeg 09\arcmin 12\arcsec$ &    219\phd\phn & 21.70 \\ 
Canes Venatici I  & $13^{\mathrm{h}} 28^{\mathrm{m}} 04^{\mathrm{s}}$ & $+33\arcdeg 33\arcmin 21\arcsec$ &    210\phd\phn & 21.62 \\ 
Ursa Minor        & $15^{\mathrm{h}} 09^{\mathrm{m}} 11^{\mathrm{s}}$ & $+67\arcdeg 12\arcmin 52\arcsec$ & \phn69\phd\phn & 19.18 \\ 
Draco             & $17^{\mathrm{h}} 20^{\mathrm{m}} 19^{\mathrm{s}}$ & $+57\arcdeg 54\arcmin 48\arcsec$ & \phn92\phd\phn & 19.84 \\ 
\enddata
\tablenotetext{a}{Extinction corrected distance modulus.}
\tablerefs{The coordinates and distance moduli for the globular clusters are given by \citet[][updated 2003, \url{http://www.physics.mcmaster.ca/~harris/mwgc.dat}]{har96}.  \citeauthor{har96} relied on data from the following sources: NGC~288, \citet{bel01}; M79, \citet{fer92}; NGC~2419, \citet{har97}; M5, \citet{bro96} and \citet{san96}; M13, \citet{pal98}; M92, \citet{car92}; M71, \citet{gef00}; NGC~7006, \citet{buo91}; M15, \citet{dur93}; M2, \citet{har75}; Pal~13, \citet{sie01}; NGC~7492, \citet{cot91}.  The dSph coordinates are adopted from \citet{mat98}, and the distances are adopted from the following sources: Sculptor, \citet{pie08}; Fornax, \citet{riz07}; Leo~I, \citet{bel04}; Sextans, \citet{lee03}; Leo~II, \citet{sie10}; Canes Venatici~I, \citet{kue08}; Ursa Minor, \citet{mig99}; Draco, \citet{bel02}.}
\end{deluxetable}

The medium-resolution abundances in this article depend on
spectroscopy obtained with the Deep Imaging Multi-Object Spectrograph
\citep[DEIMOS,][]{fab03} on the Keck II telescope and photometry from
a variety of sources.  This section explains how spectroscopic targets
were selected from photometric catalogs and how the DEIMOS
observations were performed.

Every spectroscopic target is an individual star in one of three types
of stellar system: globular cluster, the Milky Way halo, and dwarf
galaxy.  All of the target systems except the MW halo field stars are
listed in Table~\ref{tab:targets}.  The MW halo field stars are
included in Table~\ref{tab:hrscompare}.  The globular cluster and halo
field stars are interesting in their own right.  The globular cluster
stars in particular will be examined in further detail in a future
work.  However, we examine the globular cluster and halo field stars
here only to assess the accuracy of the abundance measurement
technique.  The observations most relevant to this article's
scientific focus are the individual dSph stellar spectra.

\subsection{Globular Clusters}
\label{sec:gcobs}

Stars in most globular clusters (GCs) are excellent metallicity
standards because every star in a given cluster has the nearly the
same iron abundance and the same heliocentric distance.  Exceptions
include $\omega$~Cen \citep{fre75} and M22 \citep{dac09,mar09}.  We
targeted individual stars in twelve monometallic GCs, listed in
Table~\ref{tab:targets}.  We check stellar abundances within each
cluster for internal consistency, and we also check the DEIMOS-derived
abundances against HRS abundance measurements.

\citet{ste00} has made photometry publicly available for all of these
clusters as part of his photometric standard field
database.\footnote{\url{http://www2.cadc-ccda.hia-iha.nrc-cnrc.gc.ca/}
  \url{community/STETSON/standards/}} Every star has been measured in
at least two of the Johnson-Cousins $B$, $V$, $R$, and $I$ filters.
This database does not contain every star in each field, but only
those stars suitable for photometric standardization.  Therefore,
P.~B.~Stetson (private communications, 2007, 2008) generously provided
complete photometric data for our targets in M13, M71, NGC~7006, M15,
and NGC~7492.

We supplemented Stetson's photometry with additional photometry for
five clusters, mostly to increase the field of view so that it spanned
a full DEIMOS slitmask.  For M79, we utilized the $UBV$ measurements
of \citet{kra97} and the $VI$ measurements of \citet{ros00}.
For M5, we supplemented Stetson's photometry with the $ugriz$ Sloan
Digital Sky Survey (SDSS) crowded field photometry of \citet{an08}.
For consistency with the other data in the Johnson-Cousins system, we
converted SDSS $ugriz$ to Johnson-Cousins $UBVRI$ following the
global, metallicity-independent transformations of \citet*{jor06}.
All magnitudes were corrected for extinction based on $E(B-V)$ from
\citeauthor{har96}'s (1996, updated 2003) catalog of GC properties.
In order to arrive at his values, Harris averaged measurements of
$E(B-V)$ by \citet{ree88}, \citet{web85}, \citet{zin85}, and the
authors listed in Table~\ref{tab:targets}.

Not all targets could be observed spectroscopically due to the limited
field of view and slitmask design constraints.  First, we attempted to
maximize the number of target stars previously observed with
high-resolution spectroscopy (HRS) by placing the slitmasks in regions
with a high density of HRS targets.  We selected the remaining targets
based on the clusters' color-magnitude diagrams (CMDs).  In order of
priority, we filled each slitmask with stars from the (1) upper red
giant branch (RGB), (2) lower RGB, (3) red clump, and (4) blue
horizontal branch.  For the GCs with angular sizes smaller than the
DEIMOS field of view, we filled the slits at the edges of the slitmask
far from the center of the GCs with objects having similar colors and
magnitudes to stars on the RGB.  \citet{sim07} prioritized targets for
the n1904 and n2419 slitmasks (see Table~2) differently.  We refer the
reader to their article for further information.

\subsection{Halo Field Stars}

The most metal-poor MW globular cluster known is M15 \citep[$\mathfeh
  = -2.38$;][]{sne97,sne00,pri05}.  Dwarf galaxies contain even more
metal-poor stars \citep[e.g.,][]{she01a,ful04}.  In order to verify
that the abundance measurements based on MRS are accurate at these
very low metallicities, we obtained DEIMOS spectra of metal-poor Milky
Way halo field stars having HRS data.

We chose several studies of metal-poor halo field stars: the Keck
High-Resolution Echelle Spectrograph (HIRES) and Lick/Hamilton
spectrograph measurements of \citet{joh02} at $R \sim 45000-60000$;
the Keck/HIRES measurements of \citet{ful00}, \citet{coh02}, and
\citet{car02}; the Kitt Peak National Observatory Coud{\' e}
spectrograph measurements of \citet{pil96}; and the Keck Echellette
Spectrograph and Imager (ESI) measurements of \citet{lai04,lai07} at
$R \sim 7000$.  Each of these authors selected targets from
low-resolution surveys for metal-poor stars in the Galactic halo
\citep[][and additional references from \citeauthor{ful00}
  \citeyear{ful00}]{bon80,bee85,bee92,nor99}.  These abundance
measurements are based on equivalent width analysis and/or spectral
synthesis.  Not all of the hundreds of stars in these studies could be
observed.  The highest priority was assigned to the most metal-poor
stars.

Strictly speaking, \citet{lai04,lai07} did not conduct an HRS study
because their Keck/ESI spectra had a spectral resolution of $R \sim
7000$.  We still include their sample in the comparison along with the
truly high-resolution studies.

\subsection{Dwarf Spheroidal Galaxies}

\begin{figure*}[hp!]
\centering
\includegraphics[width=0.46\textwidth]{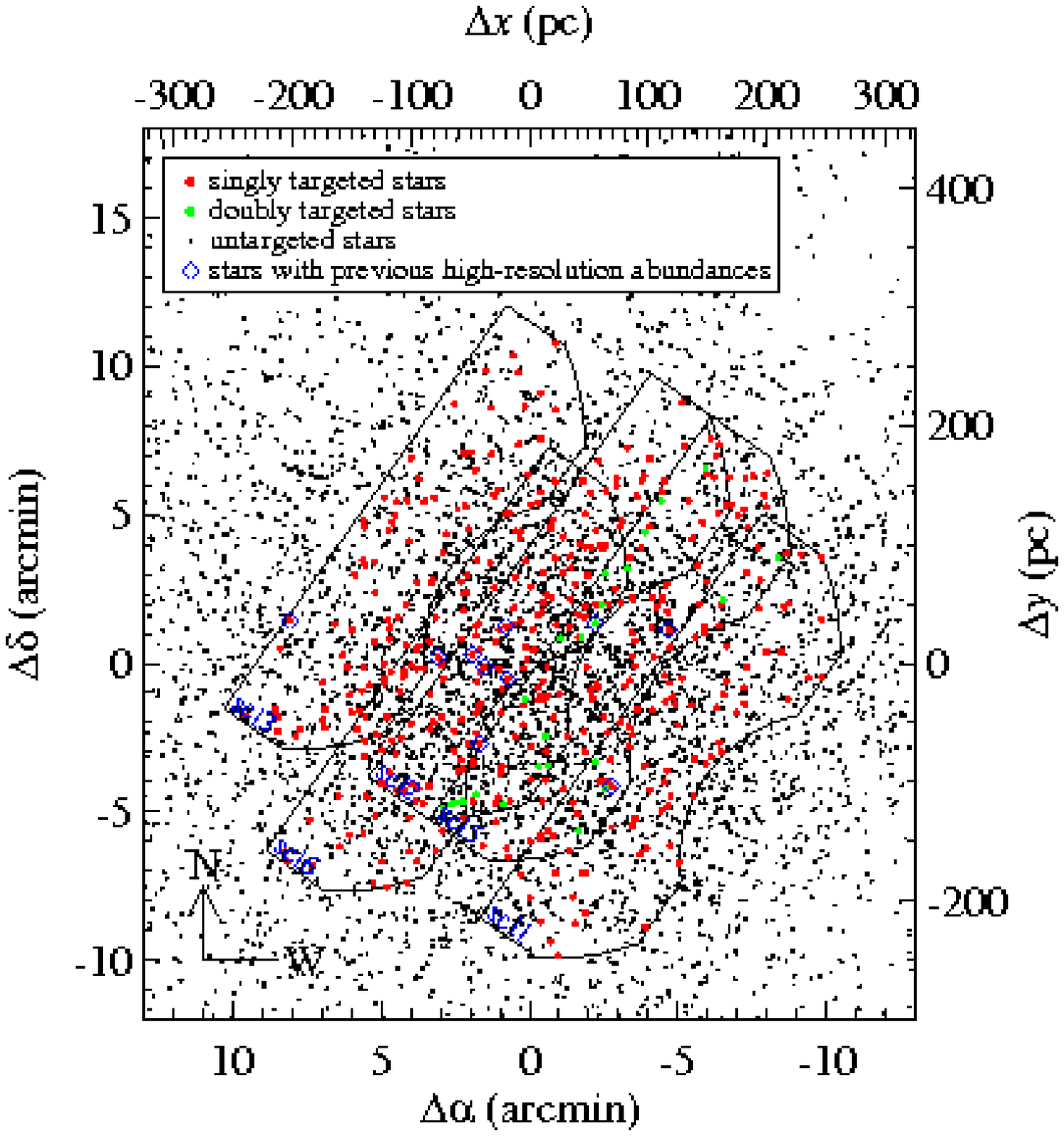}
\hfil
\includegraphics[width=0.46\textwidth]{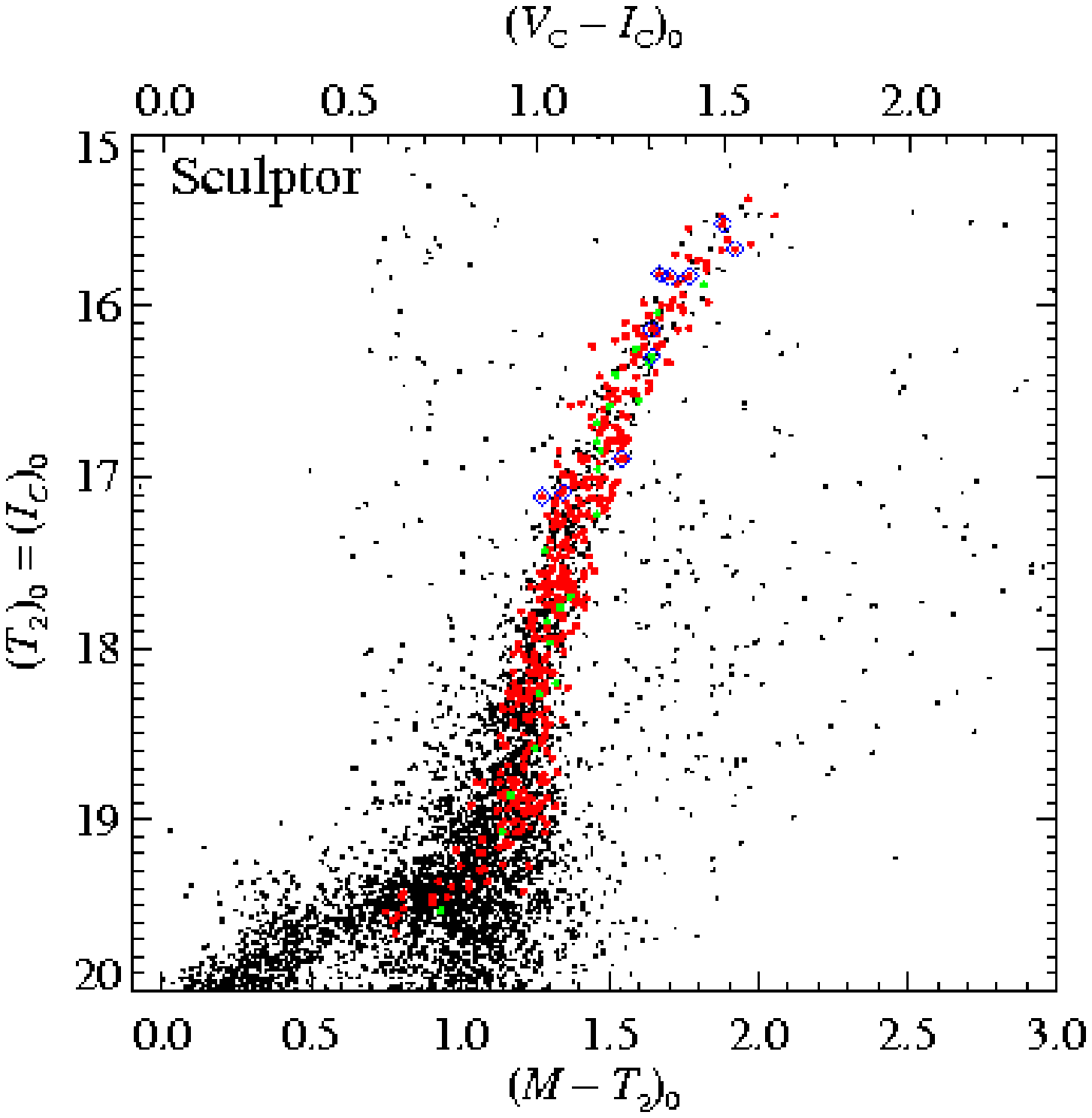}
\caption{{\bf Sculptor.}  {\it Left:} DEIMOS slitmask footprints laid
  over a map of sources from the Sculptor photometric catalog
  \citep{wes06}.  Targets selected for spectroscopy are shown in red.
  Targets observed in more than one mask are shown in green.  Blue
  diamonds enclose stars with previous HRS abundance measurements.
  Each slitmask is labeled in blue with the names given in Table~2.
  The left and bottom axis scales show the angular displacement in
  arcmin from the center of the galaxy \citep[$\alpha_0 =
    1^{\rm{h}}00^{\rm{m}}09^{\rm{s}}$, $\delta_0 = -33^{\circ}42
    \farcm 5$,][]{mat98}, and the right and top axis scales show the
  projected physical distance for an assumed distance to Sculptor of
  85.9~kpc \citep{pie08}.  {\it Right:} Extinction- and
  reddening-corrected color-magnitude diagram in the Washington and
  Cousins systems for the photometric sources within the right
  ascension and declination ranges shown at left.  The transformation
  from the Washington system ($M$ and $T_2$) to the Cousins system
  ($V_{\rm C}$ and $I_{\rm C}$) is $I_{\rm C} = T_2$ and $V_{\rm C} -
  I_{\rm C} = 0.800(M-T_2) - 0.006$ \citep{maj00}.\notetoeditor{B\&W
    figure caption: {\bf Sculptor.}  {\it Left:} DEIMOS slitmask
    footprints laid over a map of sources from the Sculptor
    photometric catalog \citep{wes06}.  Targets selected for
    spectroscopy are shown as bold, filled points.  Targets observed
    in more than one mask are shown as hollow points.  Gray diamonds
    enclose stars with previous HRS abundance measurements.  Each
    slitmask is labeled with the names given in Table~2.  The left and
    bottom axis scales show the angular displacement in arcmin from
    the center of the galaxy \citep[$\alpha_0 =
      1^{\rm{h}}00^{\rm{m}}09^{\rm{s}}$, $\delta_0 = -33^{\circ}42
      \farcm 5$,][]{mat98}, and the right and top axis scales show the
    projected physical distance for an assumed distance to Sculptor of
    85.9~kpc \citep{pie08}.  {\it Right:} Extinction- and
    reddening-corrected color-magnitude diagram in the Washington and
    Cousins systems for the photometric sources within the right
    ascension and declination ranges shown at left.  The
    transformation from the Washington system ($M$ and $T_2$) to the
    Cousins system ($V_{\rm C}$ and $I_{\rm C}$) is $I_{\rm C} = T_2$
    and $V_{\rm C} - I_{\rm C} = 0.800(M-T_2) - 0.006$ \citep{maj00}.
    (A color version of this figure is available in the online
    journal.)}\label{fig:scl}}
\end{figure*}

\begin{figure*}[hp!]
\centering
\includegraphics[width=0.46\textwidth]{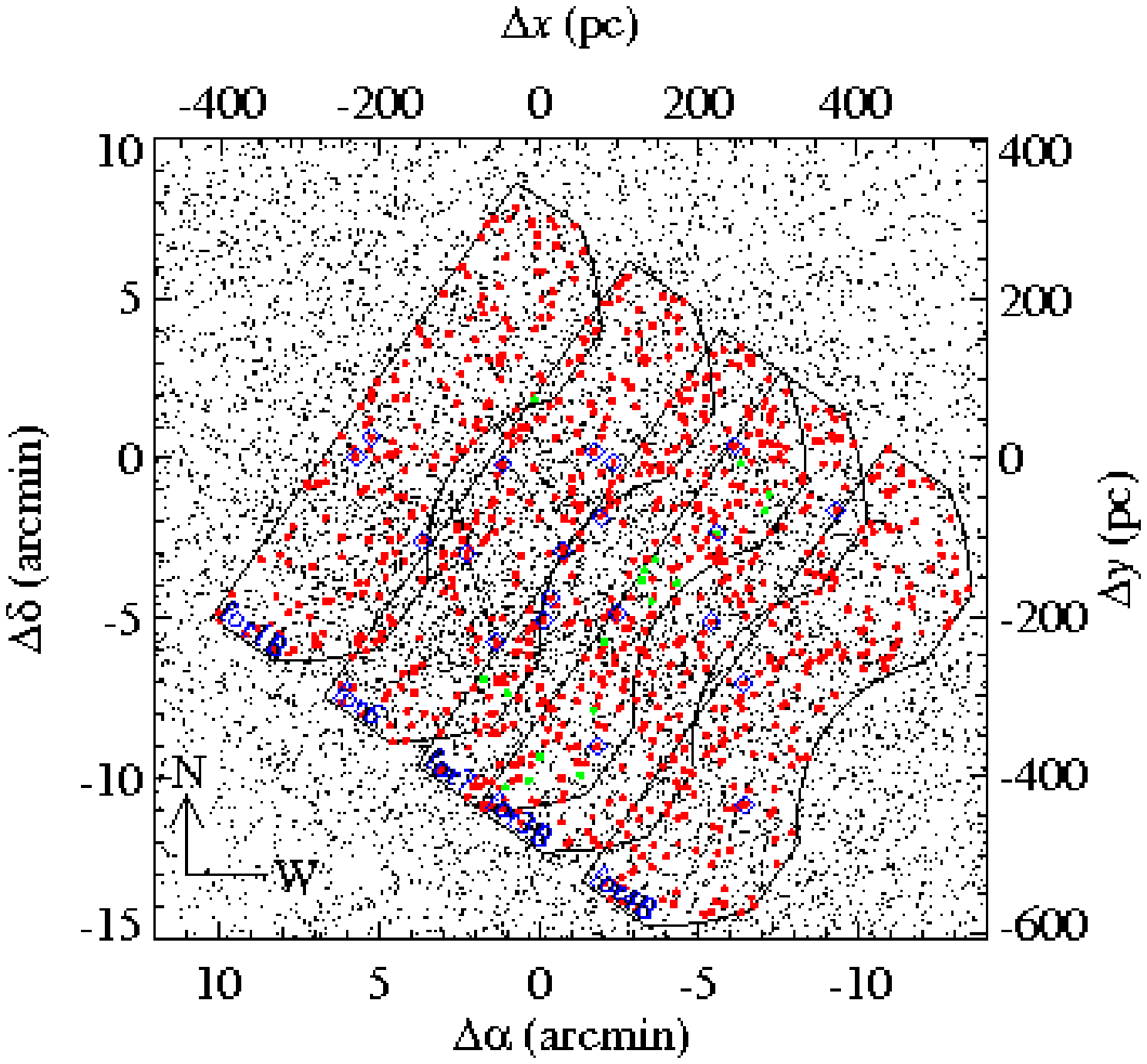}
\hfil
\includegraphics[width=0.46\textwidth]{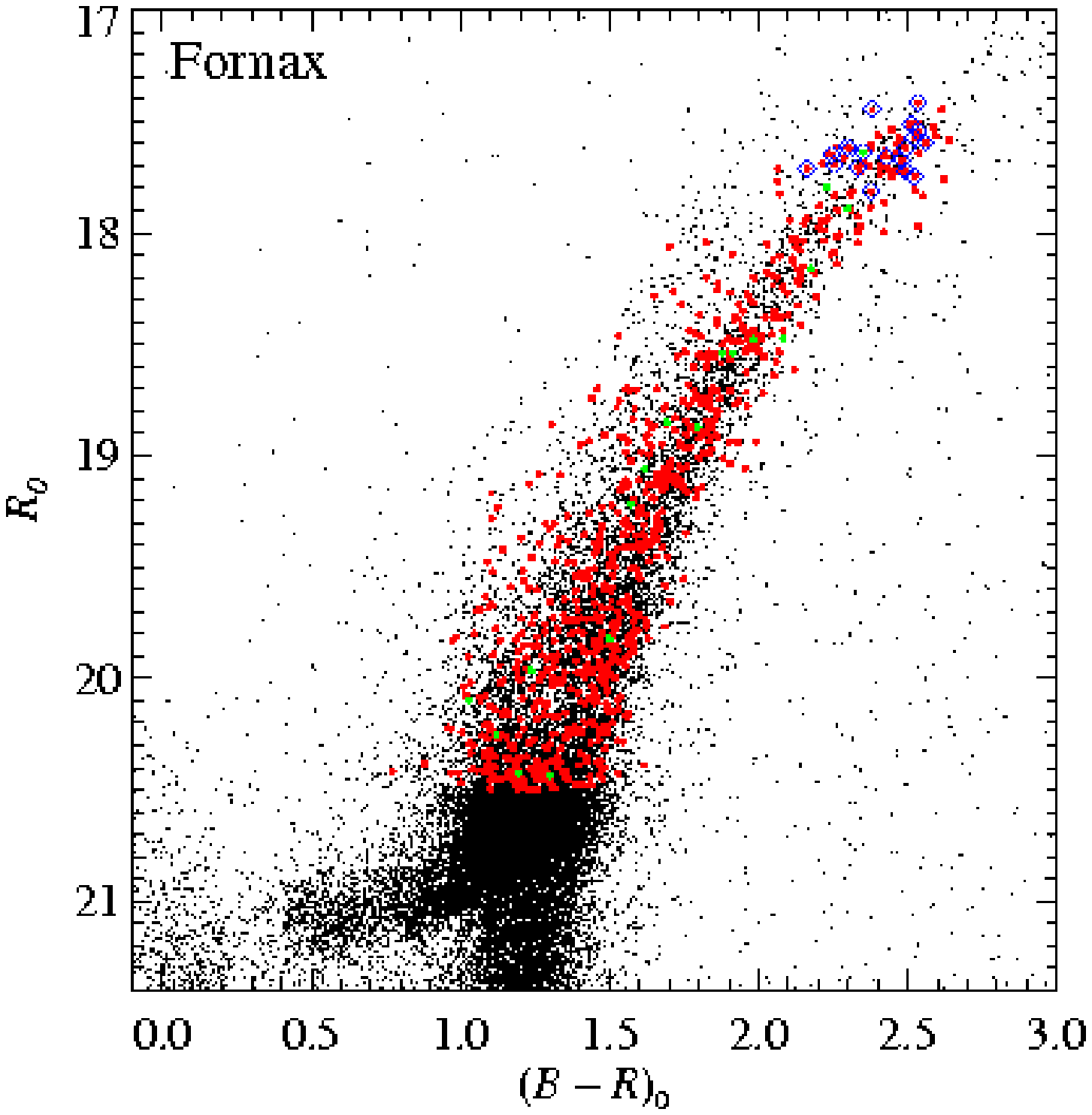}
\caption{{\bf Fornax.}  {\it Left:} DEIMOS slitmask footprints laid
  over a map of sources from the Fornax photometric catalog
  \citep{ste98} with magnitude $R_0 < 20.5$.  The center of the galaxy
  is $\alpha_0 = 2^{\rm{h}}39^{\rm{m}}59^{\rm{s}}$, $\delta_0 =
  -34^{\circ}27 \farcm 0$ \citep{mat98}, and the distance is 139~kpc
  \citep{riz07}.  {\it Right:} Color-magnitude diagram for the
  photometric sources within the right ascension and declination
  ranges shown at left.  See Fig.~\ref{fig:scl} for further
  explanation.\notetoeditor{B\&W figure caption: {\bf Fornax.}  {\it
      Left:} DEIMOS slitmask footprints laid over a map of sources
    from the Fornax photometric catalog \citep{ste98} with magnitude
    $R_0 < 20.5$.  The center of the galaxy is $\alpha_0 =
    2^{\rm{h}}39^{\rm{m}}59^{\rm{s}}$, $\delta_0 = -34^{\circ}27
    \farcm 0$ \citep{mat98}, and the distance is 139~kpc
    \citep{riz07}.  {\it Right:} Color-magnitude diagram for the
    photometric sources within the right ascension and declination
    ranges shown at left.  See Fig.~\ref{fig:scl} for further
    explanation.  (A color version of this figure is available in the
    online journal.)}\label{fig:for}}
\end{figure*}

\begin{figure*}[hp!]
\centering
\includegraphics[width=0.46\textwidth]{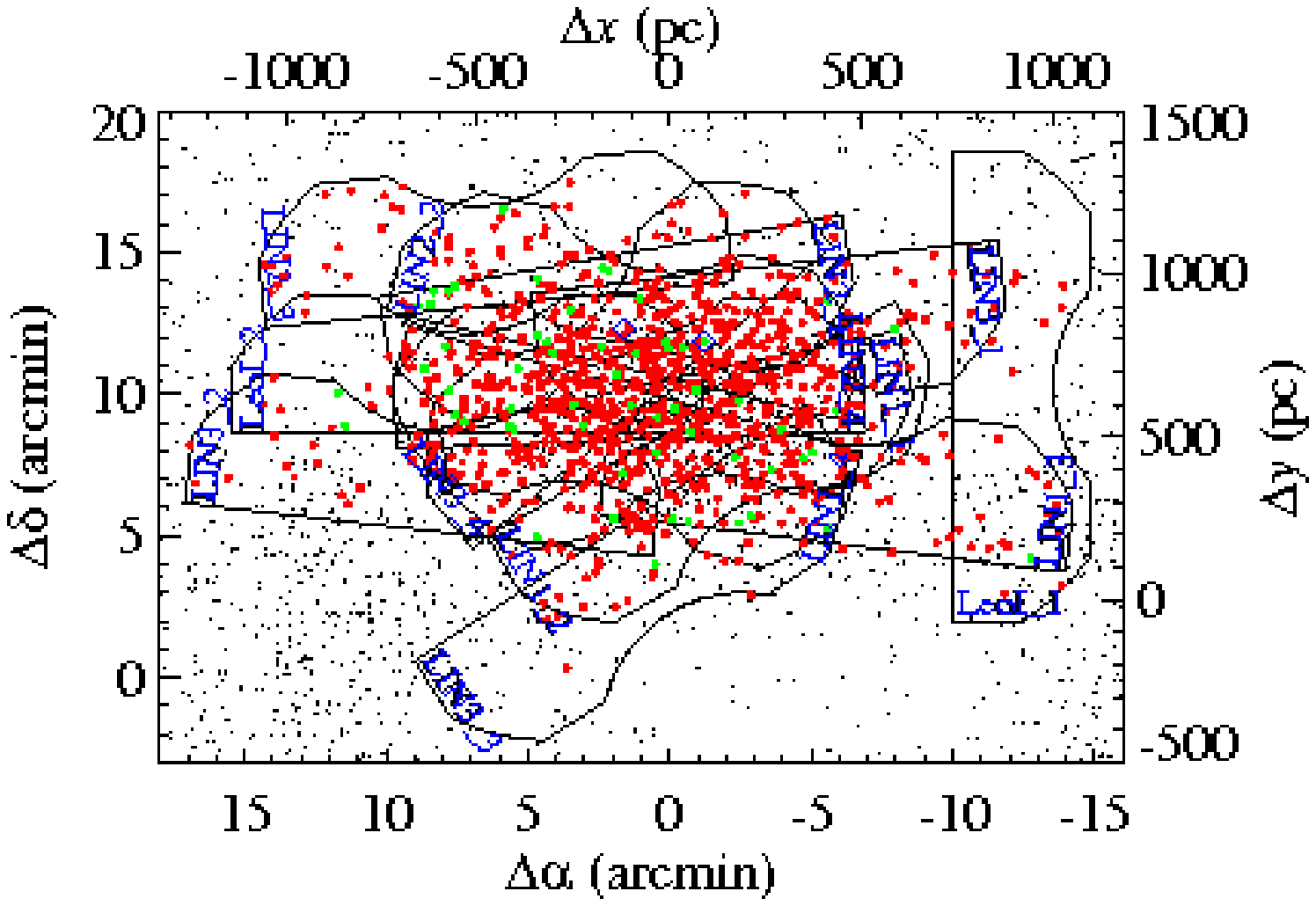}
\hfil
\includegraphics[width=0.46\textwidth]{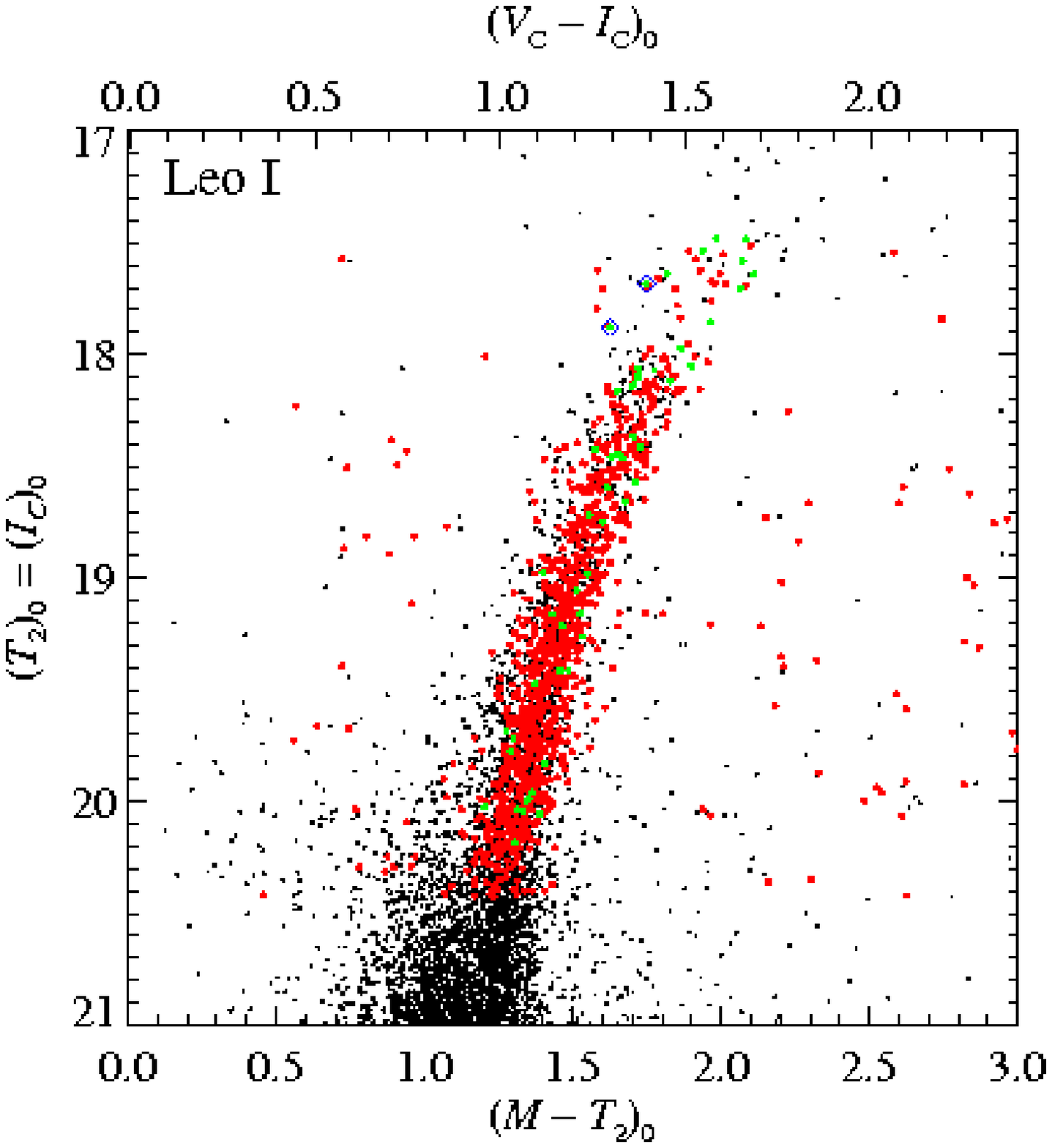}
\caption{{\bf Leo~I.}  {\it Left:} DEIMOS slitmask footprints laid
  over a map of sources from the NOMAD catalog \citep{zac04}, which is
  used only to show context of the sky area surrounding the DEIMOS
  targets.  The center of the galaxy is $\alpha_0 =
  10^{\rm{h}}08^{\rm{m}}27^{\rm{s}}$, $\delta_0 = 12^{\circ}08 \farcm
  5$ \citep{mat98}, and the distance is 254~kpc \citep{bel04}.  {\it
    Right:} Colors and magnitudes of the spectroscopic targets ({\it
    red points}).  Colors and magnitudes of the spectroscopically
  untargeted stars ({\it black points}) come from the catalog of
  \citet{bel04}, which is used only to show context in the
  color-magnitude diagram.  Only one of the two HRS targets was
  observed with DEIMOS.  See Fig.~\ref{fig:scl} for further
  explanation.\notetoeditor{B\&W figure caption: {\bf Leo~I.}  {\it
      Left:} DEIMOS slitmask footprints laid over a map of sources
    from the NOMAD catalog \citep{zac04}, which is used only to show
    context of the sky area surrounding the DEIMOS targets.  The
    center of the galaxy is $\alpha_0 =
    10^{\rm{h}}08^{\rm{m}}27^{\rm{s}}$, $\delta_0 = 12^{\circ}08
    \farcm 5$ \citep{mat98}, and the distance is 254~kpc
    \citep{bel04}.  {\it Right:} Colors and magnitudes of the
    spectroscopic targets ({\it bold, black points}).  Colors and
    magnitudes of the spectroscopically untargeted stars ({\it gray
      points}) come from the catalog of \citet{bel04}, which is used
    only to show context in the color-magnitude diagram.  Only one of
    the two HRS targets was observed with DEIMOS.  See
    Fig.~\ref{fig:scl} for further explanation.  (A color version of
    this figure is available in the online journal.)}\label{fig:leoi}}
\end{figure*}

\begin{figure*}[hp!]
\centering
\includegraphics[width=0.46\textwidth]{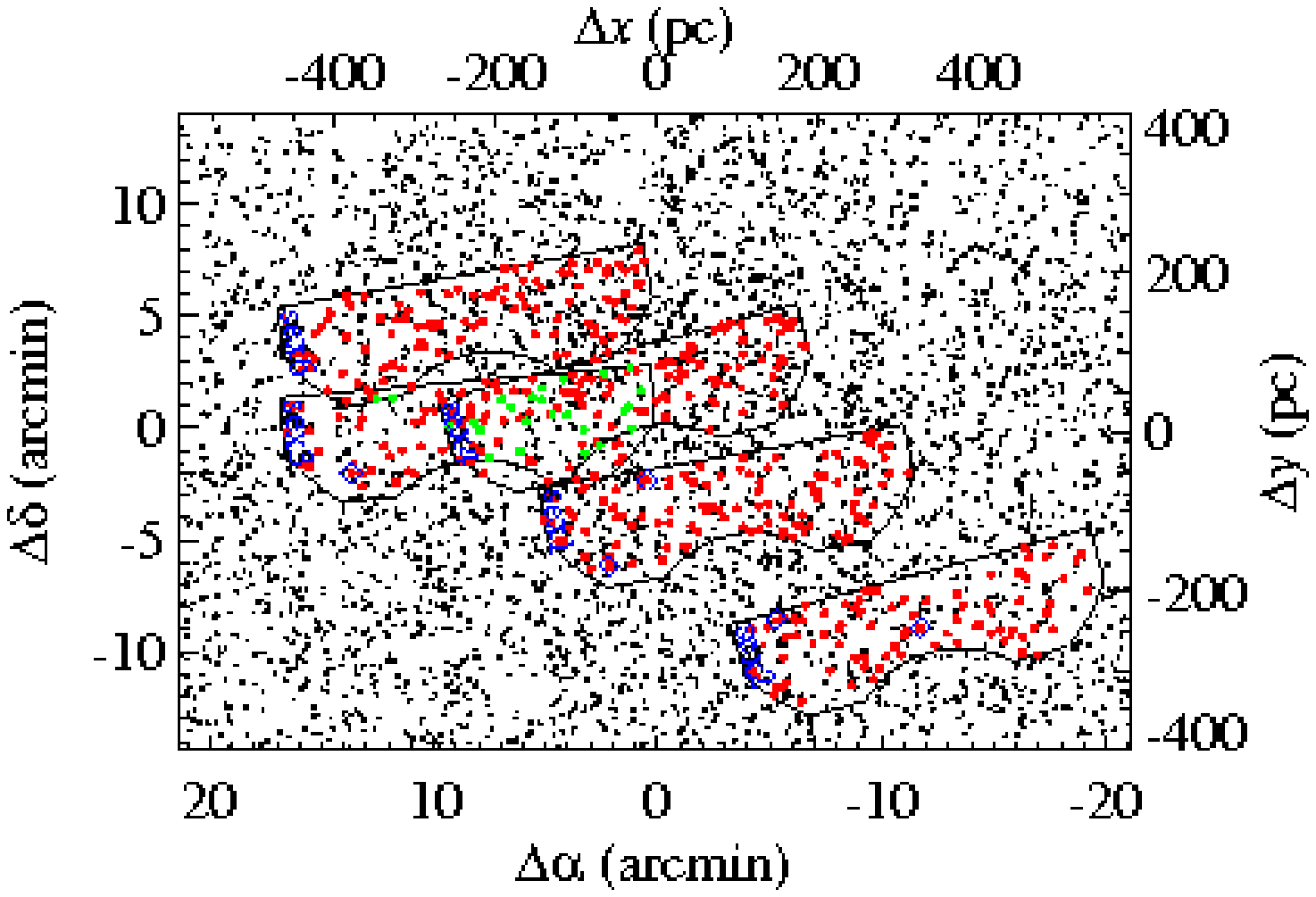}
\hfil
\includegraphics[width=0.46\textwidth]{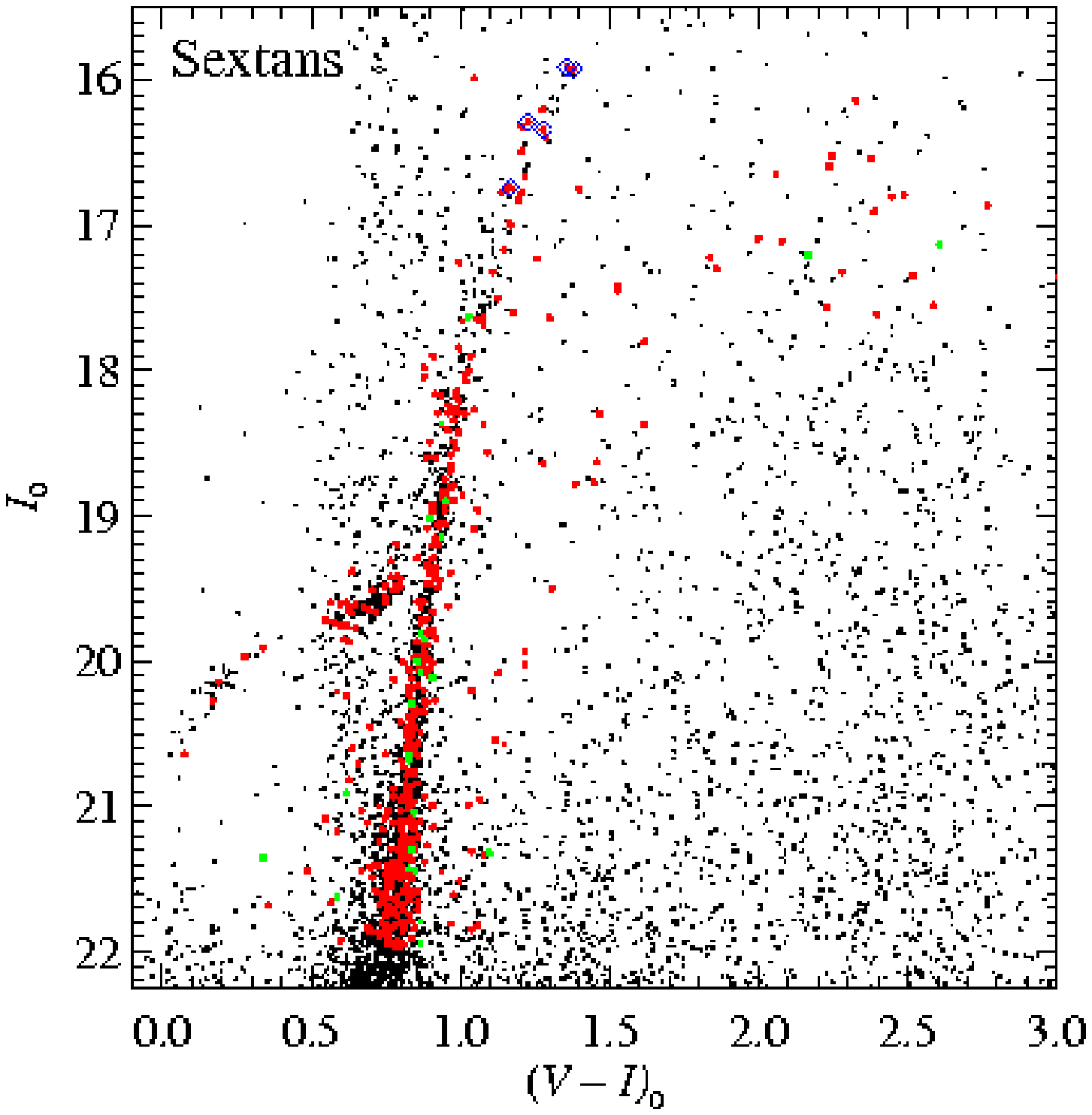}
\caption{{\bf Sextans.}  {\it Left:} DEIMOS slitmask footprints laid
  over a map of sources from the Sextans photometric catalog
  \citep{lee03} with magnitude $I < 22$.  The center of the galaxy is
  $\alpha_0 = 10^{\rm{h}}13^{\rm{m}}03^{\rm{s}}$, $\delta_0 =
  -01^{\circ}36 \farcm 9$ \citep{mat98}, and the distance is 95.5~kpc
  \citep{lee03}.  {\it Right:} Color-magnitude diagram for the
  photometric sources within the right ascension and declination
  ranges shown at left.  See Fig.~\ref{fig:scl} for further
  explanation.\notetoeditor{B\&W figure caption: {\bf Sextans.}  {\it
      Left:} DEIMOS slitmask footprints laid over a map of sources
    from the Sextans photometric catalog \citep{lee03} with magnitude
    $I < 22$.  The center of the galaxy is $\alpha_0 =
    10^{\rm{h}}13^{\rm{m}}03^{\rm{s}}$, $\delta_0 = -01^{\circ}36
    \farcm 9$ \citep{mat98}, and the distance is 95.5~kpc
    \citep{lee03}.  {\it Right:} Color-magnitude diagram for the
    photometric sources within the right ascension and declination
    ranges shown at left.  See Fig.~\ref{fig:scl} for further
    explanation.  (A color version of this figure is available in the
    online journal.)}\label{fig:sex}}
\end{figure*}

\begin{figure*}[hp!]
\centering
\includegraphics[width=0.46\textwidth]{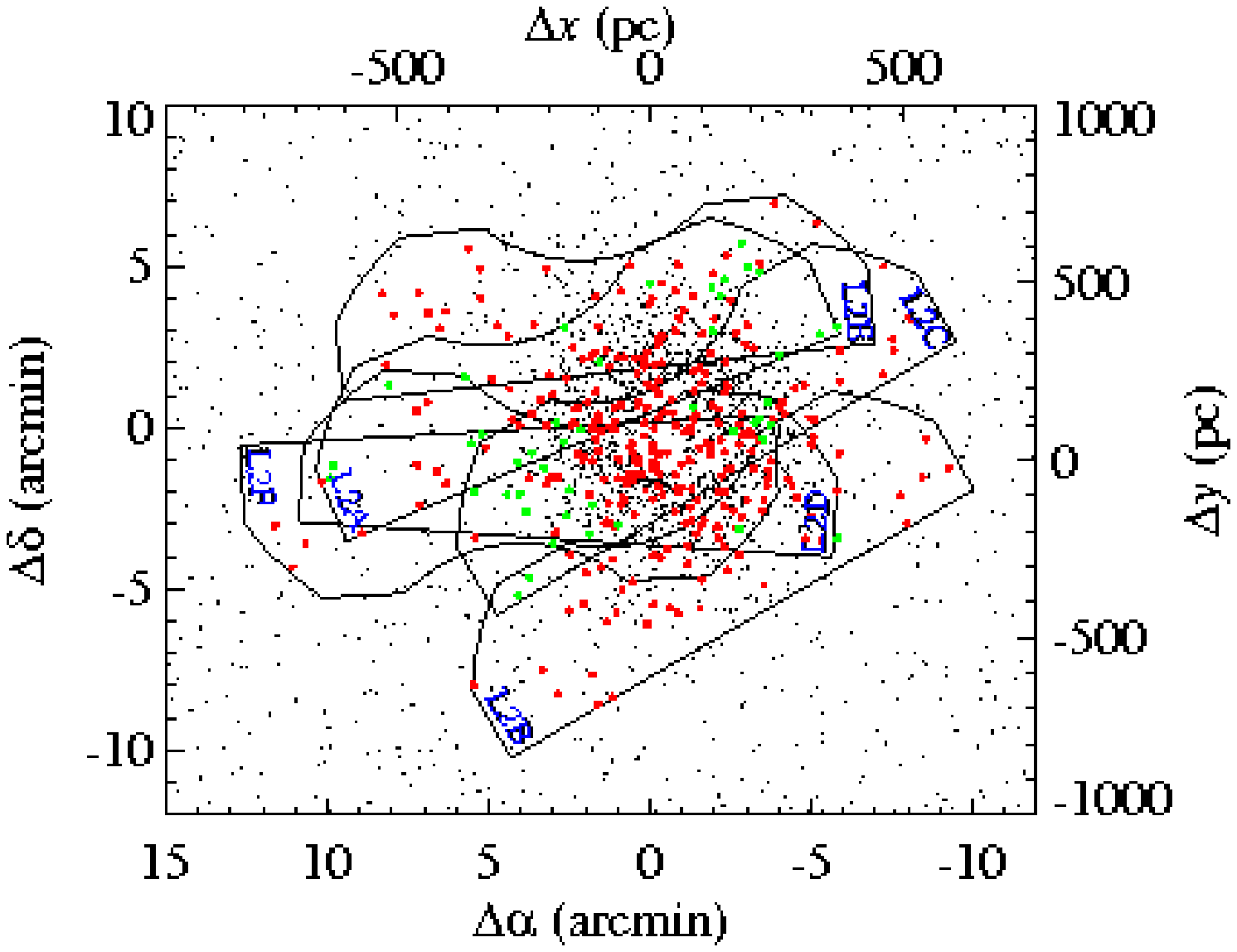}
\hfil
\includegraphics[width=0.46\textwidth]{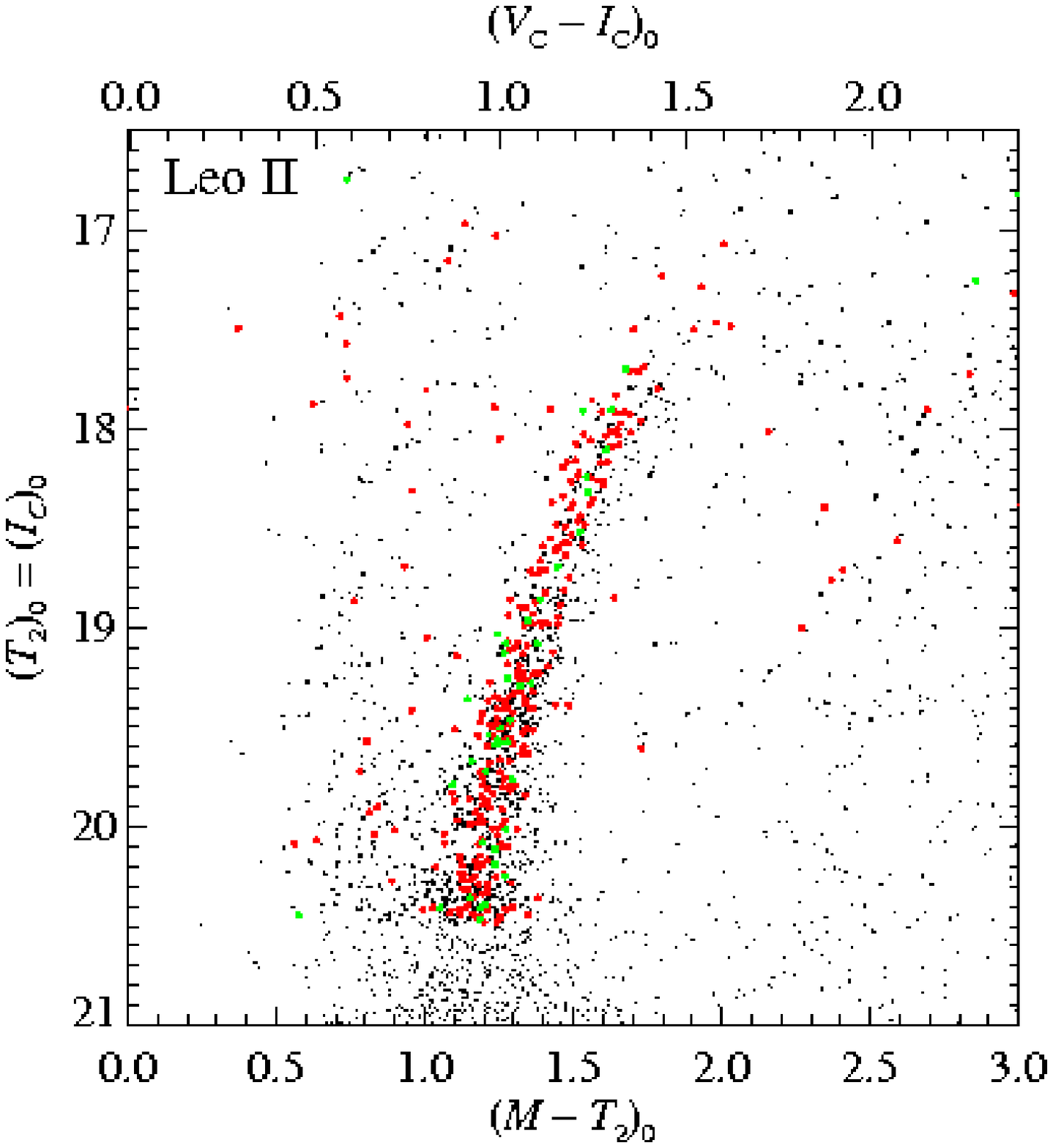}
\caption{{\bf Leo~II.}  {\it Left:} DEIMOS slitmask footprints laid
  over a map of stellar sources from the SDSS DR5 photometric catalog
  \citep{ade07}, which is used only to show context of the sky area
  surrounding the DEIMOS targets.  The center of the galaxy is
  $\alpha_0 = 11^{\rm{h}}13^{\rm{m}}29^{\rm{s}}$, $\delta_0 =
  +22^{\circ}09 \farcm 2$ \citep{mat98}, and the distance is 219~kpc
  \citep{sie10}.  {\it Right:} Colors and magnitudes of the
  spectroscopic targets ({\it red points}) from the photometric
  catalog of \citet{sie10}.  Colors and magnitudes of the
  spectroscopically untargeted stars ({\it black points}) come from
  the SDSS DR5 catalog, which is used only to show context in the
  color-magnitude diagram.  The SDSS $ugriz$ magnitudes have been
  roughly converted to the Washington $MT_2$ system, causing a shift
  between the red and black points.  See Fig.~\ref{fig:scl} for
  further explanation.\notetoeditor{B\&W figure caption: {\bf Leo~II.}
    {\it Left:} DEIMOS slitmask footprints laid over a map of stellar
    sources from the SDSS DR5 photometric catalog \citep{ade07}, which
    is used only to show context of the sky area surrounding the
    DEIMOS targets.  The center of the galaxy is $\alpha_0 =
    11^{\rm{h}}13^{\rm{m}}29^{\rm{s}}$, $\delta_0 = +22^{\circ}09
    \farcm 2$ \citep{mat98}, and the distance is 219~kpc
    \citep{sie10}.  {\it Right:} Colors and magnitudes of the
    spectroscopic targets ({\it bold, black points}) from the
    photometric catalog of \citet{sie10}.  Colors and magnitudes of
    the spectroscopically untargeted stars ({\it gray points}) come
    from the SDSS DR5 catalog, which is used only to show context in
    the color-magnitude diagram.  The SDSS $ugriz$ magnitudes have
    been roughly converted to the Washington $MT_2$ system, causing a
    shift between the black and gray points.  See Fig.~\ref{fig:scl}
    for further explanation.  (A color version of this figure is
    available in the online journal.)}\label{fig:leoii}}
\end{figure*}

\begin{figure*}[hp!]
\centering
\includegraphics[width=0.46\textwidth]{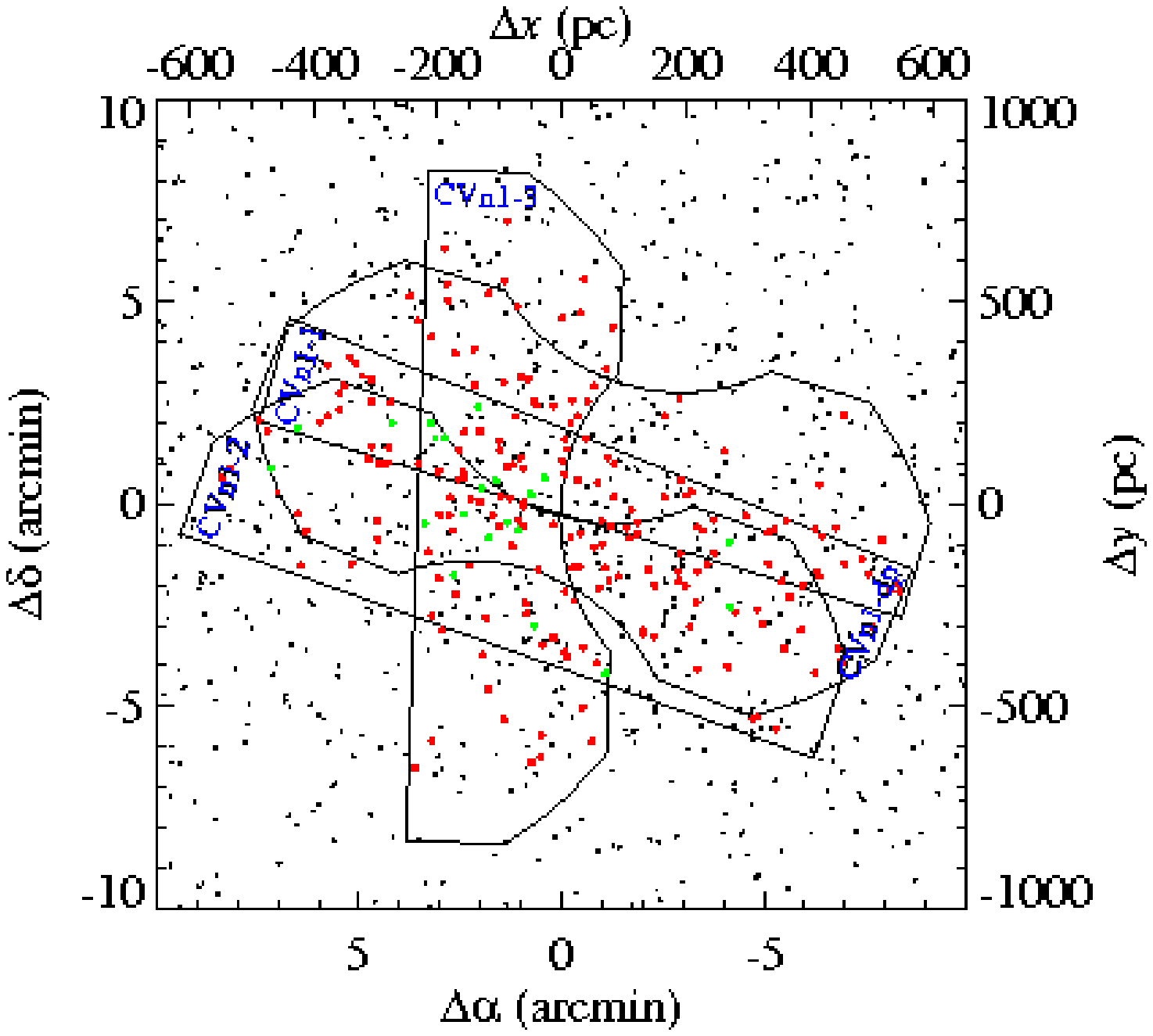}
\hfil
\includegraphics[width=0.46\textwidth]{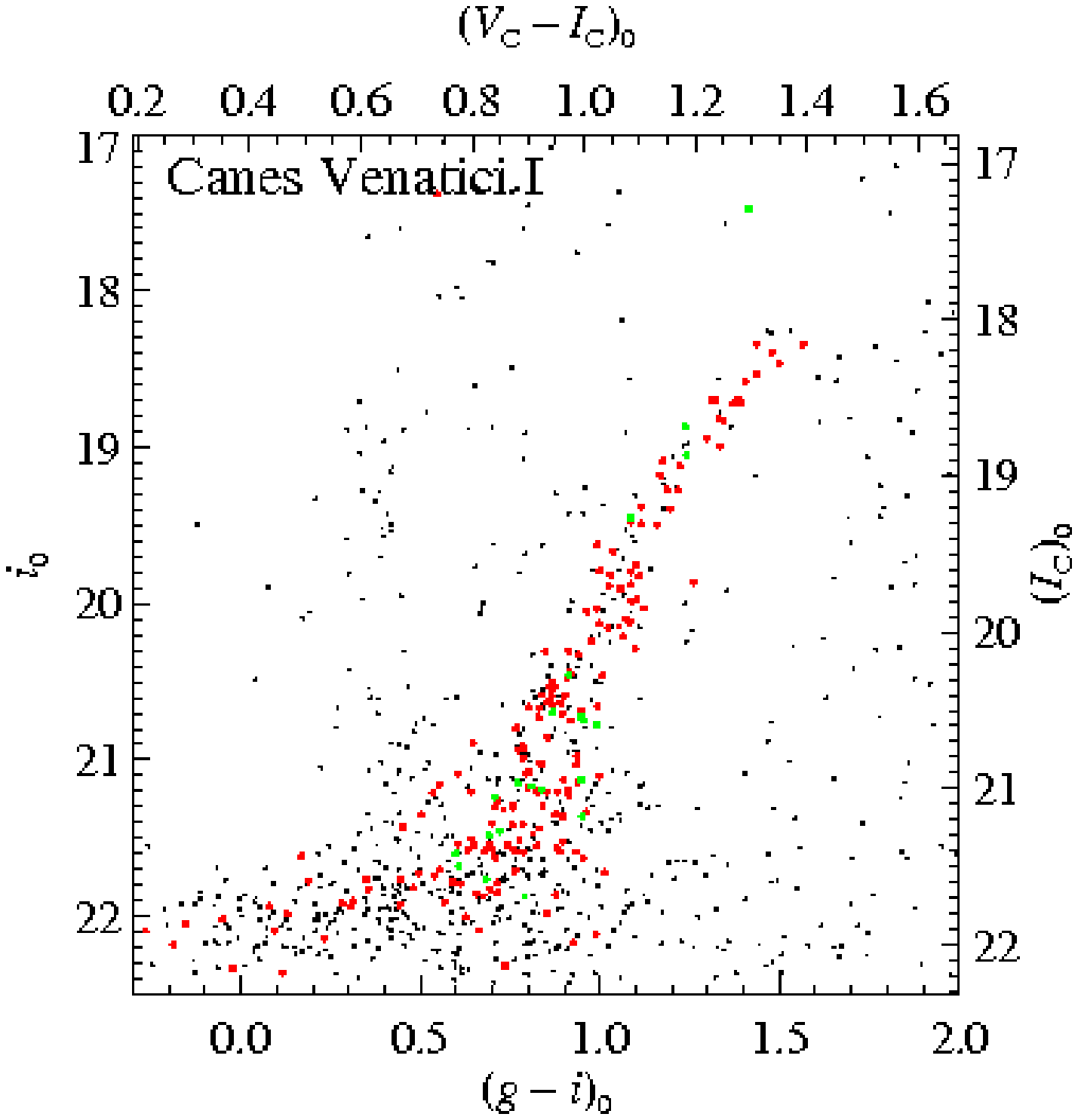}
\caption{{\bf Canes Venatici~I.}  {\it Left:} DEIMOS slitmask
  footprints laid over a map of stellar sources from the SDSS DR5
  photometric catalog \citep{ade07}.  The center of the galaxy is
  $\alpha_0 = 13^{\rm{h}}28^{\rm{m}}04^{\rm{s}}$, $\delta_0 =
  +33^{\circ}33' 21''$ \citep{zuc06}, and the distance is 210~kpc
  \citep{kue08}.  {\it Right:} Color-magnitude diagram for the
  photometric sources within the right ascension and declination
  ranges shown at left.  The top and left axes give rough Cousins $VI$
  magnitudes, assuming average colors on the RGB and following the
  transformations of \citet{jor06}.  See Fig.~\ref{fig:scl} for
  further explanation.\notetoeditor{B\&W figure caption: {\bf Canes
      Venatici~I.}  {\it Left:} DEIMOS slitmask footprints laid over a
    map of stellar sources from the SDSS DR5 photometric catalog
    \citep{ade07}.  The center of the galaxy is $\alpha_0 =
    13^{\rm{h}}28^{\rm{m}}04^{\rm{s}}$, $\delta_0 = +33^{\circ}33'
    21''$ \citep{zuc06}, and the distance is 224~kpc \citep{zuc06}.
    {\it Right:} Color-magnitude diagram for the photometric sources
    within the right ascension and declination ranges shown at left.
    The top and left axes give rough Cousins $VI$ magnitudes, assuming
    average colors on the RGB and following the transformations of
    \citet{jor06}.  See Fig.~\ref{fig:scl} for further explanation.
    (A color version of this figure is available in the online
    journal.)}\label{fig:cvni}}
\end{figure*}

\begin{figure*}[hp!]
\centering
\includegraphics[width=0.46\textwidth]{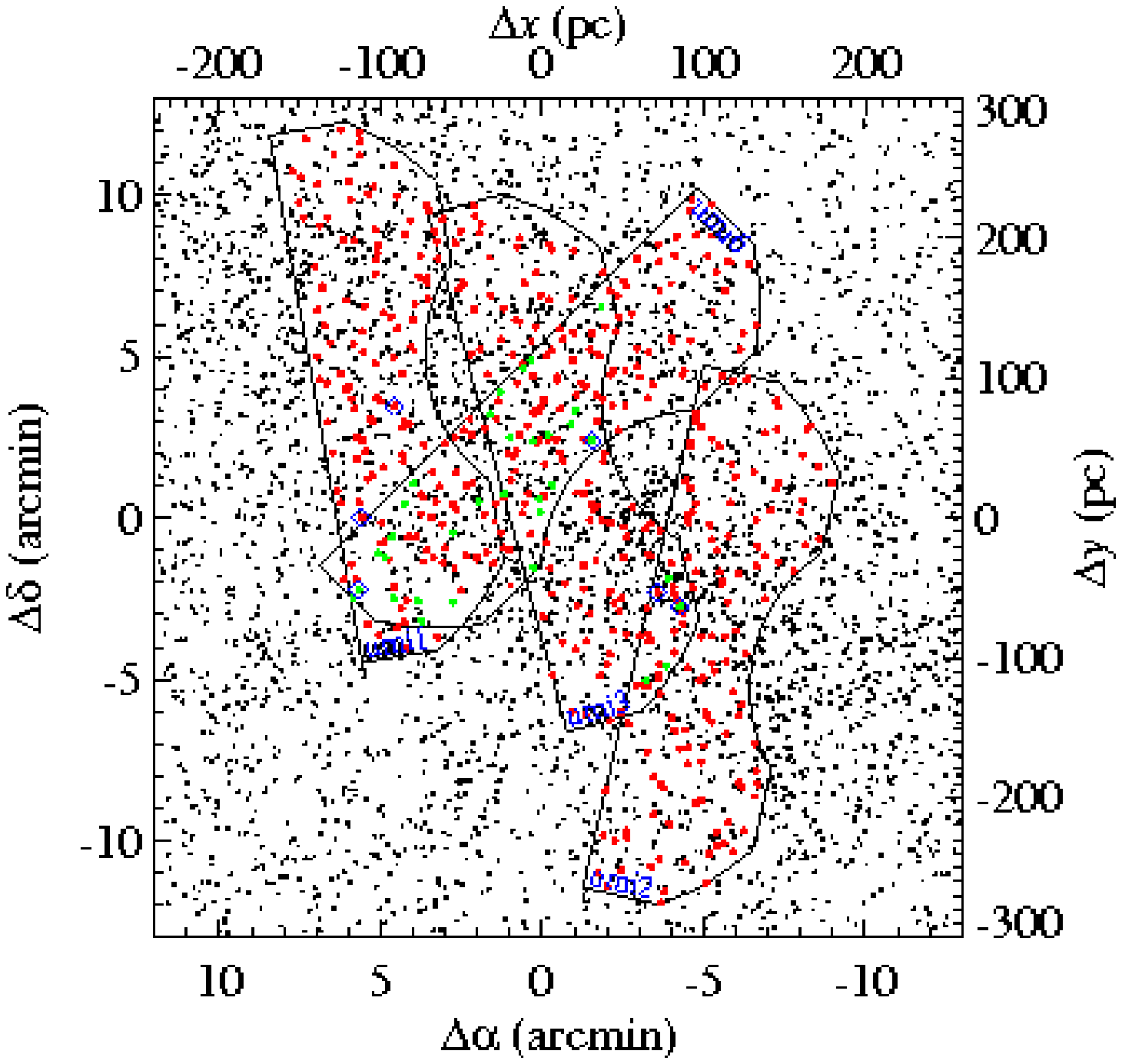}
\hfil
\includegraphics[width=0.46\textwidth]{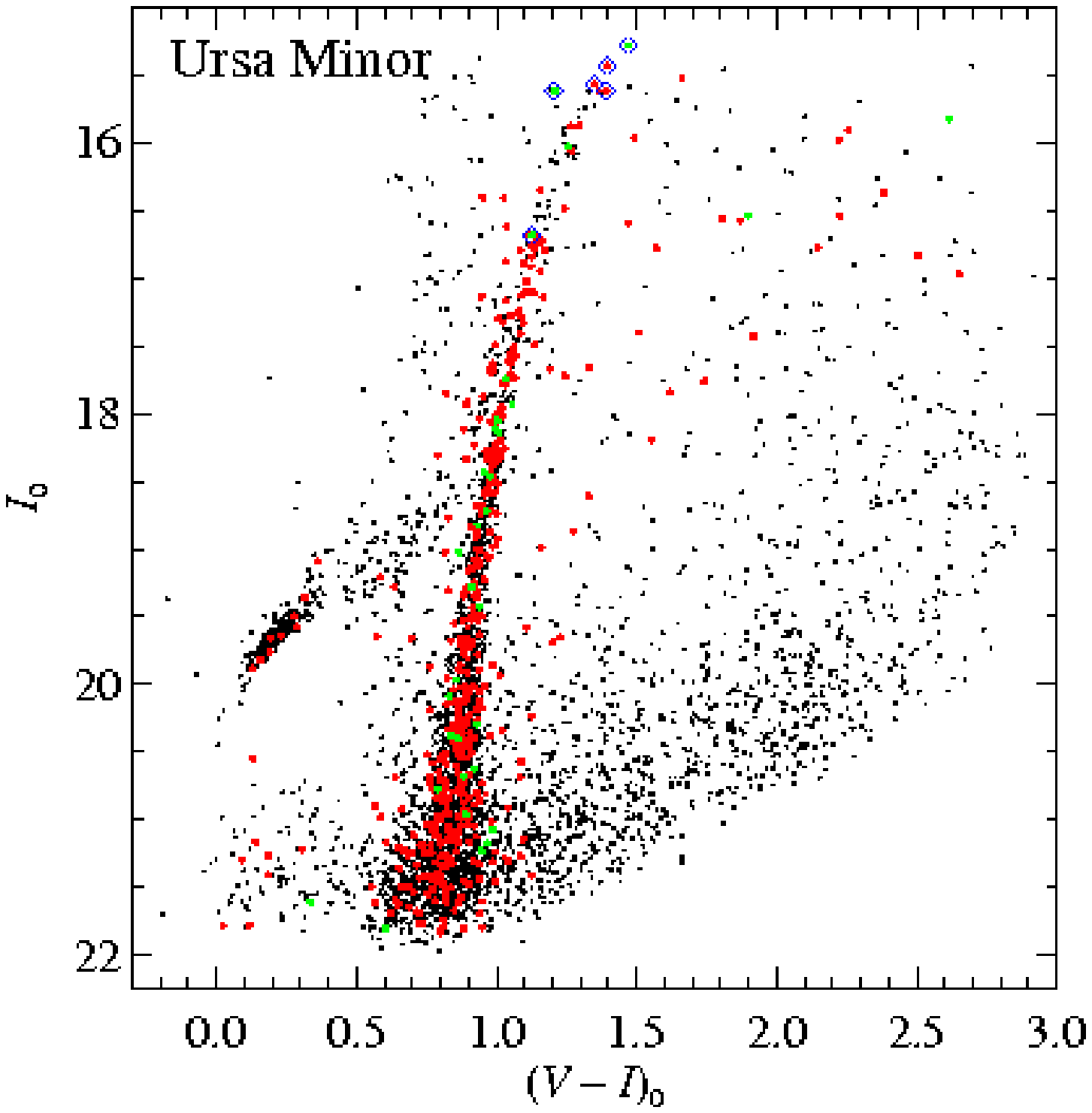}
\caption{{\bf Ursa Minor.}  {\it Left:} DEIMOS slitmask footprints
  laid over a map of sources from the Ursa Minor photometric catalog
  \citep{bel02}.  The center of the galaxy is $\alpha_0 =
  15^{\rm{h}}09^{\rm{m}}11^{\rm{s}}$, $\delta_0 = +67^{\circ}12 \farcm
  9$ \citep{mat98}, and the distance is 69~kpc \citep{mig99}.  {\it
    Right:} Color-magnitude diagram for the photometric catalog
  \citep{bel02} sources within the right ascension and declination
  ranges shown at left.  See Fig.~\ref{fig:scl} for further
  explanation.\notetoeditor{B\&W figure caption: {\bf Ursa Minor.}
    {\it Left:} DEIMOS slitmask footprints laid over a map of sources
    from the Ursa Minor photometric catalog \citep{bel02}.  The center
    of the galaxy is $\alpha_0 = 15^{\rm{h}}09^{\rm{m}}11^{\rm{s}}$,
    $\delta_0 = +67^{\circ}12 \farcm 9$ \citep{mat98}, and the
    distance is 69~kpc \citep{mig99}.  {\it Right:} Color-magnitude
    diagram for the photometric catalog \citep{bel02} sources within
    the right ascension and declination ranges shown at left.  See
    Fig.~\ref{fig:scl} for further explanation.  (A color version of
    this figure is available in the online journal.)}\label{fig:umi}}
\end{figure*}

\begin{figure*}[hp!]
\centering
\includegraphics[width=0.46\textwidth]{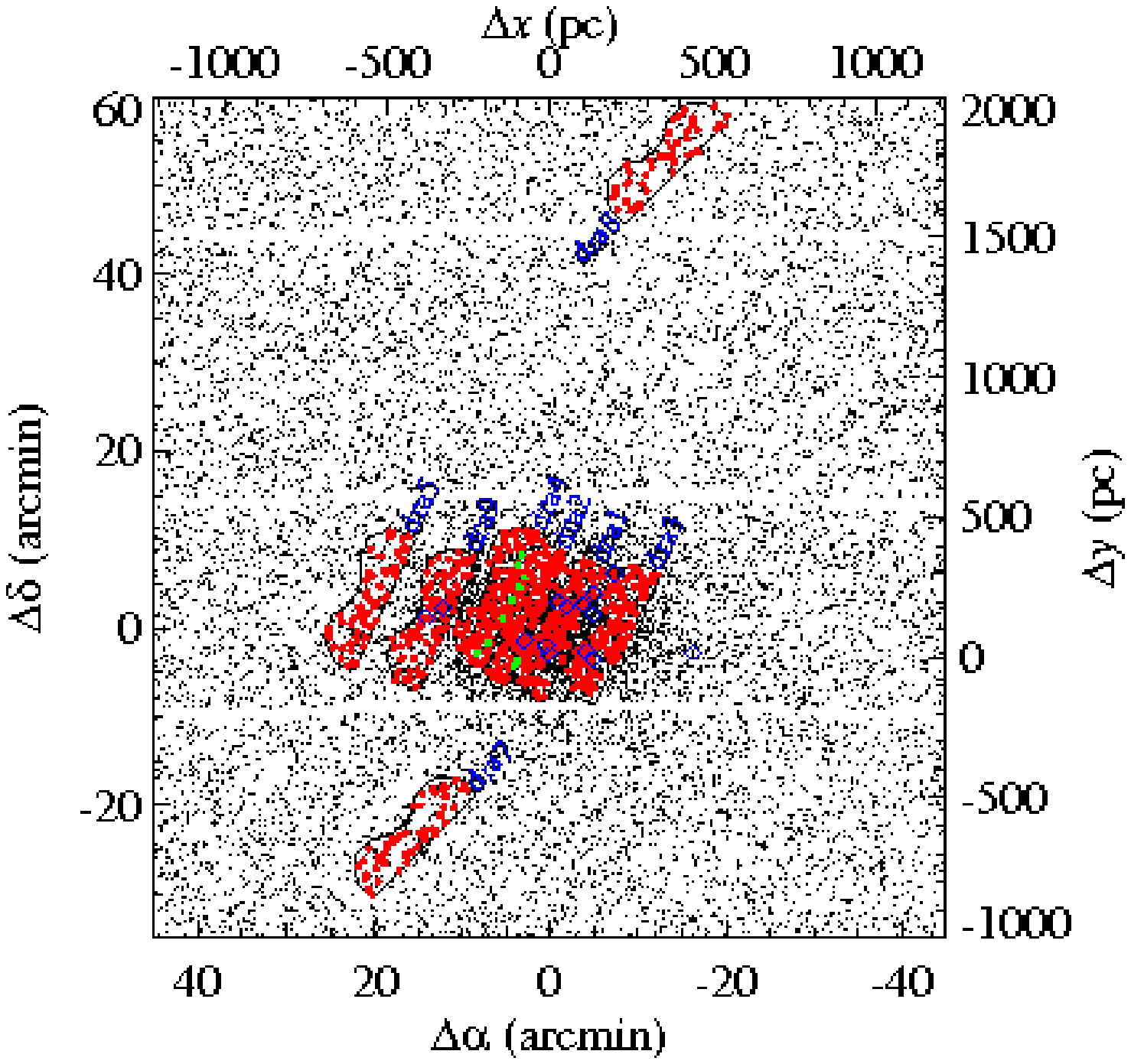}
\hfil
\includegraphics[width=0.46\textwidth]{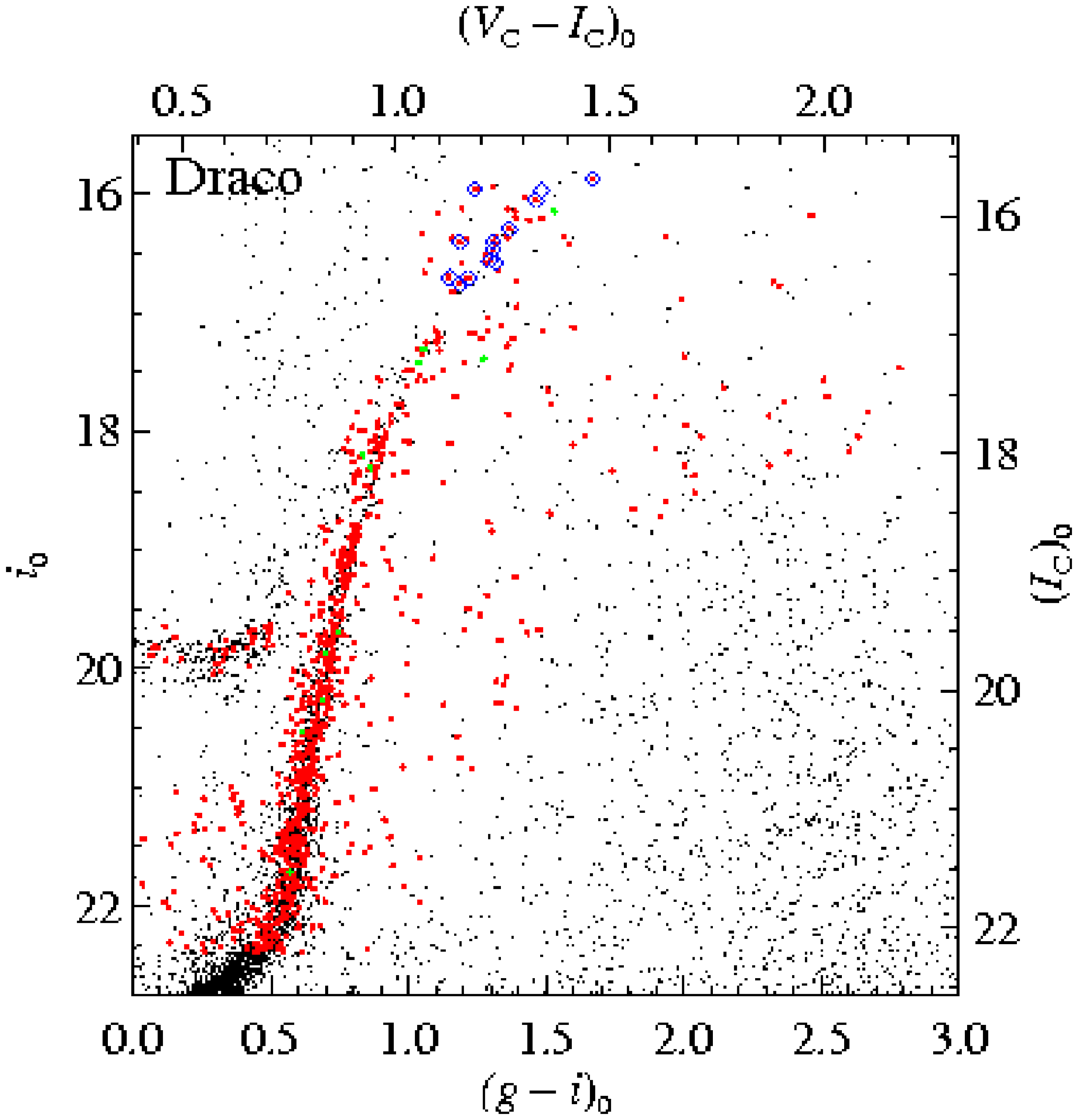}
\caption{{\bf Draco.}  {\it Left:} DEIMOS slitmask footprints laid
  over a map of sources with $i < 22.5$ and $g-i < 1.7$ from the Draco
  photometric catalog \citep{seg07}.  The center of the galaxy is
  $\alpha_0 = 17^{\rm{h}}20^{\rm{m}}19^{\rm{s}}$, $\delta_0 =
  +57^{\circ}54 \farcm 8$ \citep{mat98}, and the distance is 92.9~kpc
  \citep{bel02}.  {\it Right:} Color-magnitude diagram for the
  photometric sources within the outline of the slitmasks shown at
  left.  The top and left axes give rough Cousins $VI$ magnitudes,
  assuming average colors on the RGB and following the transformations
  of \citet{tuc06} and \citet{jor06}.  See Fig.~\ref{fig:scl} for
  further explanation.\notetoeditor{B\&W figure caption: {\bf Draco.}
    {\it Left:} DEIMOS slitmask footprints laid over a map of sources
    with $i < 22.5$ and $g-i < 1.7$ from the Draco photometric catalog
    \citep{seg07}.  The center of the galaxy is $\alpha_0 =
    17^{\rm{h}}20^{\rm{m}}19^{\rm{s}}$, $\delta_0 = +57^{\circ}54
    \farcm 8$ \citep{mat98}, and the distance is 92.9~kpc
    \citep{bel02}.  {\it Right:} Color-magnitude diagram for the
    photometric sources within the outline of the slitmasks shown at
    left.  The top and left axes give rough Cousins $VI$ magnitudes,
    assuming average colors on the RGB and following the
    transformations of \citet{tuc06} and \citet{jor06}.  See
    Fig.~\ref{fig:scl} for further explanation.  (A color version of
    this figure is available in the online journal.)}\label{fig:dra}}
\end{figure*}

\subsubsection{Sculptor}

\citeauthor*{kir09} describes in detail the target selection for stars
in the Sculptor dSph.  For convenience, we reproduce here the map of
targets on the sky and the Sculptor CMD (Fig.~\ref{fig:scl}).

\subsubsection{Fornax}
\label{sec:fornax}

We adopted the $BR$ photometric catalog of \citet{ste98}, and we
selected targets based on position in the CMD\@.  Magnitudes were
corrected star by star for extinction determined from the
\citet{sch98} dust maps.  (The same dust maps were not used for GCs
because GCs subtend a smaller solid angle than dSphs.  Therefore, we
used a single value for each GC rather than interpolation in the
\citeauthor{sch98}\ dust maps.)  Targets were drawn from seven
polygons surrounding the RGB and horizontal branch.  In order from
highest to lowest priority, the targets were selected (1) from $R_0
\le 18.3$ to the tip of the RGB, (2) $18.3 \le R_0 < 19.0$, (3) $19.0
\le R_0 < 19.5$, (4) $19.5 \le R_0 < 20.0$, and (5) $20.0 \le R_0 <
20.5$.  The red and blue edges of the polygons extended until the
stellar density reached the background level, with a typical width of
$\Delta(B-R)_0 = 0.9$ centered on the RGB.

Figure~\ref{fig:for} shows the Fornax field with the spectroscopic
targets highlighted in red.  To minimize confusion, only targets with
$R_0 < 20.5$ are plotted.  The five DEIMOS slitmasks are for1B, for3B,
for4B, for6, and for7 (see Table~2).  Each slitmask except for4B
included at least one of 18 duplicate targets included on other
slitmasks.

Figure~\ref{fig:for} also shows the CMD of the targets within the
right ascension and declination ranges of the axes in the left panel
of Fig.~\ref{fig:for}.  Some stars on the extreme blue end of the RGB
did not pass the spectroscopic selection criteria.  These stars may be
asymptotic giant branch (AGB) stars, or they may be extremely young or
extremely metal-poor red giants.  This potential selection bias should
be kept in mind when considering the derived metallicity distribution
of Fornax.

Twenty stars, listed in Table~\ref{tab:hrscompare}, have previously
published HRS abundance measurements \citep{she03,let10}, and all 20
were observed.  One of these stars, M12 (not shown in the right panel
of Fig.~\ref{fig:for}), was absent from the $BR$ photometric catalog.
For calculation of photometric temperature and surface gravity, we
adopted the same extinction-corrected $VI$ magnitudes used by
\citet{she03}.

\subsubsection{Leo~I}
\label{sec:leoi}

We used \citeauthor{soh07}'s (\citeyear{soh07}) DEIMOS spectra of
individual stars in Leo~I and $MT_2$ magnitudes of the spectroscopic
targets.  We converted $M$ and $T_2$ to Cousins $V$ and $I$ magnitudes
in the same manner as for the Sculptor photometric catalog.
Magnitudes were corrected star by star for extinction determined from
the \citet{sch98} dust maps.  For other details on the photometric
catalog and spectroscopic target selection, we refer to
\citeauthor{soh07}'s published paper on the DEIMOS slitmasks LeoI\_1
and LeoI\_2. 
slitmasks.

Figure~\ref{fig:leoi} shows the sky chart and CMD for spectroscopic
targets.  The black points in the right panel of the figure show other
stars from the $VI$ catalog of \citet{bel04} for context, but the
spectroscopic targets were not drawn from this catalog.  \citet{she03}
measured spectroscopic abundances of two stars in Leo~I.  One of
these, M5, was observed on three different DEIMOS slitmasks.

\subsubsection{Sextans}

We selected spectroscopic targets from the deep, wide-field $BVI$
catalog of \citet{lee03}.  They assumed a constant reddening of
$E(B-V) = 0.01$, and we corrected all of the magnitudes and colors
accordingly.  We selected RGB members by overlaying Yonsei-Yale
isochrones \citep{dem04} between 2 and 14~Gyr, $\mathfeh = -3.76$ and
$+0.05$, and $\mathafe = 0.0$ and $+0.3$ on the CMD\@.  All stars that
lay between the bluest and reddest of these isochrones within
photometric errors were considered for spectroscopic selection.  An
additional 0.05~mag was allowed on the blue edge of the bluest
isochrones to account for possible systematic error in the isochrones
which might have excluded extremely metal-poor stars.  When forced to
choose between multiple RGB candidates, we selected the brightest one.

Figure~\ref{fig:sex} shows the Sextans field with the spectroscopic
targets highlighted in red.  Only targets with $I < 22$ are plotted to
minimize confusion.  The sex3 slitmask contained 25 targets also
included on other slitmasks.

Figure~\ref{fig:sex} also shows the CMD of the targets within the
right ascension and declination ranges of the axes in the left panel
of Fig.~\ref{fig:sex}.  Some extremely red stars were targeted in
order to fill the slitmask with targets.  Although these stars are
unlikely to be Sextans members, they were included because they could
potentially be metal-rich RGB stars in Sextans.  Additionally, Sextans
is near enough to permit a significant number of horizontal branch
stellar spectra, although the spectroscopic abundance measurement
technique used here does not yet work for horizontal branch stars.
Five stars, listed in Table~\ref{tab:hrscompare}, have previously
published HRS abundance measurements \citep{she01a}, and all five were
observed.  \citet{aok09} published measurements of an additional six
stars at high resolution after we designed the DEIMOS slitmasks.
These stars were not included in our slitmasks.

\subsubsection{Leo~II}

DEIMOS observations of Leo~II were conducted in the same program as
observations of Leo~I.  The photometry and spectroscopy were treated
identically.  For additional information, see Sec.~\ref{sec:leoi}.
Figure~\ref{fig:leoii} shows the spectroscopic target selection.

\subsubsection{Canes Venatici~I}

We used \citeauthor{sim07}'s (\citeyear{sim07}) DEIMOS spectra for
individual stars in Canes Venatici~I.  Photometry and extinction
corrections were taken from the SDSS Data Release 5 \citep{ade07}.  In
order to use the same isochrones for determining effective
temperatures and surface gravities as we used for other galaxies, we
converted SDSS $ugriz$ to Johnson-Cousins $UBVRI$ following the
global, metallicity-independent transformations of
\citet{jor06}.\footnote{\citet{kir08b} chose the conversions given by
  \citet{cho08}.  However, \citeauthor{cho08} suggest that their
  purpose was to derive photometric zero points, not to find relations
  valid for astrophysical sources.  Therefore, we relied on
  \citeauthor{jor06}'s (\citeyear{jor06}) transformations, which are
  valid for astrophysical sources.}  We refer to \citeauthor{sim07}'s
article for the details of the spectroscopic target selection, shown
in Figure~\ref{fig:cvni}.  No stars in Canes Venatici~I have been
observed yet at high spectral resolution.

\subsubsection{Ursa Minor}

We selected spectroscopic targets from the $VI$ catalog of
\citet{bel02}.  Magnitudes were corrected star by star for extinction
determined from the \citet{sch98} dust maps.  We selected RGB members
by following the same procedure as for Sextans, except that we allowed
an additional 0.1~mag on the blue side of the RGB instead of 0.05~mag
because the extra contamination was negligible.

Figure~\ref{fig:umi} shows the Ursa Minor field with the spectroscopic
targets highlighted in red.  Every slitmask contained several of the
33 targets also included on other slitmasks.

Figure~\ref{fig:umi} also shows the CMD of the targets within the
right ascension and declination ranges of the axes in the left panel
of Fig.~\ref{fig:umi}.  The discussion of the Sextans CMD similarly
applies to the Ursa Minor CMD\@.  Six stars, listed in
Table~\ref{tab:hrscompare}, have previously published HRS abundance
measurements \citep{she01a,sad04}, and all six were observed in our
study.


\subsubsection{Draco}
\label{sec:dra}

\citet{seg07} devoted an extremely wide-field photometric survey to
Draco.  They observed Draco with three different telescopes, but we
used only their $gri$ catalog from the CFHT MegaCam Camera for its
large field of view.  In a manner similar to the target selection for
Sextans, we chose targets from the $(g,\ g-i)$ CMD between the bluest
and reddest Padova isochrones \citep{gir02} between 2 and 14~Gyr,
$\mathfeh = -2.23$ and $+0.05$, and $\mathafe = 0.0$ and $+0.3$.  We
allowed an additional 0.05~mag on the blue side of the CMD to account
for stars more metal-poor than the most metal-poor Padova isochrone
($\mathfeh = -2.23$).

Figure~\ref{fig:dra} shows the large field on which eight Draco
slitmasks were placed.  Only stars with $i < 22.5$ and $g-i < 1.7$ are
shown to minimize confusion.  The dra7 and dra8 slitmasks were located
on the periphery of the field to search for Draco members close to the
tidal radius.  The dra2 slitmask included 11 targets also observed on
the dra1 and dra4 slitmasks.

Figure~\ref{fig:dra} also shows the CMD of targets that lay within the
outlines of the slitmasks shown in the left panel of
Fig.~\ref{fig:dra}.  The discussion of the Sextans CMD similarly
applies to the Draco CMD\@.  Fourteen stars in Draco have been
observed with high-resolution spectroscopy
\citep{she98,she01a,ful04,coh09}.  We observed 12 of these.  They are
listed in Table~\ref{tab:hrscompare}.

We wished to use the same isochrones to determine effective
temperature and surface gravities that we used for other galaxies.
\citet{cle08} showed that CFHT $gri$ magnitudes are indistinguishable
from the USNO standard $g'r'i'$ magnitudes.  Therefore, we transformed
CFHT $gri$ magnitudes to SDSS $gri$ magnitudes following the
prescription of \citet[]{tuc06}.  Then, we transformed SDSS $gri$
magnitudes to Johnson-Cousins $BVRI$ magnitudes \citep{jor06}.
\citet{reg09} give color transformations from $BVRI$ to CFHT $griz$
but not absolute zero points for that conversion.  Therefore, we used
the two-step transformation.  Because the CFHT and SDSS filter
transmission curves are very similar, the first transformation from
the CFHT system to the SDSS system introduced negligible error.

\subsubsection{Additional Galaxies}

The galaxies Ursa Major~II, Leo~IV, Coma Berenices, and Hercules are
not included in the catalog presented here (Table~\ref{tab:catalog}).
However, some stars in these four galaxies have been observed both
with HRS and MRS.  Like Canes Venatici~I, these galaxies were observed
with DEIMOS by \citet{sim07}.  \citet{fre10b} observed at
high resolution three stars each in Ursa Major~II and Coma Berenices;
\citet{sim10} observed at high resolution one star in Leo~IV; and
\citet{koc08} observed at high resolution two stars in Hercules, but
only one overlaps our sample.  We used these stars in the comparison
between HRS and MRS abundance measurements
(Sec.~\ref{sec:hrscompare}).

\subsection{Slitmask Design}

We designed the DEIMOS slitmasks with the
\texttt{dsimulator}\footnote{\url{http://www.ucolick.org/~phillips/deimos_ref/masks.html}}
IRAF\footnote{IRAF is distributed by the National Optical Astronomy
  Observatories, which are operated by the Association of Universities
  for Research in Astronomy, Inc., under cooperative agreement with
  the National Science Foundation.} software module.  Each slitmask
subtended approximately $16' \times 5'$.  In order to subtract night
sky emission lines adequately, we required a minimum slit length of
$4''$.  The minimum space between slits was $0 \farcs 35$.  Although
this spacing is less than the typical seeing FWHM, light contamination
between slitlets was negligible.  Because slits had the freedom to be
placed along the dispersion axis of the slitmask, the end of a slit
was rarely located within 1\arcsec\ of a neighboring slit.  Even in
those rare instances, the stars were located far enough from the ends
of their slits that light spillage into a neighboring slit did not
affect the spectral extraction or sky subtraction.

When the slitmask design constraints forced the selection of one among
multiple possible red giant candidates, we invoked priorities
explained in Secs.~\ref{sec:gcobs}--\ref{sec:dra} (usually the
brightest candidate).  For most dSph slitmasks, the slitmasks were
designed to be observed at approximately the parallactic angle at the
anticipated time of observation.  This choice minimized the small
light losses due to differential atmospheric refraction.  Most
slitmasks' sky position angle (PA) was offset by $\sim 10^{\circ}$
from the slit PA.  The resulting tilt of the night sky emission lines
relative to the CCD pixel grid increased the subpixel wavelength
sampling and improved sky subtraction.  The dSph slitmasks contained
many duplicate observations, which provide estimates of uncertainties
in abundance measurements (Sec.~\ref{sec:duplicate}).

The spectral coverage of each slit is not the same.  The minimum and
maximum wavelengths of spectra of targets near the long, straight edge
of the DEIMOS footprint can be up to 400~\AA\ lower than for targets
near the irregularly shaped edge of the footprint (upper left and
lower right of the slitmask footprints in Fig.~\ref{fig:scl},
respectively).  Furthermore, spectra of targets near either extreme of
the long axis of the slitmask suffered from vignetting, which reduced
the spectral range.  It is important to keep these differences of
spectral range in mind when interpreting the differences of
measurements derived from duplicate observations.

\subsection{Spectroscopic Configuration and Exposures}
\label{sec:exposures}

\begin{figure*}[ht!]
\centering
\includegraphics[width=\textwidth]{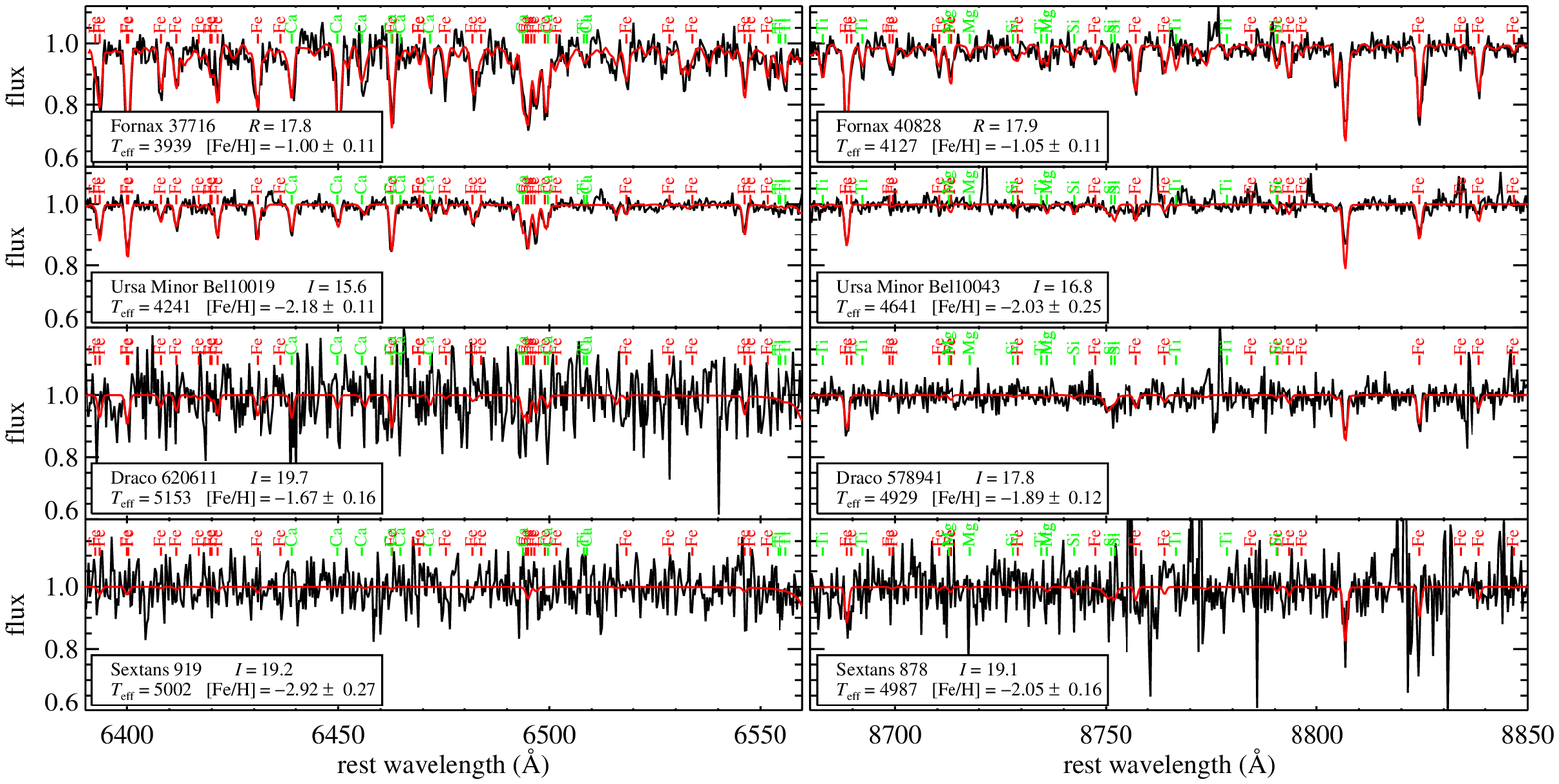}
\caption{Examples of DEIMOS spectra of eight different dSph stars in
  two spectral regions.  Variations in the lower and upper wavelengths
  from star to star prohibit showing both spectral regions for the
  same star.  The average total spectral range (2800~\AA) is about
  eight times larger than shown here.  The figure caption gives the
  dSph name, star name, magnitude, and best-fit \teff\ and \feh.  The
  best-fit synthetic spectrum is overplotted in red.  Some of the
  absorption lines (from Table~\ref{tab:linelist}) used in the
  abundance measurements are labeled.\notetoeditor{B\&W figure
    caption: Examples of DEIMOS spectra of eight different dSph stars
    in two spectral regions.  Variations in the lower and upper
    wavelengths from star to star prohibit showing both spectral
    regions for the same star.  The average total spectral range
    (2800~\AA) is about eight times larger than shown here.  The
    figure caption gives the dSph name, star name, magnitude, and
    best-fit \teff\ and \feh.  The best-fit synthetic spectrum is
    overplotted in gray.  Some of the absorption lines (from
    Table~\ref{tab:linelist}) used in the abundance measurements are
    labeled.  (A color version of this figure is available in the
    online journal.)}\label{fig:snexamples}}
\end{figure*}

Our observing strategy was nearly identical to that of \citet{guh06}
and \citet{sim07}.  In summary, we used the 1200 lines~mm$^{-1}$
grating at a central wavelength of $\sim 7800$~\AA.  The slit widths
were $0\farcs 7$ (except for $1\farcs 0$ slit widths for M79,
NGC~2419, and Leo~I and II), yielding a spectral resolution of $\sim
1.2$~\AA\ FWHM (resolving power $R \sim 7000$ at 8500~\AA).  The OG550
filter blocked diffraction orders higher than $m=1$.  The spectral
range was about 6400--9000~\AA\ with variation depending on the slit's
location along the dispersion axis.  Exposures of Kr, Ne, Ar, and Xe
arc lamps provided wavelength calibration, and exposures of a quartz
lamp provided flat fielding.  Table~2 lists the slitmasks observed,
the number of targets on each slitmask, the dates of observations, the
airmass, the approximate seeing (if known), and the exposure times.
For the multi-slit masks, seeing was estimated by measuring the FWHM
of a Gaussian fit to the one-dimensional profiles of alignment stars.


Because halo stars are not clustered on the sky, only one star could
be observed in each exposure.  Therefore, these observations did not
require custom slitmasks.  Instead, the spectra were obtained through
longslits.  The instrument configuration for the longslit exposures
was identical to the configuration for custom multi-slit masks.  To
estimate the seeing, we measured the FWHM from the guider camera of
the star used to the focus the telescope.  All but two of the spectra
of halo field stars presented here were obtained during nights of poor
transparency in 2008 Apr and May.  Star VII-18 in the GC M92 was also
observed in this manner.

The raw frames were reduced into one-dimensional spectra using version
1.1.4 of \texttt{spec2d}, the DEIMOS data reduction pipeline developed
by the DEEP Galaxy Redshift
Survey\footnote{\url{http://astro.berkeley.edu/~cooper/deep/spec2d/}}.
In addition to the procedure described by \citet{guh06}, KGS08, and in
\citeauthor*{kir09}, we tweaked the arc lamp wavelength solution by
using the pipeline's optional \texttt{SKYTWEAK\_1D} procedure.  The
procedure evaluates small wavelength shifts between the observed and
expected wavelengths of night sky emission lines in 100~\AA\ regions.
A polynomial fit to the individual shifts adjusts the wavelength array
of the observed spectrum.  For any given pixel, a typical wavelength
shift is 0.1~\AA.

The Fornax slitmasks for1B, for6, and for7 (see Table~2) required
special attention.  Because exposures for the same masks were taken
months apart, the heliocentric velocity corrections changed enough
that frames from the different months could not be stacked without
affecting spectral line widths significantly.  In order to overcome
the velocity shift, we extracted one set of 1-D spectra for each
night.  The wavelength arrays for the second epoch were shifted by the
difference in the heliocentric velocity correction between the two
nights.  The velocity-shifted spectra were rebinned to match the
wavelength arrays of the spectra from the first epoch.  Then, the 1-D
spectra were coadded.  The rest of the analysis proceeded in the same
way as for the slitmasks observed on single nights.

Figure~\ref{fig:snexamples} shows small spectral regions for eight
different dSph stars.  The stars were chosen to show a range of
signal-to-noise ratio (S/N) and metallicity.  Some absorption lines
used in the abundance determination are labeled.  Some prominent lines
are not labeled because they are not included in the abundance
measurements.  For example, \ion{Mg}{1}~$\lambda 8807$ and its blended
neighbor, \ion{Fe}{1}~$\lambda 8805$, are not labeled because the Mg
line is too strong for a local thermodynamic equilibrium abundance
measurement.

\subsection{Membership}

We determined membership in each dSph by the radial velocities of the
stars.  (Table~\ref{tab:catalog} does not include radial velocities.
Geha et al\@. will publish the radial velocities in a separate work.)
In order to measure radial velocities, we cross-correlated each
one-dimensional spectrum with several different stellar templates
\citep[see][]{sim07}.  We adopted the velocity of the
cross-correlation peak of the best-fitting template.  Then, we fit a
Gaussian to the velocity distribution of each dSph.  Every star within
$\pm 3\sigma$ of the peak (99.7\% of the true members, assuming that
they are normally distributed in velocity) was considered a member.
Stars outside of this range were discarded, and they are not included
in the catalog in Sec.~\ref{sec:catalog}.  Finally, horizontal branch
stars and stars obviously not belonging the red giant branch of the
dSph were excluded.  These steps reduce contamination by foreground
and background MW stars, but it is impossible to eliminate
contamination entirely.  The small remaining contamination may
influence the interpretation of the abundance measurements, but it
does not our affect our estimation of their accuracy and precision
(Sec.~\ref{sec:accuracy}).  The number of unique member stars across
all eight dSphs is \ndsphstars.


\section{Abundance Measurements}
\label{sec:measurements}

KGS08 and \citeauthor*{kir09} discuss the details of the MRS technique
used to create the catalog in this article.  Here we present only a
summary and some modifications to the method.  Other than the changes
described in Sections~\ref{sec:grid}--\ref{sec:offsets}, the details
of the procedure remain identical to the method of KGS08 with the
modifications in \citeauthor*{kir09}.

\subsection{Summary of the Technique}
\label{sec:techniquesummary}

\addtocounter{table}{1}

\begin{deluxetable}{lccc}
\tablewidth{0pt}
\tablecolumns{4}
\tablecaption{Important Atomic Spectral Lines\label{tab:linelist}}
\tablehead{\colhead{Species} & \colhead{Wavelength (\AA)} & \colhead{Excitation Potential (eV)} & \colhead{$\log (gf)$}} \\
\startdata
\smion{Mg}{I} & 6319.237 & 5.110 & $-2.150$ \\
\smion{Mg}{I} & 6319.495 & 5.108 & $-2.630$ \\
\smion{Mg}{I} & 7691.553 & 5.753 & $-0.800$ \\
\smion{Mg}{I} & 7811.133 & 5.946 & $-1.550$ \\
\smion{Mg}{I} & 7811.141 & 5.946 & $-1.550$ \\
\smion{Mg}{I} & 8047.720 & 5.932 & $-1.970$ \\
\smion{Mg}{I} & 8098.719 & 5.946 & $-1.120$ \\
\smion{Mg}{I} & 8098.727 & 5.946 & $-1.450$ \\
\smion{Mg}{I} & 8213.034 & 5.753 & $-0.509$ \\
\smion{Mg}{I} & 8346.106 & 5.946 & $-1.080$ \\
\enddata
\tablecomments{Table~\ref{tab:linelist} is published in its entirety
  in the electronic edition of the Astrophysical Journal.  This is not
  a complete line list.  It is a subset of Table~4 of KGS08.}
\end{deluxetable}

Each observed spectrum was compared to a large grid of synthetic
spectra at a variety of effective temperatures (\teff), surface
gravities (\logg), metallicities (represented by \feh, but all
elements heavier than He were modulated), and alpha enhancements
(\afe, an additional modulation for Mg, Si, Ca, and Ti).
Section~\ref{sec:grid} gives additional details on the grid.  The
spectra were synthesized with the local thermodynamic equilibrium,
plane-parallel spectrum synthesis code MOOG \citep{sne73}, ATLAS9
model atmospheres \citep{kur93,sbo04,sbo05}, and atomic and molecular
transition data from the Vienna Atomic Line Database
\citep[VALD,][]{kup99}, modified as described by KGS08.
Table~\ref{tab:linelist} gives a subset of the line list for the lines
visible in a high S/N DEIMOS spectrum of a red giant with $\mathfeh
\sim -0.5$, such as Arcturus.  A Levenberg-Marquardt algorithm (the
IDL code \texttt{MPFIT} by \citeauthor{mark09} \citeyear{mark09})
minimized $\chi^2$ to find the best-fitting synthetic spectrum in
several steps.  In the first step, \teff, \feh, and \afe\ were
determined iteratively.  Surface gravity was fixed by photometry.
Only regions sensitive to Fe absorption were used to determine Fe and
only regions sensitive to Mg, Si, Ca, or Ti absorption were used to
determine \afe\ (see \citeauthor*{kir09}).  The observed spectrum's
continuum was refined between each iteration by removing a high-order
spline fit to the quotient of the observed spectrum and the
best-fitting synthetic spectrum of the previous iteration.  The
temperature was allowed to vary, but within the bounds of a
photometrically determined temperature based on available photometry.
(The spectroscopic refinement of the temperature leads to more
accurate abundances when compared to HRS measurements.  Because lines
of ionized species are sparse in these far-red spectra of red giants,
spectroscopic refinement of surface gravity is not possible.)  In the
next steps, the individual Mg, Si, Ca, and Ti abundances were
determined by fitting the spectral regions sensitive to absorption
from each element.

Special consideration was made for the MW halo field star sample.
Because the distances to the halo field stars are poorly known,
surface gravities could not be fixed by photometry.  However, every
star in the MW halo field star sample was chosen because previous
authors had already observed it with HRS.  We adopted the
high-resolution measurements of surface gravities for all of the MW
halo field stars.


\subsection{Expansion of the Spectral Grid}
\label{sec:grid}

In \citeauthor*{kir09}, we announced the discovery of one star in
Sculptor at $\mathfeh = -3.80$, which \citet*{fre10a} confirmed with
the Magellan/MIKE high-resolution spectrograph.  The lower \feh\ limit
of the spectral grid of KGS08 and \citeauthor*{kir09} was $\mathfeh =
-4.0$.  If any star reached the lower \feh\ limit, we discarded it
under the presumption that the S/N was inadequate to yield an
\feh\ measurement of useful precision.  However, the confirmed
existence of a star very close to the grid limit prompted us to expand
the spectral grid to $\mathfeh = -5.0$.  Although we have not
recovered any ultra metal-poor stars ($-5 < \mathfeh < -4$), the
expansion gives us confidence that we have not discarded any such
stars because of a limitation in the grid.

Expanding the grid required computing model atmospheres at $-5 <
\mathfeh < -4$ with \teff, \logg, and \afe\ on the grid described by
Table~3 of \citeauthor*{kir09}.  For each value of \teff\ and
\logg\ in that table, we computed ODFNEW opacity distribution
functions \citep{cas04} using R.~L.~Kurucz's DFSYNTHE code
\citep[described by][]{cas05} at $\mathfeh = -4.5$ and $-5.0$ for each
of 21 steps in \afe\ (0.1~dex between $\mathafe = -1.2$ and $+0.8$)
and 2 steps in microturbulent velocity ($\xi$).  The values of $\xi$
were chosen to be the two velocities that bracket the microturbulent
velocity appropriate for the star's surface gravity (Eq.~2 of
\citeauthor*{kir09}).  The values of \feh\ and \xfe\ in the
computation of the opacity distribution functions are given relative
to the solar abundances of \citet{and89}, except that the abundance of
iron is $12 + \log \epsilon (\mathrm{Fe}) = 7.52$ \citep[see][for an
  explanation]{sne92}.  We computed the model atmospheres with
\citeauthor{kur93}'s (\citeyear{kur93}) ATLAS9 code ported into Linux
\citep{sbo04,sbo05}.  Convective overshooting was turned off, and the
mixing length parameter for convection was $l/H_p = 1.25$.  These
convection parameters are the same as for \citeauthor{cas04}'s
(\citeyear{cas04}) grid of ATLAS9 atmospheres.  We linearly
interpolated the atmospheres in the \feh\ and $\xi$ dimensions to
populate the grid described in \citeauthor*{kir09}.

\subsection{Correction to Continuum Division}
\label{sec:continuum}

Following \citet{she09}, we modified the method of KGS08 to perform
better continuum determination in \citeauthor*{kir09}.  We iteratively
determined the continuum by fitting a high-order spline to the
quotient of the observed spectrum and the best-fitting synthetic
spectrum of the previous iteration.  In \citeauthor*{kir09}, we
incorporated this technique into the Levenberg-Marquardt algorithm
that determines the best-fitting synthetic spectrum.  During each
Levenberg-Marquardt iteration, any remaining fluctuations in the
continuum level were removed before the $\chi^2$ between the observed
spectrum and a trial synthetic spectrum was evaluated.  Instead, we
decided that it was more appropriate to determine the best-fitting
synthetic spectrum without modifying the observed spectrum in the
minimization of $\chi^2$.  The continuum refinement occurred between
separate Levenberg-Marquardt determinations of the best-fitting
synthetic spectrum.  The continuum refinement was considered converged
when \feh\ and the bulk \afe\ changed by less than 0.001~dex and
\teff\ changed by less than 1~K.

\subsection{Sigma Clipping}
\label{sec:sigmaclip}

After the continuum refinement, we added a new step not used by KGS08
or in \citeauthor*{kir09}: masking pixels whose absolute difference
from the best-fit synthetic spectrum exceeded 2.5 times their
variance.  The remaining steps to measure \feh\ were performed on the
clipped spectrum.  The sigma clipping excludes spectral regions with
artifacts, such as improperly removed cosmic rays.  It also insulates
the abundance measurements from Fe absorption lines with incorrect
oscillator strengths.  The comparison between MRS and HRS abundances
(Sec.~\ref{sec:hrscompare}) improved slightly as a result of including
sigma clipping.  We experimented with the threshold value, and we
found that $2.5\sigma$ included $\ga 90\%$ of the pixels while
reducing the mean and standard deviations of the differences between
MRS and HRS abundances.

\subsection{New Approach for [$\alpha$/Fe]}
\label{sec:alphafe}

Previously (\citeauthor*{kir09}), we measured [Mg/Fe], [Si/Fe],
[Ca/Fe], and [Ti/Fe] by searching the spectral grid at fixed,
previously determined \teff\ and \feh.  Because each element was not a
separate dimension, abundances of all of the $\alpha$ elements (O, Ne,
Mg, Si, S, Ar, Ca, and Ti) varied together.  Individual elements were
measured by isolating the spectral regions containing those lines.
This resulted in a different value of \afe\ for the model atmospheres
used in the measurement of each element.  However, the parameter
\afe\ controls not only the strength of $\alpha$ element lines, but
also important atmospheric quantities, such as the free electron
fraction.  The resulting changes in the continuous opacity in the red
spectral region affect absorption lines of all elements.  For red
giants, increasing the atmospheric value of \afe\ at fixed
\feh\ raises the continuous opacity by increasing the number density
of H$^{-}$ ions.  Metal lines become weaker.

In the present catalog, we have determined the individual $\alpha$
element ratios by fixing the atmospheric value of \afe, which we call
${\rm [\alpha/Fe]}_{\rm atm}$.  We synthesized a subgrid of spectra
with an additional dimension, ${\rm [\alpha/Fe]}_{\rm abund}$.  The
new dimension is the \afe\ ratio for the abundances of Mg, Si, Ca, and
Ti at fixed ${\rm [\alpha/Fe]}_{\rm atm}$.  Like ${\rm
  [\alpha/Fe]}_{\rm atm}$, ${\rm [\alpha/Fe]}_{\rm abund}$ is spaced
at 0.1~dex from $-0.8$ to $+1.2$.  It was not necessary to synthesize
spectra with ${\rm [\alpha/Fe]}_{\rm atm} = {\rm [\alpha/Fe]}_{\rm
  abund}$ because the existing grid already contained such spectra.
To save computation time, the new subgrid included only the spectral
regions used in the measurement of Mg, Si, Ca, and Ti.

We determined ${\rm [\alpha/Fe]}_{\rm atm}$ by fitting to the Mg, Si,
Ca, and Ti lines simultaneously.  Then, the values of [Mg/Fe],
[Si/Fe], [Ca/Fe], and [Ti/Fe] were determined by variation of ${\rm
  [\alpha/Fe]}_{\rm abund}$ with \teff, \logg, \feh, and ${\rm
  [\alpha/Fe]}_{\rm atm}$ held fixed.

The complete list of steps to determine the abundances is largely the
same as in Sec.~4.7 of \citeauthor*{kir09}.  The following is the
updated list.

\begin{enumerate}
\item \teff\ and \feh, first pass: An observed spectrum was compared
  to a synthetic spectrum with \teff\ and \logg\ determined
  photometrically.  Only spectral regions most susceptible to Fe
  absorption were considered.  \teff\ was loosely constrained by
  photometry.  Photometry alone determined \logg.

\item ${\rm [\alpha/Fe]}_{\rm atm}$, first pass: \teff, \logg, and
  \feh\ were fixed, but ${\rm [\alpha/Fe]}_{\rm atm}$ and ${\rm
    [\alpha/Fe]}_{\rm abund}$ varied together.  Only the spectral
  regions susceptible to absorption by Mg, Si, Ca, or Ti were
  considered.

\item Continuum refinement: The continuum-divided, observed spectrum
  was divided by the synthetic spectrum with the parameters determined
  in steps~1 and 2.  The result approximated a flat noise spectrum.
  To better determine the continuum, we fit a B-spline with a
  breakpoint spacing of 150~pixels to the residual spectrum.  We
  divided the observed spectrum by the spline fit.

\item Steps~1--3 were repeated until \teff\ changed from the previous
  iteration by less than 1~K and \feh\ changed by less than 0.001~dex.

\item \feh, second pass: We repeated step~1 with the revised,
  sigma-clipped spectrum, but \teff\ was held fixed at the previously
  determined value.

\item ${\rm [\alpha/Fe]}_{\rm atm}$, second pass: We repeated step~2
  with the revised spectrum and the value of \feh\ from step~5.

\item \feh, third pass: We repeated step~5 with the value of ${\rm
  [\alpha/Fe]}_{\rm atm}$ from step~6.

\item{[Mg/Fe]: We determined [Mg/Fe] by varying ${\rm
    [\alpha/Fe]}_{\rm abund}$ at fixed ${\rm [\alpha/Fe]}_{\rm atm}$.
  Only spectral regions subject to Mg absorption were considered.}

\item{[Si/Fe]: We repeated step~9 for Si instead of Mg.}

\item{[Ca/Fe]: We repeated step~9 for Ca instead of Mg.}

\item{[Ti/Fe]: We repeated step~9 for Ti instead of Mg.}
\end{enumerate}

\subsection{Offsets to the Abundance Measurements}
\label{sec:offsets}

In \citeauthor*{kir09}, we found that adding 0.15~dex to the MRS
measurement of \feh\ was necessary to bring $\mathfeh_{\mathrm{MRS}}$
into agreement with $\mathfeh_{\mathrm{HRS}}$ for nine stars with both
MRS and HRS measurements in Sculptor.  The modification to the
continuum division and sigma clipping mitigated the need for the
artificial offset in \feh.  There is still an offset of
$\allfehdiffmean$~dex, but it is small enough that we do not correct
it.

An offset of 0.15~dex between ${\mathrm{[Si/Fe]}}_{\mathrm{MRS}}$ and
${\mathrm{[Si/Fe]}}_{\mathrm{HRS}}$ still exists, and we added
0.15~dex to ${\mathrm{[Si/Fe]}}_{\mathrm{MRS}}$ for better agreement.
This offset is unique to Si, and does not affect other \afe\ ratios.
Therefore, the offset does not arise from continuum division
uncertainties.  We share eight \ion{Si}{1} lines in common with the
high-resolution studies of Cohen et al.\ \citep[e.g.,][]{coh09,coh10}.
Our oscillator strengths for these eight lines are all larger, with an
average difference of 0.16~dex.  Therefore, our Si measurements are
lower than Cohen et al.'s Si measurements.  Rather than adjust the
oscillator strengths and recompute the spectral grid at large
computational expense, we added 0.15~dex to our [Si/Fe] measurements.

\subsection{Applicability to Dwarf Stars}
\label{sec:dwarfs}

The maximum \logg\ in the dSph catalog is 3.5.  Some of the GC masks
observed in 2009 (see Table~2) included stars below the main sequence
turn-off.  We excluded GC and MW halo stars with $\mathlogg > 3.5$ in
order to restrict the comparison sample to stars of the same spectral
type as the dSph catalog.

\addtocounter{table}{1}

\begin{deluxetable}{lr}
\tablewidth{0pt}
\tablecolumns{2}
\tablecaption{Abundance Error Floors\label{tab:syserr}}
\tablehead{\colhead{Element Ratio} & \colhead{$\delta_{\rm sys}$}}
\startdata
\protect[Fe/H]  & \fehsyserr  \\
\protect[Mg/Fe] & \mgfesyserr \\
\protect[Si/Fe] & \sifesyserr \\
\protect[Ca/Fe] & \cafesyserr \\
\protect[Ti/Fe] & \tifesyserr \\
\enddata
\end{deluxetable}

\subsection{The Catalog}
\label{sec:catalog}

Table~\ref{tab:catalog} gives the measurements of \ndsphstars\ unique
stars (376 in Sculptor, 675 in Fornax, 827 in Leo~I, 141 in Sextans,
258 in Leo~II, 174 in Canes Venatici~I, 212 in Ursa Minor, and 298 in
Draco).  The table lists the right ascension, declination,
extinction-corrected $BVRI$ magnitudes where available
\citep[sometimes converted from $ugriz$ following][]{jor06},
temperature, surface gravity, microturbulent velocity, \feh, and the
[Mg/Fe], [Si/Fe], [Ca/Fe], and [Ti/Fe] ratios where measurable.
Uncertainties are also given for all of the measurable quantities
except the space coordinates and microturbulent velocity.  For \teff,
two types of uncertainty are given.  The first type, $\delta
T_{\rm{eff,spec}}$, is the uncertainty from random spectral noise.
(In detail, it is the square root of the diagonal element of the
covariance matrix corresponding to $\delta\mathteff$ when both
\teff\ and \feh\ are allowed to vary.)  The second type, $\delta
T_{\rm{eff,phot}}$, is the uncertainty on the photometric estimate of
\teff, accounting for random photometric uncertainty and systematic
error from isochrone modeling (Eq.~6 of \citeauthor*{kir09}).  Because
\logg\ is determined from photometry alone, the uncertainty on
\logg\ is derived analogously to $\delta T_{\rm{eff,phot}}$, but the
error on the distance modulus to each stellar system is included in
the error estimate of \logg\ as well.  The uncertainties on the
abundance measurements are the $1 \sigma$ uncertainties from the
spectral fit added in quadrature with the constant error floors, given
in Table~\ref{tab:syserr}.

As an example, consider the first star in Table~\ref{tab:catalog}.
The first column identifies its origin as Sculptor, and the second
column gives the star's identification number.  In the online version,
the next two columns give its coordinates.  $V$ and $I$ magnitudes are
given, but $B$ and $R$ magnitudes are not available.  We have measured
$\mathteff = 5085$~K with an uncertainty of 101~K based on spectral
noise.  From photometry alone, the uncertainty is 258~K.  We have also
measured a photometric surface gravity of $\mathlogg = 2.02 \pm 0.08$.
The microturbulent velocity, deduced solely from the surface gravity,
is 1.66~km~s$^{-1}$.  The remaining columns give \feh, [Mg/Fe],
[Si/Fe], [Ca/Fe], and [Ti/Fe] and their estimated uncertainties.



\section{Accuracy}
\label{sec:accuracy}

The range of abundances and size of our sample demand rigorous checks
to validate the accuracy and precision of our measurements.  In this
section, we estimate measurement uncertainties, check those estimates
against repeat measurements of the same stars, and check the accuracy
of our measurements against previously published high-resolution
spectroscopic abundance measurements.

\subsection{Estimation of Measurement Uncertainty}
\label{sec:error}

We estimated abundance measurement uncertainties from individual stars
in globular clusters.  Because we have excluded GCs suspected to
exhibit Fe abundance variations, such as M22 \citep{dac09,mar09}, we
assumed that the spread of \feh\ in each cluster is much smaller than
the measurement uncertainty, but we made no such assumption for other
elements.  Although no monometallic cluster has been found to have a
measurable spread of Ca or Ti abundances except NGC~2419
\citep{cohetal10}, many clusters exhibit star-to-star Mg variations
\citep[see the review by][]{gra04}.

KGS08 and \citeauthor*{kir09} explicitly described the procedure for
determining measurement uncertainties.  For each elemental abundance
measurement, they added an extra error component in quadrature with
the random error from spectral noise.  This error floor for \feh\ was
determined from the excess spread in the measured \feh\ in each GC
after accounting for random error due to spectral noise.  For this
purpose, we excluded M92 because we observed only one star in it, and
we excluded NGC~2419 because we suspect that it may contain a spread
in heavy element abundances \citep{cohetal10}.  The error floor in the
four \xfe\ ratios was determined from comparison to high-resolution
measurements.  NGC~2419 was included for this purpose.  The error
floor is the value needed to account for differences between the MRS
and HRS measurements after removing the MRS random error and the
published value of the HRS measurement uncertainty.
Table~\ref{tab:syserr} gives the updated values for the error floors
for each element.  The differences between this table and Table~5 of
\citeauthor*{kir09} arise from the modifications described in
Sec.~\ref{sec:measurements} and the inclusion of GC observations
obtained since the publication of KGS08.

\begin{figure}[t!]
\centering
\includegraphics[width=0.4\textwidth]{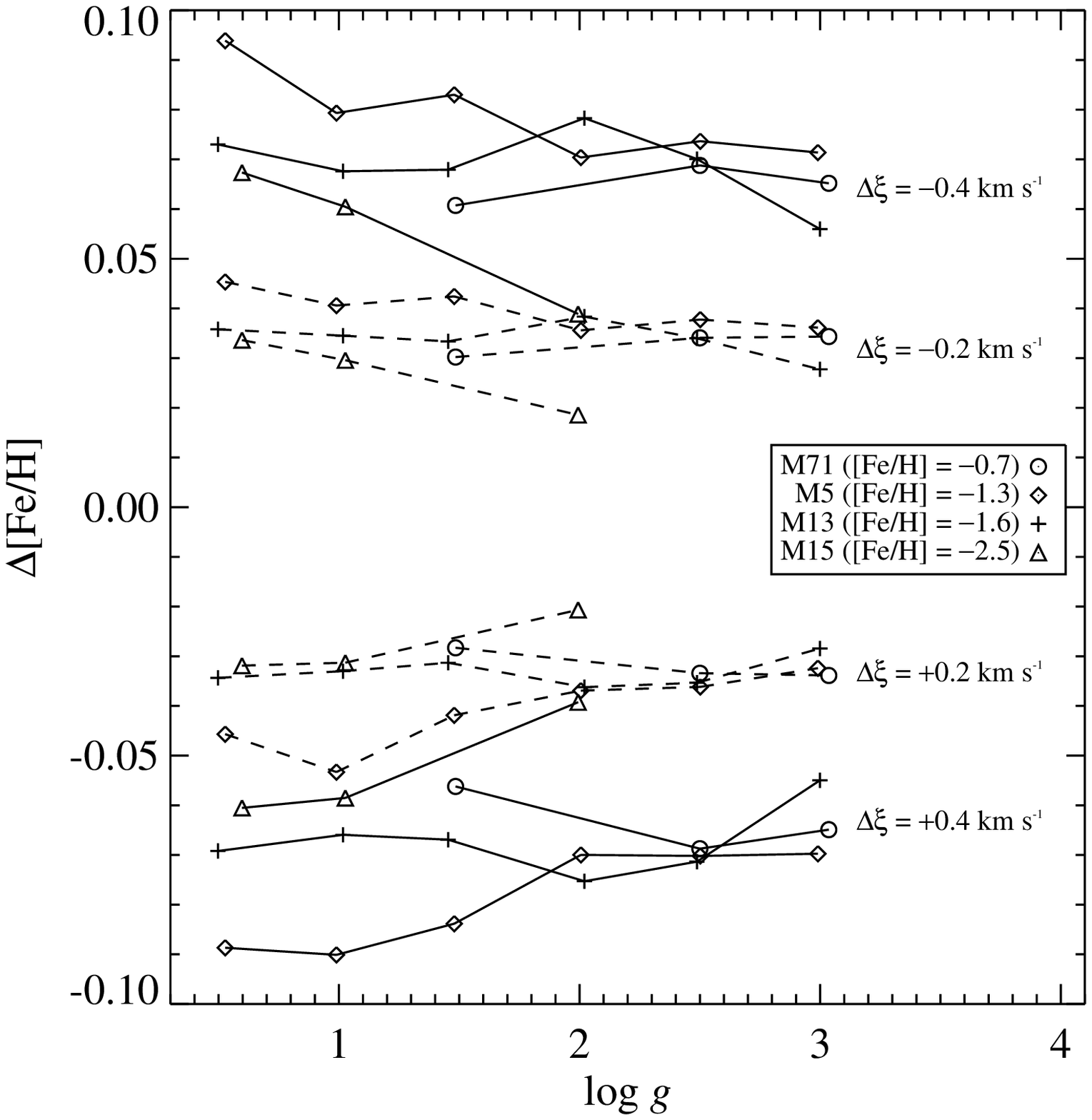}
\caption{Differences in \feh\ from changes in microturbulent velocity
  ($\xi$) of $\pm 0.2~\rm{km~s}^{-1}$ ({\it dashed lines}) and $\pm
  0.4~\rm{km~s}^{-1}$ (solid lines) vs.\ surface gravity.  The 20
  stars analyzed are selected from four different globular clusters
  over a range of metallicities and from a range of surface gravities.
  The figure legend gives the mean \feh\ for each GC from our own
  medium-resolution measurements.\label{fig:vttest}}
\end{figure}

\begin{figure*}[t!]
\centering
\begin{minipage}[t]{0.4\textwidth}
\centering
\includegraphics[width=\textwidth]{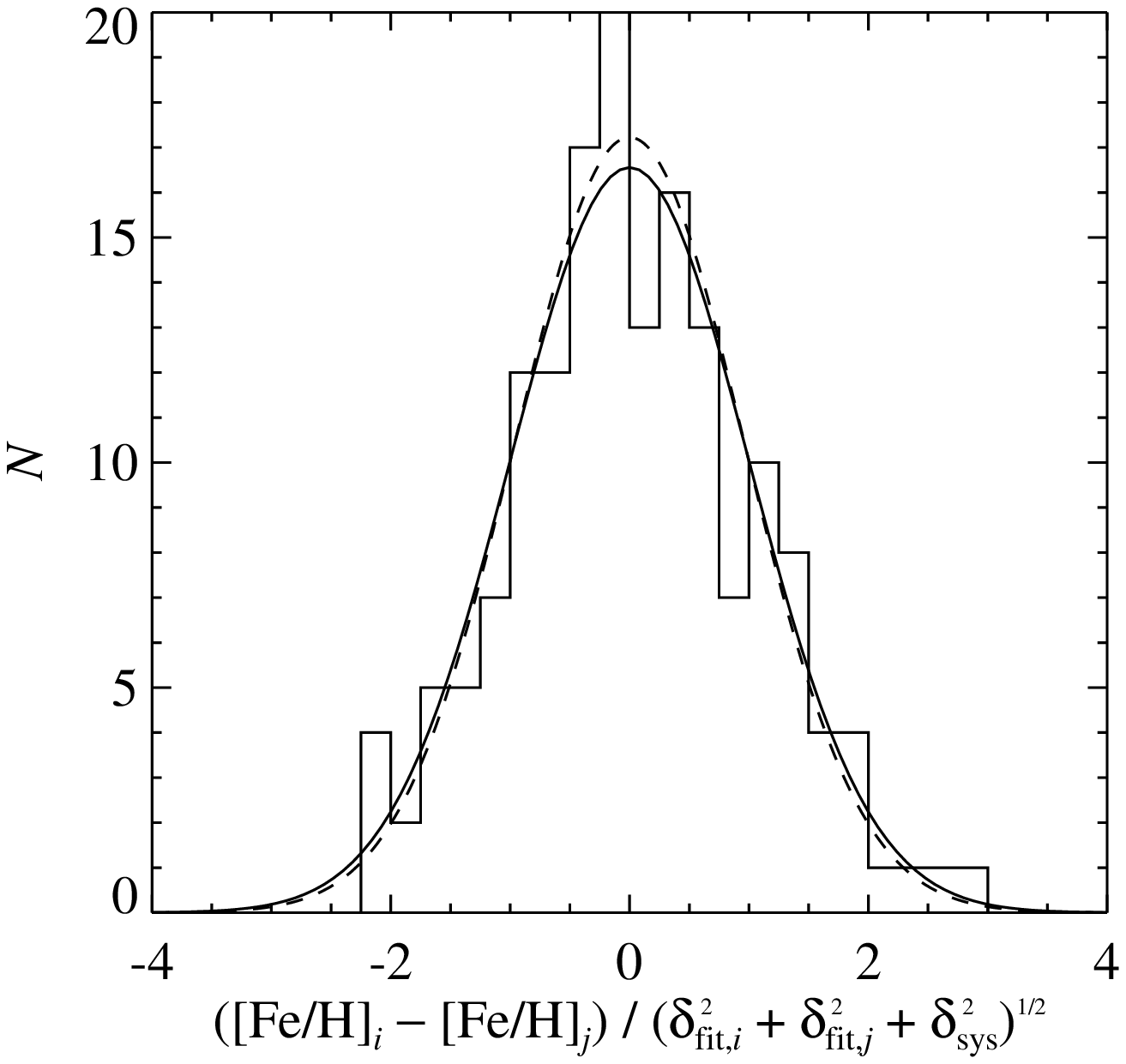}
\caption{Distribution of differences between the repeat measurements
  of \feh\ for \ndup\ stars divided by the estimated error of the
  difference.  The solid curve is a unit Gaussian with $\sigma = 1$.
  The best-fit Gaussian ({\it dashed line}) has $\sigma =
  \fehdupsigma$.\label{fig:fehdup}}
\end{minipage}
\hspace{0.1\textwidth}
\begin{minipage}[t]{0.4\textwidth}
\centering
\includegraphics[width=\textwidth]{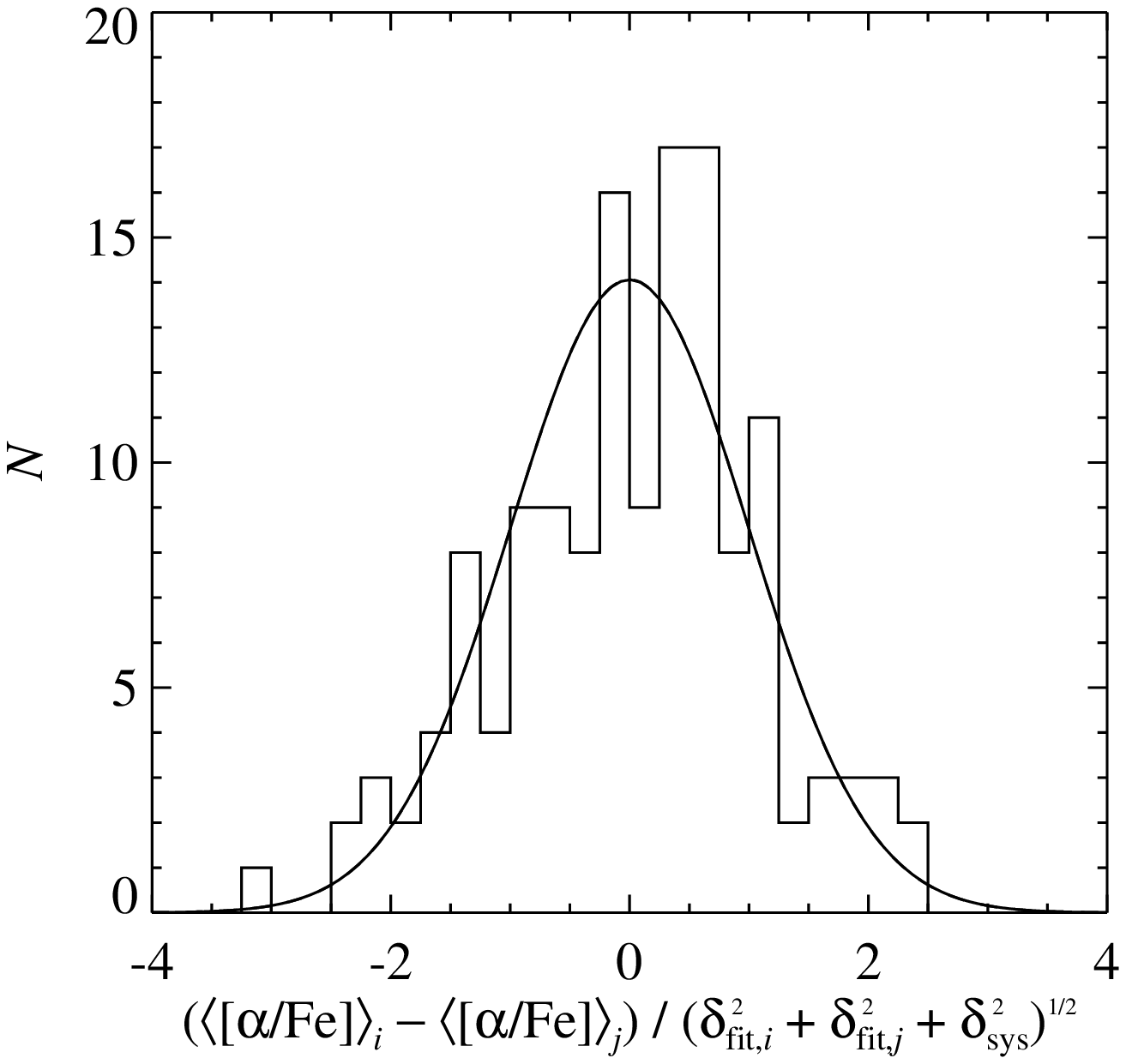}
\caption{Distribution of differences between the repeat measurements
  of $\langle[\alpha/\rm{Fe}]\rangle$, which is the average of
  [Mg/Fe], [Si/Fe], [Ca/Fe], and [Ti/Fe], for \nalphadup\ stars
  divided by the estimated error of the difference.  The solid curve
  is a unit Gaussian with $\sigma = 1$.  The best-fit Gaussian ({\it
    dashed line}, exactly overlying the solid line) has $\sigma =
  \alphadupsigma$.\label{fig:alphadup}}
\end{minipage}
\end{figure*}

\subsection{Errors from Atmospheric Parameters}
\label{sec:atmerr}

Errors in \teff, \logg, and $\xi$ have the potential to change the
measured abundances significantly.  The error floor for $\mathfeh_{\rm
  MRS}$ accounts for uncertainty in the atmospheric parameters, but we
also quantify the response of \feh\ to forced changes in \teff\ and
\logg.  Differences in \teff, \logg, and $\xi$ affect the $\alpha$
element abundance ratios less than they affect \feh.  To first order,
the measurement of an $\alpha$ element's abundance responds to the
atmospheric parameters in the same way as the measurement of the Fe
abundance.  As a result, the division of one element's abundance by
the Fe abundance somewhat masks errors in atmospheric parameters.

We recomputed abundances at 125~K and 250~K above and below the
best-fitting \teff\ for every star in the GC, MW halo, and dSph
samples.  The new abundance was determined by searching the grid for
the synthetic spectrum that best matched the observed spectrum with
\teff\ and \logg\ held fixed.  We also recomputed abundances with
\logg\ held fixed at 0.3~dex and 0.6~dex above and below the
best-fitting \logg.  The measured abundance of \feh\ is higher for a
larger value of \teff\ or a larger value of \logg.  Conversely, the
measured abundance of \feh\ is lower for a lower value of \teff\ or a
lower value of \logg.  (See Sec.~6.3 of KGS08 for a more thorough
discussion.)

The globular cluster star sample provides some insight on the general
response of \feh\ to changes in atmospheric parameters.  The average
change in \feh\ is $\pm 0.092~{\rm dex}$ per $\pm 100~{\rm K}$ change
in \teff.  The average change in \feh\ is $\pm 0.039~{\rm dex}$ per
$\pm 1~{\rm dex}$ change in \logg.  The \feh\ measurements are
relatively insensitive to changes in \logg\ because the abundance
analysis uses many \ion{Fe}{1} lines but virtually no \ion{Fe}{2}
lines, which are few and weak in red giants.  \ion{Fe}{1} lines vary
little with \logg.  Ideally, we would use only \ion{Fe}{2} lines
because they are not subject to overionization problems
\citep{the99,iva01}.  However, we must use \ion{Fe}{1} lines because
\ion{Fe}{2} lines are not visible in DEIMOS spectra of red giants.
The slopes of $\Delta \mathfeh / \Delta \mathteff$ and $\Delta
\mathfeh / \Delta (\mathlogg)$ are approximately linear within $\pm
250~{\rm K}$ and $\pm 0.6~{\rm dex}$.

Table~\ref{tab:atmparerr} gives the response of \feh\, [Mg/Fe],
[Si/Fe], [Ca/Fe], and [Ti/Fe] to a change of 125~K and 250~K in
\teff\ and 0.3~dex and 0.6~dex in \logg.  The values given are the
averages of the absolute values of the increase and decrease in
\teff\ or \logg.  For example, consider a star with $\mathfeh = -2.00$
and $\mathteff = 4500~K$.  If the column for $\Delta \mathfeh(125~{\rm
  K})$ lists 0.12, then it can be assumed that the measured abundance
of the star would be $\mathfeh = -1.88$ if \teff\ were truly 4625~K,
and the measured abundance would be $\mathfeh = -2.12$ if \teff\ were
truly 4375~K.

We could not estimate the effect of errors in microturbulent velocity
as easily as for \teff\ and \logg\ because our grid of spectra did not
have a separate dimension for $\xi$.  Instead, we fixed $\xi$ based on
a calibration to \logg\ (Eq.~2 of \citeauthor*{kir09}).  To
investigate the effect of errors in $\xi$ on \feh, we reanalyzed 20
stars from four GCs at different metallicities (M71, M5, M13, and
M15).  We chose the stars with \logg\ nearest to 0.5, 1.0, 1.5, 2.0,
2.5, and 3.0, where available.  We synthesized a new, small grid of
spectra for each star.  We generated new atmospheres by linear
interpolation in our ATLAS9 grid (Sec.~\ref{sec:grid}).  The values of
\teff\ and \logg\ were chosen to be the best-fit values we already
determined for each star.  Values of \feh\ were spaced at every
0.1~dex from at least 0.4~dex below to at least 0.4~dex above the
best-fit value.  Values of $\xi$ were $\pm 0.2$ and $\pm
0.4~\rm{km~s}^{-1}$ around the value of $\xi$ from the surface gravity
calibration.  We computed spectra with MOOG.  For each of the four
displacements of $\xi$, we found the value of \feh\ that minimized
$\chi^2$ between the new grid of spectra and the observed spectrum.
As before, we used a Levenberg-Marquardt minimization algorithm with
linear interpolation to compute $\chi^2$ between \feh\ grid points.

Figure~\ref{fig:vttest} shows the changes in the measurement of
\feh\ induced by changes in $\xi$ for each of the 20 GC stars.
Reducing $\xi$ increases \feh\ because the absorption lines saturate
at lower abundance.  A larger abundance is necessary to compensate for
the weaker lines.  The abundance change is most severe for stars of
low surface gravity (low temperature) and high metallicity because
those stars have strong lines and are most susceptible to line
saturation.  The absolute change in \feh\ does not exceed 0.1~dex,
even for changes of $\pm 0.4~\rm{km~s}^{-1}$.  This result agrees with
our later assessment of the effect of microturbulent velocity error
(Fig.~\ref{fig:covarvtfeh}), where we deduce a slope of
$\Delta\mathfeh/\Delta\xi = \vtfehslope~\rm{dex}/(\rm{km~s}^{-1})$.
For comparison, the standard deviation between $\xi_{\rm MRS}$ and
$\xi_{\rm HRS}$ is $\allvtdiffsigma~\rm{km~s}^{-1}$ (see
Sec.~\ref{sec:hrscompare} and Table~\ref{tab:hrsdiff}).

\subsection{Duplicate Observations}
\label{sec:duplicate}

The repeat observations of \ndup\ dSphs stars provided insight into
the effect of random error on the measurements of \feh\ and
$\langle[\alpha/\rm{Fe}]\rangle$ (an average of [Mg/Fe], [Si/Fe],
[Ca/Fe], and [Ti/Fe]).  Repeat measurements of
$\langle[\alpha/\rm{Fe}]\rangle$ were possible for both stars in
\nalphadup\ pairs.  Figures~\ref{fig:fehdup} and \ref{fig:alphadup}
summarize the comparisons of measurements of different spectra of the
same stars.  They show the distributions of the absolute difference
between the measured \feh\ and $\langle[\alpha/\rm{Fe}]\rangle$ for
each pair of spectra divided by the expected error of the difference.
In calculating the expected error of the difference, we apply the
error floor to only one of the two stars.  Even though the same
technique is used to measure abundances in both stars---including the
same photometric estimate of surface gravity----the extra error, which
accounts for sources of error beyond random spectral noise, is
appropriate because the wavelength range within a pair of spectra
differs by up to 400~\AA.  The different Fe lines in these ranges span
a different range of excitation potentials, and the
Levenberg-Marquardt algorithm converges on different solutions.

Figures~\ref{fig:fehdup} also shows the best-fit Gaussian ($\sigma =
\fehdupsigma$) and a Gaussian with unit variance.  The best-fit
Gaussian in Fig.~\ref{fig:alphadup} has $\sigma = \alphadupsigma$.
The areas of the Gaussians are normalized to the number of stars.  Our
estimate of uncertainty is accurate because the variances of the
best-fit Gaussians are close to unity.


\subsection{Comparison to High-Resolution Measurements}
\label{sec:hrscompare}

\addtocounter{table}{1}
\addtocounter{table}{1}
\begin{deluxetable}{lr}
\tablewidth{200pt}
\tablecolumns{2}
\tablecaption{Adopted Solar Composition\label{tab:solar}}
\tablehead{\colhead{Element} & \colhead{$12 + \log \epsilon$}}
\startdata
Mg & 7.58 \\
Si & 7.55 \\
Ca & 6.36 \\
Ti & 4.99 \\
Fe & 7.52 \\
\enddata
\tablecomments{This composition is adopted from \protect\citet{and89},
  except for Fe.  For justification of the adopted Fe solar abundance,
  see \citet{sne92}.  The abundance of an element X is defined as its
  number density relative to hydrogen: $12 + \log \epsilon_{\rm{X}} =
  12 + \log (n_{\rm{X}}) - \log (n_{\rm{H}})$.}
\end{deluxetable}

\begin{deluxetable*}{lccccc}
\tablewidth{500pt}
\tablecolumns{6}
\tablecaption{Previously Published HRS Abundance Methods\label{tab:hrsmethod}}
\tablehead{\colhead{Reference} & \colhead{\phantom{\tablenotemark{a}}Atmospheres\tablenotemark{a}} & \colhead{\phantom{\tablenotemark{b}}Code\tablenotemark{b}} & \colhead{\phantom{\tablenotemark{c}}\teff\tablenotemark{c}} & \colhead{\phantom{\tablenotemark{d}}\logg\tablenotemark{d}} & \colhead{\phantom{\tablenotemark{e}}$\xi$\tablenotemark{e}}} \\ 
\startdata
\cutinhead{Globular Clusters}
\protect \citet{coh05a,coh05b} & ATLAS9 & MOOG & phot  & phot & \phantom{\tablenotemark{f}}spec\tablenotemark{f} \\
\protect \citet{gra89} & \protect \citet{bel76} & WIDTH2 & phot & phot & spec \\
\protect \citet{iva01} & MARCS & MOOG & \phantom{\tablenotemark{g}}spec\tablenotemark{g} & \phantom{\tablenotemark{g}}spec\tablenotemark{g} & spec \\
\protect \citet{mis03} & ATLAS9 & WIDTH9 & spec & spec & spec \\
\protect \citet{ram02,ram03} & ATLAS9 & MOOG & phot & phot & \phantom{\tablenotemark{h}}spec\tablenotemark{h} \\
\protect \citet{sne97,sne00} & MARCS & MOOG & spec & spec & spec \\
\protect \citet{sne04} & MARCS & MOOG & spec & phot & spec \\
\cutinhead{Halo Field Stars}
\protect \citet{car02} & ATLAS9 & \phantom{\tablenotemark{i}}unknown\tablenotemark{i} & phot & phot & spec \\
\protect \citet{coh06,coh08} & ATLAS9 & MOOG & phot & phot & \phantom{\tablenotemark{j}}phot\tablenotemark{j} \\
\protect \citet{ful00} & ATLAS9 & MOOG & spec & spec & spec \\
\protect \citet{joh02} & ATLAS9 & MOOG & spec & spec & spec \\
\protect \citet{lai04,lai07} & ATLAS9 & MOOG & phot & phot & evolutionary \\
\protect \citet{pil96} & MARCS & MOOG & phot & phot & spec \\
\cutinhead{dSphs}
\protect \citet{coh09,coh10} & ATLAS9 & MOOG & spec  & spec & spec \\
\protect \citet{fre10a,fre10b} & ATLAS9 & MOOG & spec & spec & spec \\
\protect \citet{ful04} & ATLAS9 & MOOG & phot & phot & spec \\
\protect \citet{gei05} & MARCS & MOOG & phot & spec & spec \\
\protect \citet{koc08} & ATLAS9 & MOOG & spec & spec & spec \\
\protect \citet{let10} & MARCS & CALRAI & phot & phot & spec \\
\protect \citet{sad04} & ATLAS9 & SPTOOL & spec & spec & spec \\
\protect \citet{she01a,she03,she09} & MARCS & MOOG & spec & spec & spec \\
\protect \citet{sim10} & ATLAS9 & MOOG & \phantom{\tablenotemark{k}}phot\tablenotemark{k} & \phantom{\tablenotemark{k}}phot\tablenotemark{k} & spec \\
\enddata
\tablenotetext{a}{ATLAS9: \protect \citet{kur93} or \protect \citet{cas04}, also \protect \url{http://kurucz.harvard.edu/grids.html}; MARCS: \protect \citet{gus75,gus03,gus08}}
\tablenotetext{b}{MOOG: \protect \citet{sne73}; WIDTH2: \protect \citet{gra82}; WIDTH9: \protect \citet{kur05}; CALRAI: \protect \citet{spi67}; SPTOOL: Y.~Takeda (unpublished, but based on WIDTH9)}
\tablenotetext{c}{phot: empirical color-\teff\ relation or model isochrones; spec: Fe~I excitation equilibrium}
\tablenotetext{d}{phot: determined from model isochrones or \teff, with luminosity based on bolometric corrections, and assumption of the stellar mass; spec: Fe~I and Fe~II ionization balance}
\tablenotetext{e}{spec: based on removing abundance trends with equivalent width; evolutionary: $\xi$ assigned based on position in the color-magnitude diagram}
\tablenotetext{f}{Because $\xi$ for all stars was similar, \protect \citet{coh05b} assumed 2.0~km~s$^{-1}$ for all stars in NGC~7492.}
\tablenotetext{g}{Photometric values were also derived, but we have adopted the spectroscopic values and corresponding abundances.}
\tablenotetext{h}{A $\mathteff-\xi$ relation derived from the brighter stars was assumed for the fainter stars.}
\tablenotetext{i}{Although the paper does not mention it, these authors typically use MOOG.}
\tablenotetext{j}{The value of $\xi$ was set to 1.6--1.8~km~s$^{-1}$ with variation depending on \teff.}
\tablenotetext{k}{Spectroscopic values were also derived, but the authors preferred the photometric values.}
\end{deluxetable*}

\begin{deluxetable*}{ccccc}
\tablewidth{375pt}
\tablecolumns{5}
\tablecaption{Differences Between MRS and HRS Atmospheric Parameters and Abundances\label{tab:hrsdiff}}
\tablehead{\colhead{Quantity} & \colhead{GCs} & \colhead{MW Halo} & \colhead{dSphs} & \colhead{All}}
\startdata
$\langle\delta\mathteff\rangle$ (K)           & $\gcteffdiffmean$ ($\gcteffdiffsigma$)       & \phn$\haloteffdiffmean$ ($\haloteffdiffsigma$)   & \phn$\dsphteffdiffmean$ ($\dsphteffdiffsigma$) & \phn$\allteffdiffmean$ ($\allteffdiffsigma$)       \\
$\langle\delta\mathlogg\rangle$ (cm~s$^{-2}$) & $\gcloggdiffmean$ ($\gcloggdiffsigma$)       & \phantom{\tablenotemark{a}}\nodata\tablenotemark{a} & $\dsphloggdiffmean$ ($\dsphloggdiffsigma$)         & $\allloggdiffmean$ ($\allloggdiffsigma$)       \\
$\langle\delta\xi\rangle$ (km~s$^{-1}$)       & \phn$\gcvtdiffmean$ ($\gcvtdiffsigma$)\phn   & \phn$\halovtdiffmean$ ($\halovtdiffsigma$)\phn   & \phn$\dsphvtdiffmean$ ($\dsphvtdiffsigma$)\phn     & \phn$\allvtdiffmean$ ($\allvtdiffsigma$)\phn   \\
$\langle\delta\mathfeh\rangle$                & $\gcfehdiffmean$ ($\gcfehdiffsigma$)         & $\halofehdiffmean$ ($\halofehdiffsigma$)         & $\dsphfehdiffmean$ ($\dsphfehdiffsigma$)           & $\allfehdiffmean$ ($\allfehdiffsigma$)         \\
$\langle\delta {\rm [Mg/Fe]}\rangle$          & $\gcmgfediffmean$ ($\gcmgfediffsigma$)       & $\halomgfediffmean$ ($\halomgfediffsigma$)       & $\dsphmgfediffmean$ ($\dsphmgfediffsigma$)         & $\allmgfediffmean$ ($\allmgfediffsigma$)       \\
$\langle\delta {\rm [Si/Fe]}\rangle$          & $\gcsifediffmean$ ($\gcsifediffsigma$)       & $\halosifediffmean$ ($\halosifediffsigma$)       & $\dsphsifediffmean$ ($\dsphsifediffsigma$)         & $\allsifediffmean$ ($\allsifediffsigma$)       \\
$\langle\delta {\rm [Ca/Fe]}\rangle$          & $\gccafediffmean$ ($\gccafediffsigma$)       & $\halocafediffmean$ ($\halocafediffsigma$)       & $\dsphcafediffmean$ ($\dsphcafediffsigma$)         & $\allcafediffmean$ ($\allcafediffsigma$)       \\
$\langle\delta {\rm [Ti/Fe]}\rangle$          & $\gctifediffmean$ ($\gctifediffsigma$)       & $\halotifediffmean$ ($\halotifediffsigma$)       & $\dsphtifediffmean$ ($\dsphtifediffsigma$)         & $\alltifediffmean$ ($\alltifediffsigma$)       \\
\phantom{\tablenotemark{b}}$\langle\delta\langle\mathafe\rangle\rangle$\tablenotemark{b} & $\gcalphafediffmean$ ($\gcalphafediffsigma$) & $\haloalphafediffmean$ ($\haloalphafediffsigma$) & $\dsphalphafediffmean$ ($\dsphalphafediffsigma$)   & $\allalphafediffmean$ ($\allalphafediffsigma$) \\
\enddata
\tablecomments{Positive numbers indicate that the MRS measurements are
  larger than the HRS measurements.  Numbers in parentheses are
  standard deviations.}
\tablenotetext{a}{For MW halo stars, for which distances are poorly known, $(\mathlogg)_{\rm MRS}$ was set equal to $(\mathlogg)_{\rm HRS}$.}
\tablenotetext{b}{$\langle[\alpha/\rm{Fe}]\rangle$ is an average of the measurements of [Mg/Fe], [Si/Fe], [Ca/Fe], and [Ti/Fe], but only for \xfe\ measurements with estimated uncertainties of less than 0.5~dex.}
\end{deluxetable*}

\begin{figure*}[p]
\centering
\begin{minipage}[t]{0.48\textwidth}
\centering
\includegraphics[width=0.7\textwidth]{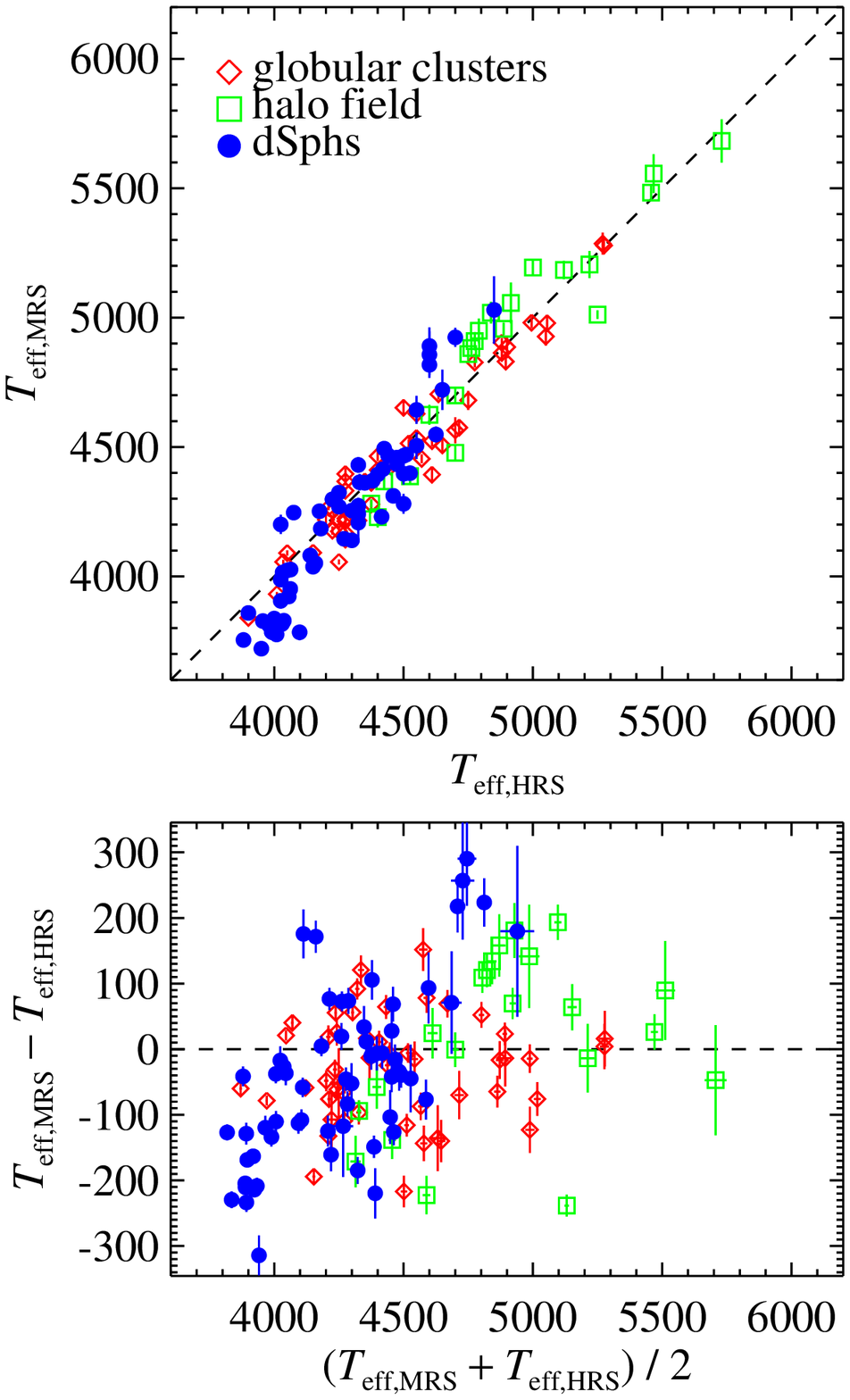}
\caption{{\it Top:} Comparison between effective temperature (\teff)
  used in previous HRS abundance analyses and \teff\ used for this
  work's MRS abundance analysis.  The dashed line is one-to-one.  The
  shape and color of the plotting symbol indicates the type of stellar
  system of which the star is a member.  The error bars represent the
  error on the MRS fit to \teff, and they are not intended to
  represent all sources of error.  {\it Bottom:} Residuals between
  $T_{\mathrm{eff,MRS}}$ and $T_{\mathrm{eff,HRS}}$ vs.\ the average
  of $T_{\mathrm{eff,MRS}}$ and
  $T_{\mathrm{eff,HRS}}$.\notetoeditor{B\&W figure caption: {\it Top:}
    Comparison between effective temperature (\teff) used in previous
    HRS abundance analyses and \teff\ used for this work's MRS
    abundance analysis.  The dashed line is one-to-one.  The plotting
    symbol indicates the type of stellar system of which the star is a
    member.  The error bars represent the error on the MRS fit to
    \teff, and they are not intended to represent all sources of
    error.  {\it Bottom:} Residuals between $T_{\mathrm{eff,MRS}}$ and
    $T_{\mathrm{eff,HRS}}$ vs.\ the average of $T_{\mathrm{eff,MRS}}$
    and $T_{\mathrm{eff,HRS}}$.  (A color version of this figure is
    available in the online journal.)}\label{fig:teff}}
\end{minipage}
\hfill
\begin{minipage}[t]{0.48\textwidth}
\centering
\includegraphics[width=0.7\textwidth]{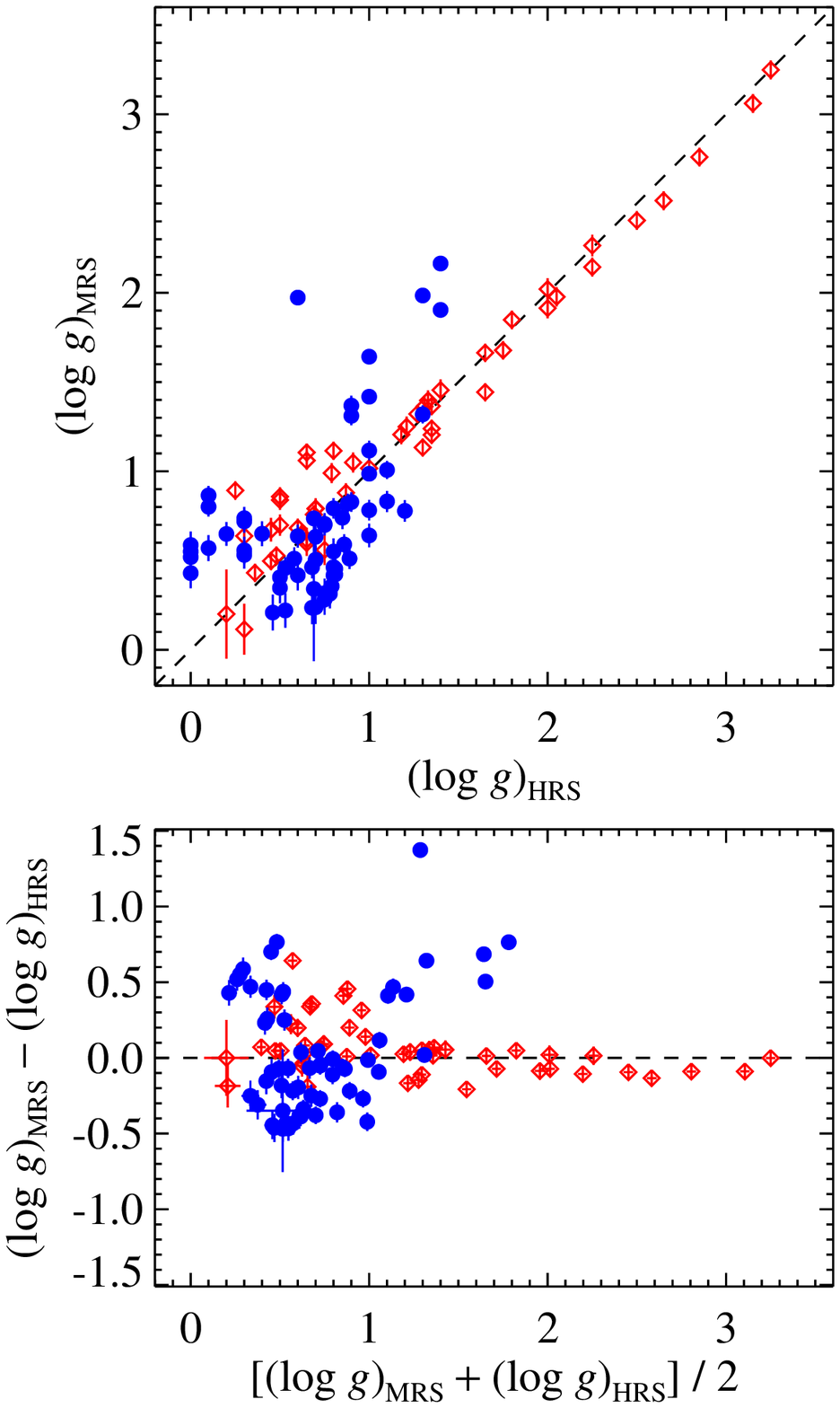}
\caption{Same as Fig.~\ref{fig:teff} except for surface gravity
  (\logg) instead of \teff.  The error bars represent photometric
  uncertainties and isochrone modeling errors.\notetoeditor{B\&W
    figure caption: Same as Fig.~\ref{fig:teff} except for surface
    gravity (\logg) instead of \teff.  The error bars represent
    photometric uncertainties and isochrone modeling errors.  (A color
    version of this figure is available in the online
    journal.)}\label{fig:logg}}
\end{minipage}
\end{figure*}

\begin{figure*}[p]
\centering
\begin{minipage}[t]{0.48\textwidth}
\centering
\includegraphics[width=0.7\textwidth]{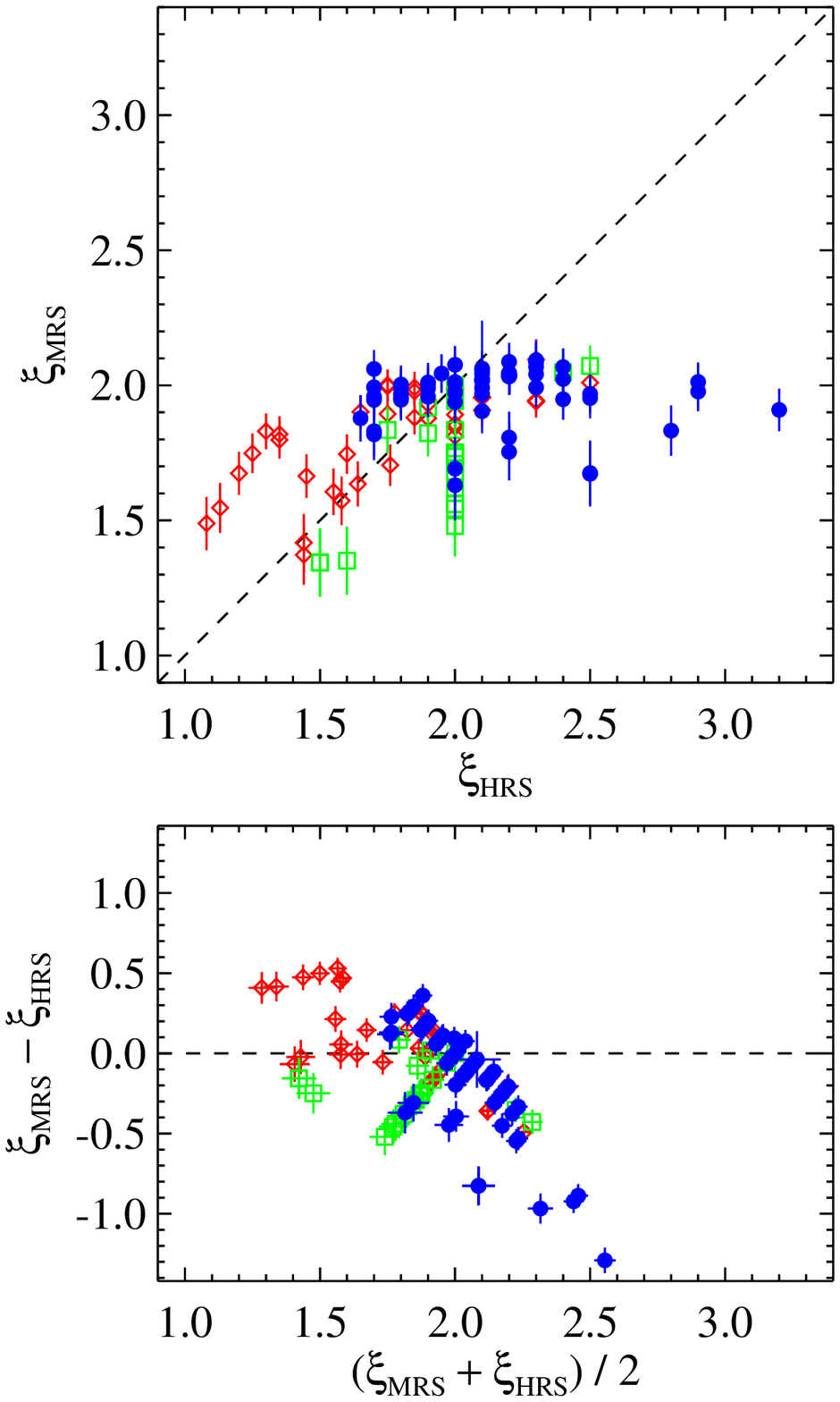}
\caption{Same as Fig.~\ref{fig:teff} except for microturbulent
  velocity ($\xi$) instead of \teff.  The error bars are found by
  propagating the error on \logg\ through the equation to determine
  $\xi$ from \logg\ (Eq.~2 of \citeauthor*{kir09}).\notetoeditor{B\&W
    figure caption: Same as Fig.~\ref{fig:teff} except for
    microturbulent velocity ($\xi$) instead of \teff.  The error bars
    are found by propagating the error on \logg\ through the equation
    to determine $\xi$ from \logg\ (Eq.~2 of \citeauthor*{kir09}).  (A
    color version of this figure is available in the online
    journal.)}\label{fig:vt}}
\end{minipage}
\hfill
\begin{minipage}[t]{0.48\textwidth}
\centering
\includegraphics[width=0.7\textwidth]{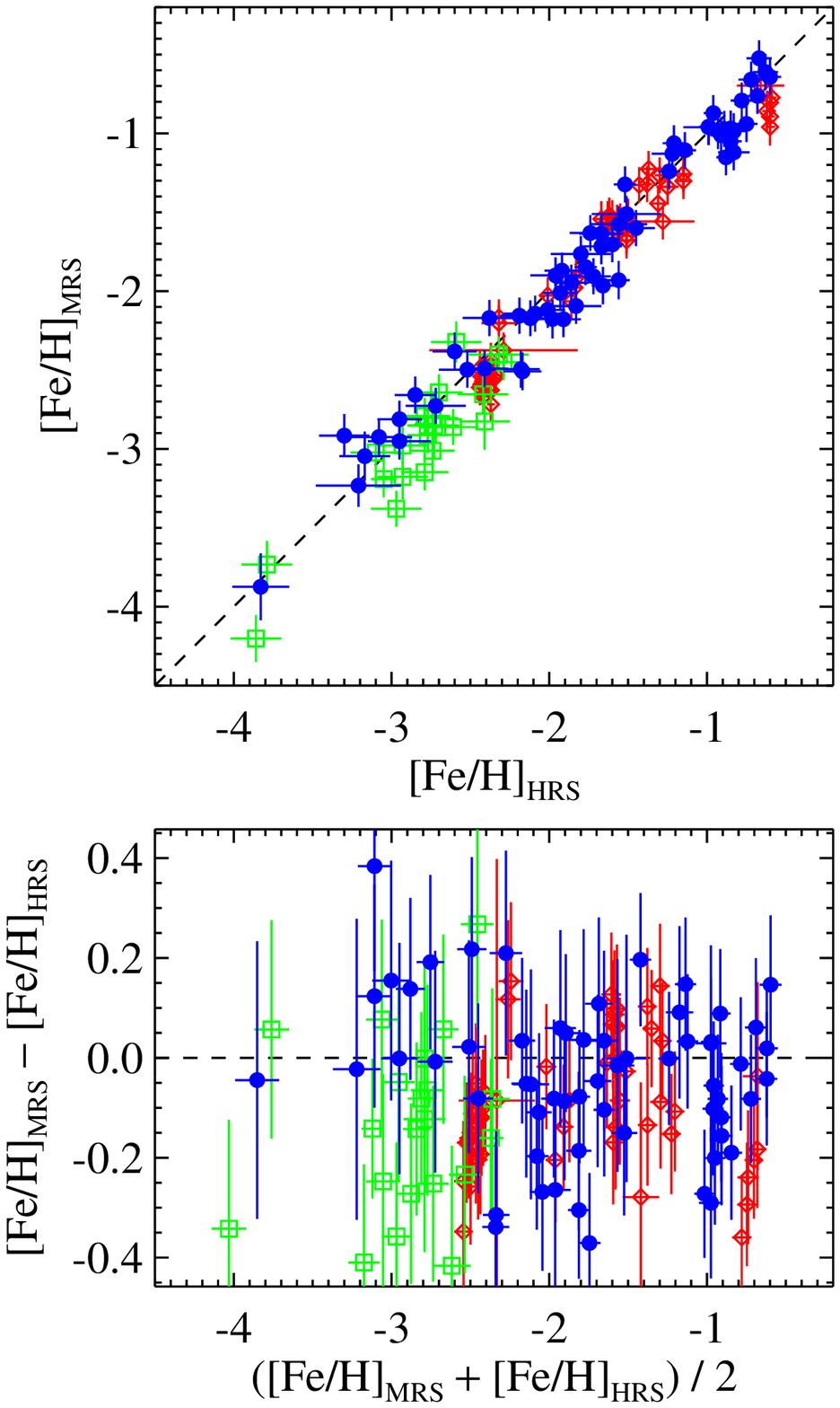}
\caption{Comparison between \feh\ derived from previous HRS abundance
  analyses and \feh\ derived from this work's MRS abundance analysis.
  Colors are the same as in Fig.~\ref{fig:teff}.\notetoeditor{B\&W
    figure caption: Comparison between \feh\ derived from previous HRS
    abundance analyses and \feh\ derived from this work's MRS
    abundance analysis.  Colors are the same as in
    Fig.~\ref{fig:teff}.  (A color version of this figure is available
    in the online journal.)}\label{fig:feh}}
\end{minipage}
\end{figure*}

\begin{figure*}[p!]
\centering
 \columnwidth=.45\columnwidth
 \includegraphics[width=0.4\textwidth]{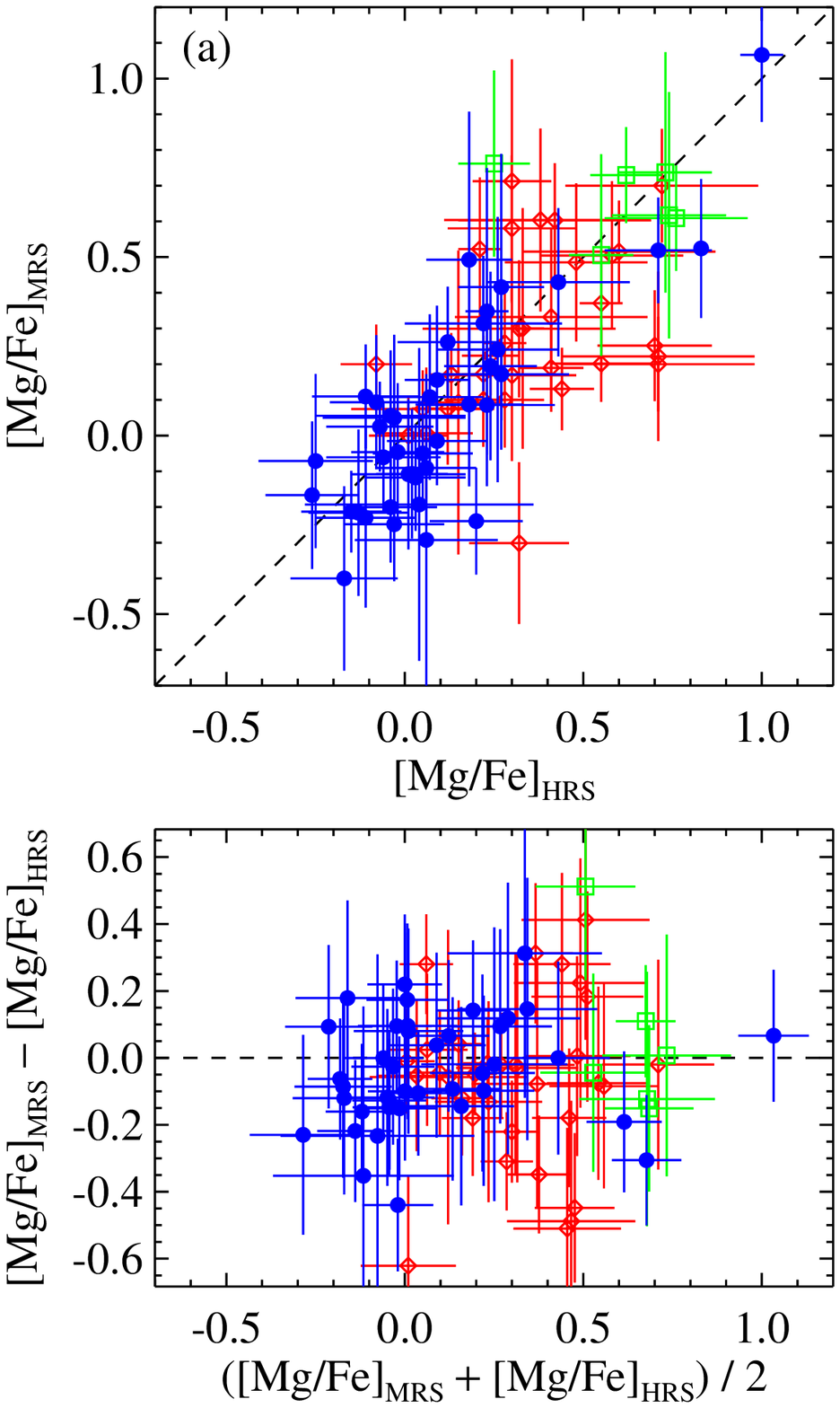}
 \hfil
 \includegraphics[width=0.4\textwidth]{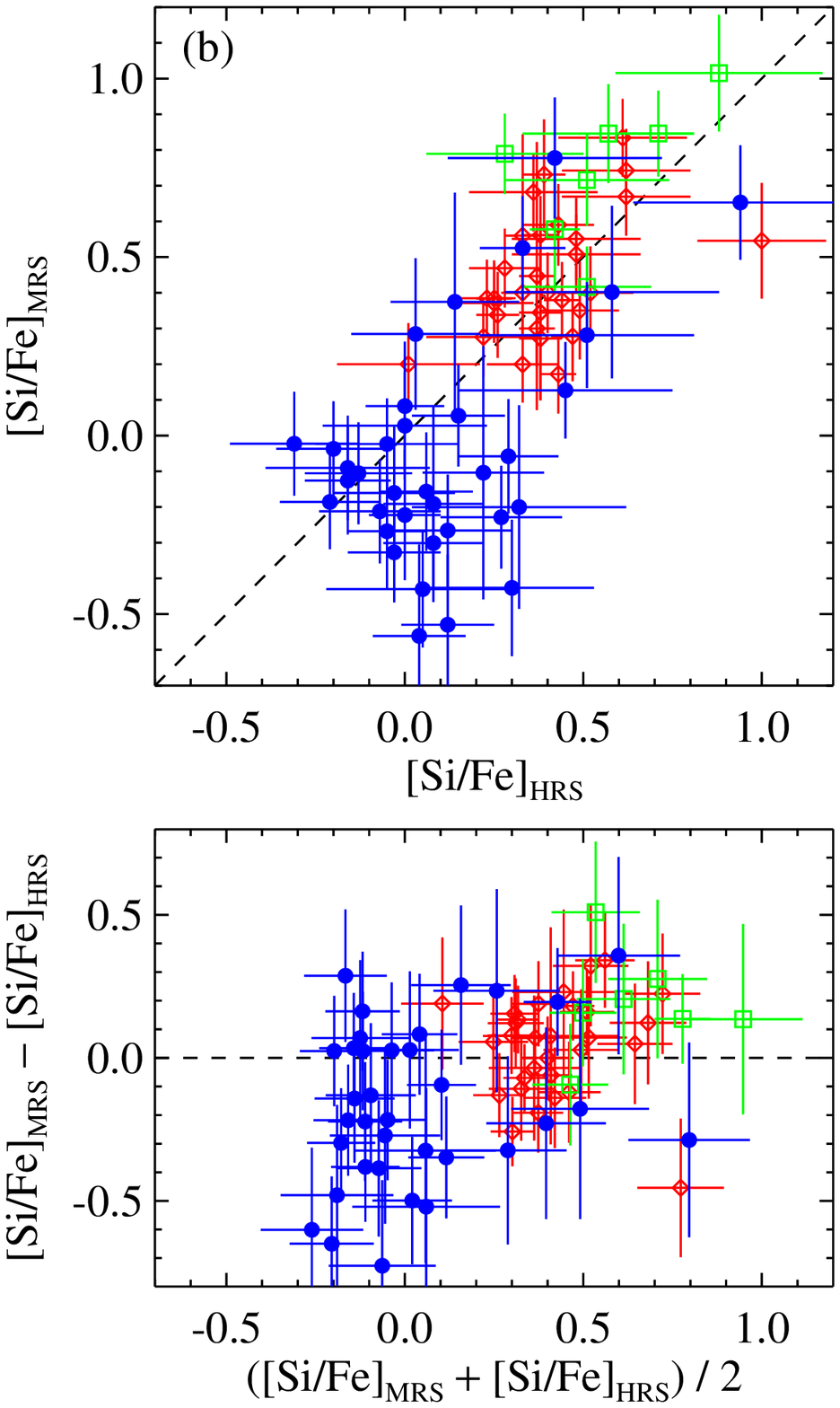}
 \hfil
 \includegraphics[width=0.4\textwidth]{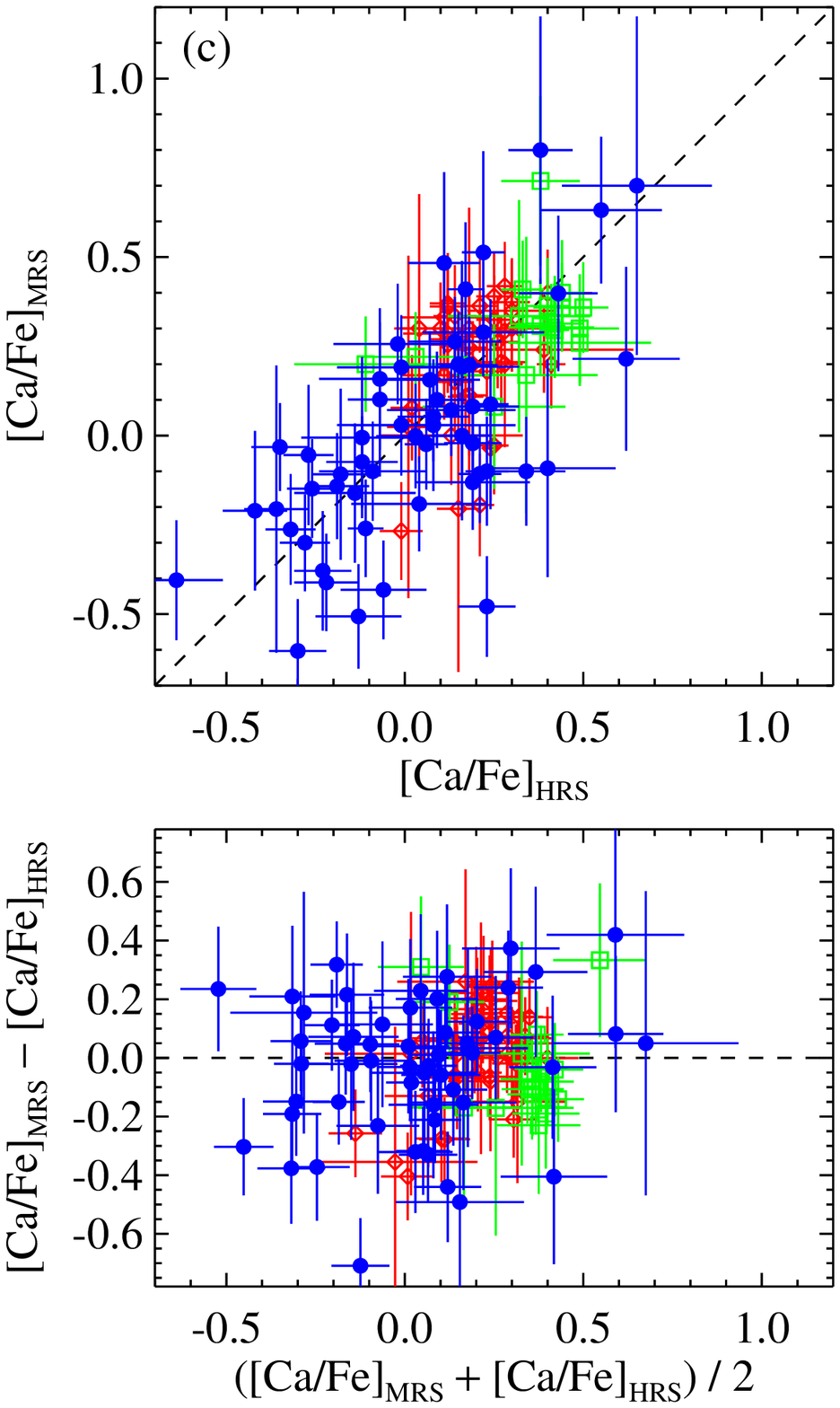}
 \hfil
 \includegraphics[width=0.4\textwidth]{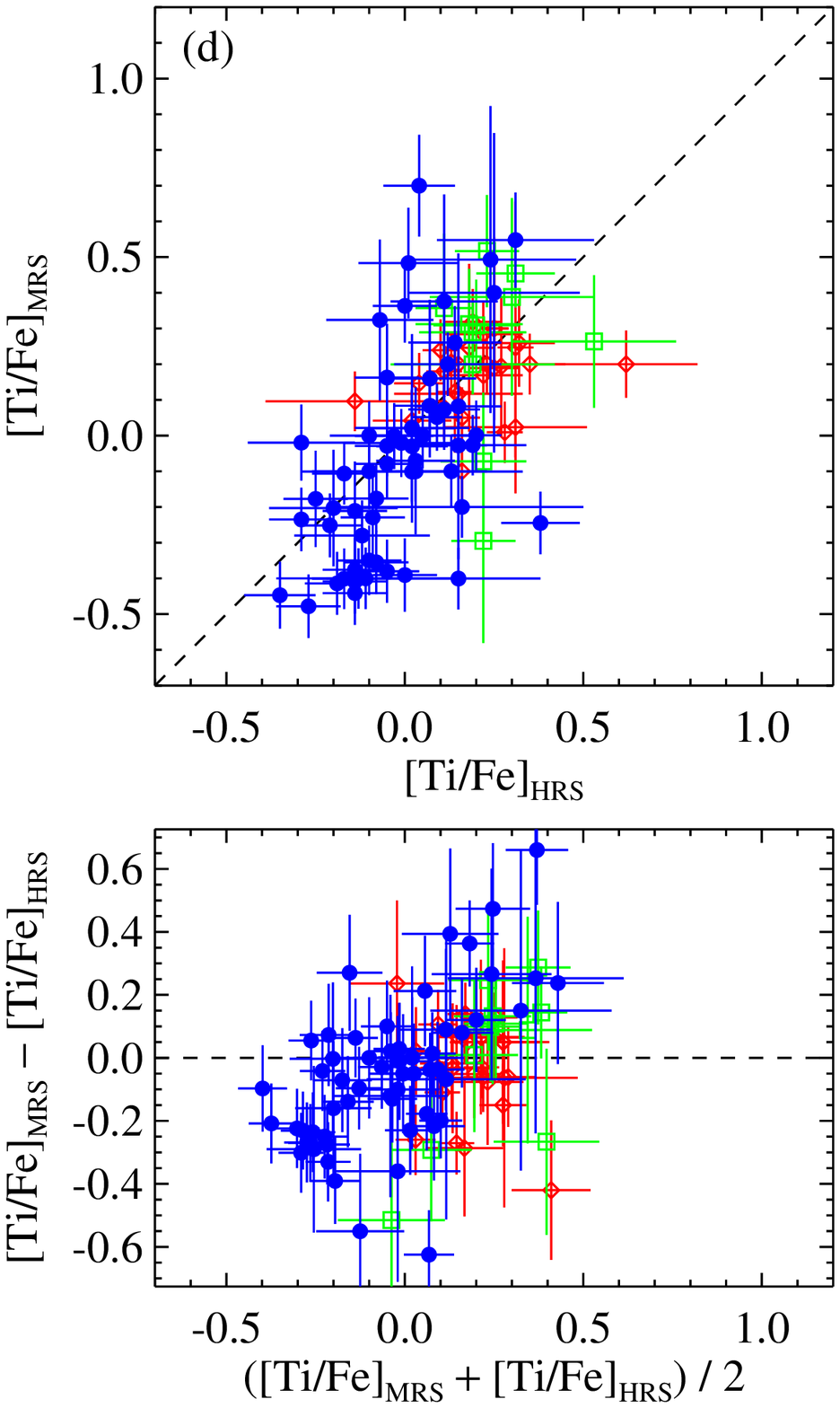}
\caption{Same as Fig.~\ref{fig:feh} except for (a) [Mg/Fe], (b)
  [Si/Fe], (c) [Ca/Fe], and (d) [Ti/Fe].\notetoeditor{B\&W figure
    caption: Same as Fig.~\ref{fig:feh} except for (a) [Mg/Fe], (b)
    [Si/Fe], (c) [Ca/Fe], and (d) [Ti/Fe].  (A color version of this
    figure is available in the online journal.)}\label{fig:alpha}}
\end{figure*}

\begin{figure}[t!]
\centering
\includegraphics[width=0.4\textwidth]{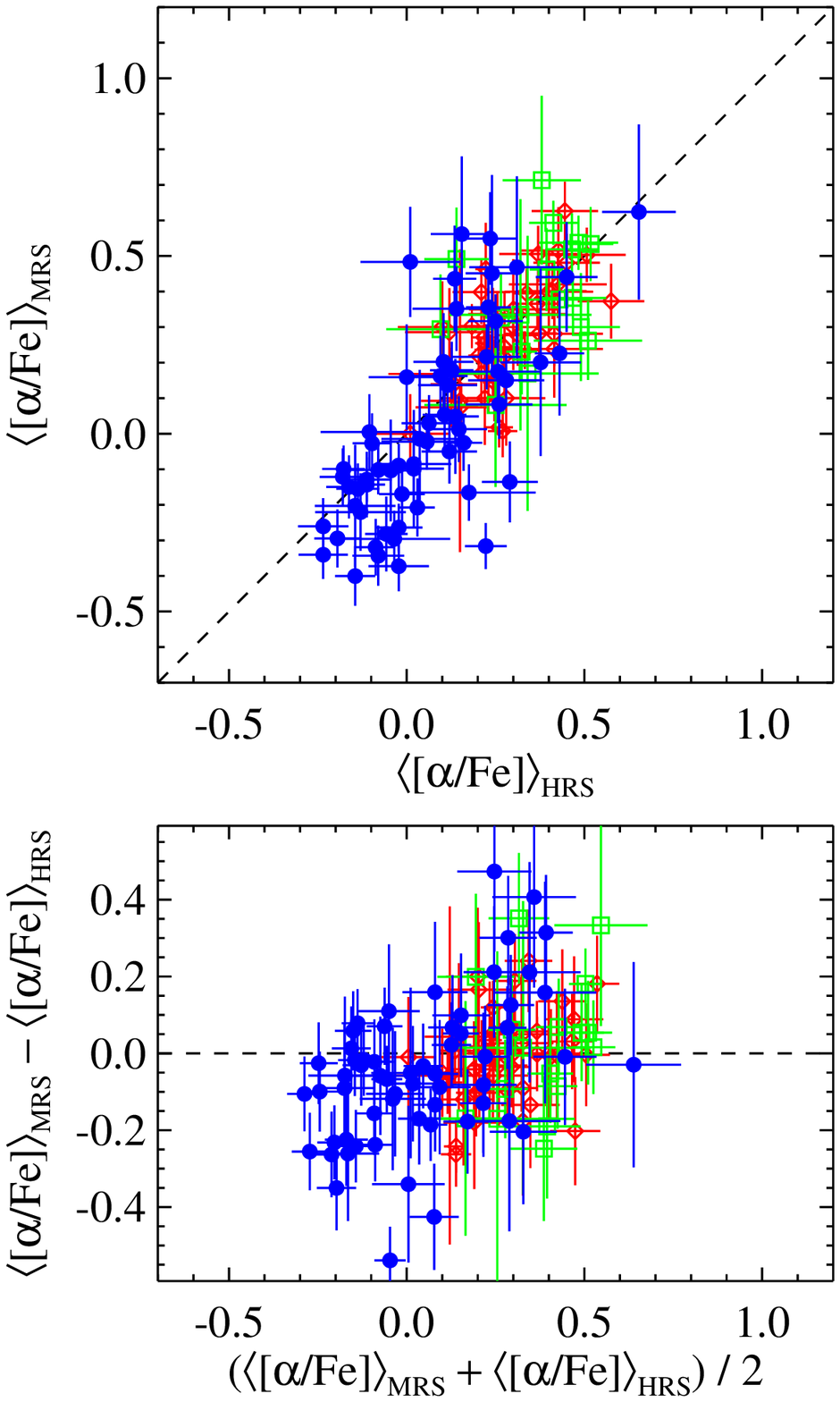}
\caption{Same as Fig.~\ref{fig:alpha} except for an average of the
  $\alpha$ elements.  The average includes only those
  \xfe\ measurements with estimated uncertainties less than
  0.5~dex.\notetoeditor{B\&W figure caption: Same as
    Fig.~\ref{fig:alpha} except for an average of the $\alpha$
    elements.  The average includes only those \xfe\ measurements with
    estimated uncertainties less than 0.5~dex.  (A color version of
    this figure is available in the online
    journal.)}\label{fig:alphaavg}}
\end{figure}

\begin{figure*}[p]
\centering
\begin{minipage}[t]{0.49\textwidth}
\centering
\includegraphics[width=\textwidth]{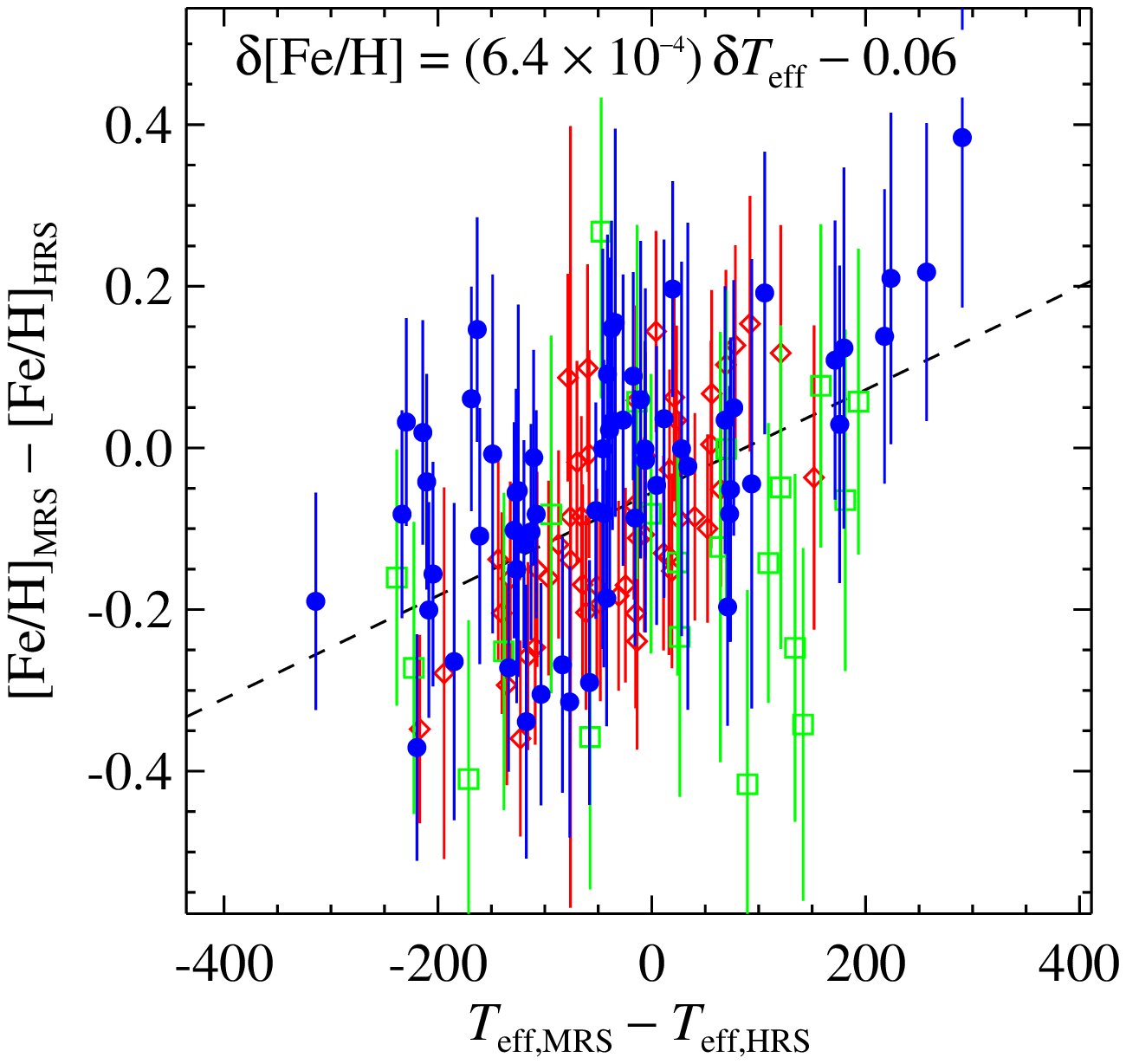}
\caption{Covariance between errors in \teff\ and \feh.  The $x$-axis
  shows the difference between the MRS and HRS measurements of \teff,
  and the $y$-axis shows the same for \feh.  The colors and shapes of
  the points are the same as in Fig.~\ref{fig:feh}.  The dashed line
  is a least-squares fit.  The strong covariance between these
  differences illustrates the degeneracy between \teff\ and
  \feh\ inherent in stellar spectral analysis.\notetoeditor{B\&W
    figure caption: Covariance between errors in \teff\ and \feh.  The
    $x$-axis shows the difference between the MRS and HRS measurements
    of \teff, and the $y$-axis shows the same for \feh.  The shapes of
    the points are the same as in Fig.~\ref{fig:feh}.  The dashed line
    is a least-squares fit.  The strong covariance between these
    differences illustrates the degeneracy between \teff\ and
    \feh\ inherent in stellar spectral analysis.  (A color version of
    this figure is available in the online
    journal.)}\label{fig:covartefffeh}}
\end{minipage}
\hfill
\begin{minipage}[t]{0.49\textwidth}
\centering
\includegraphics[width=\textwidth]{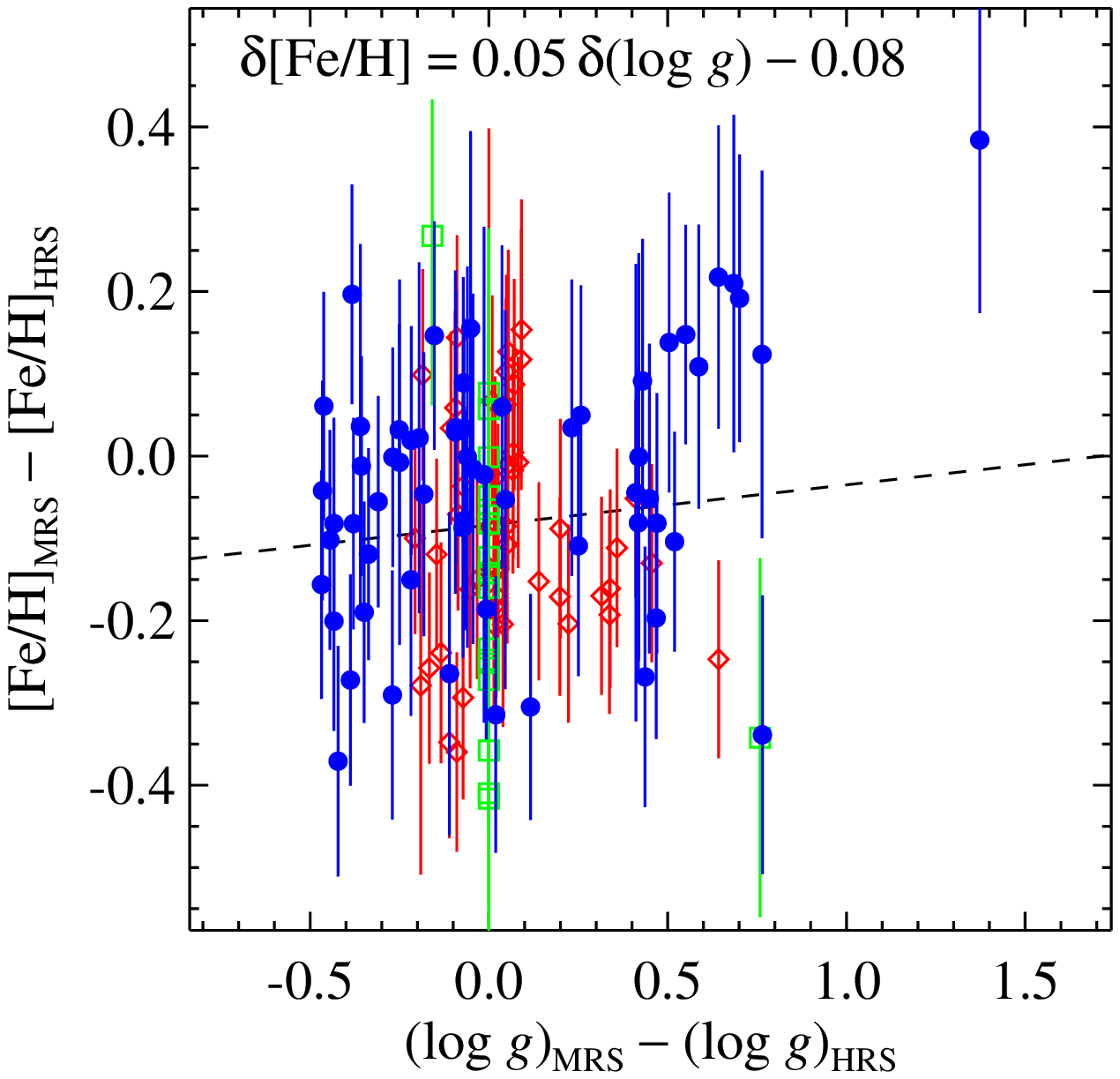}
\caption{Covariance between errors in \logg\ and \feh.  See
  Fig.~\ref{fig:covartefffeh} for further explanation.  The weak
  covariance between these differences illustrates the slight
  degeneracy between \logg\ and \feh\ inherent in stellar atmosphere
  analysis of mostly neutral metal lines.  The combination of very
  discrepant \teff, \logg, and $\xi$ causes the outlier with
  $\mathrm{[Fe/H]}_{\mathrm MRS} - \mathrm{[Fe/H]}_{\mathrm MRS} =
  +0.4$.\notetoeditor{B\&W figure caption: Covariance between errors
    in \logg\ and \feh.  See Fig.~\ref{fig:covartefffeh} for further
    explanation.  The weak covariance between these differences
    illustrates the slight degeneracy between \logg\ and
    \feh\ inherent in stellar atmosphere analysis of mostly neutral
    metal lines.  The combination of very discrepant \teff, \logg, and
    $\xi$ causes the outlier with $\mathrm{[Fe/H]}_{\mathrm MRS} -
    \mathrm{[Fe/H]}_{\mathrm MRS} = +0.4$.  (A color version of this
    figure is available in the online
    journal.)}\label{fig:covarloggfeh}}
\end{minipage}
\end{figure*}

\begin{figure*}[p]
\centering
\begin{minipage}[t]{0.49\textwidth}
\centering
\includegraphics[width=\textwidth]{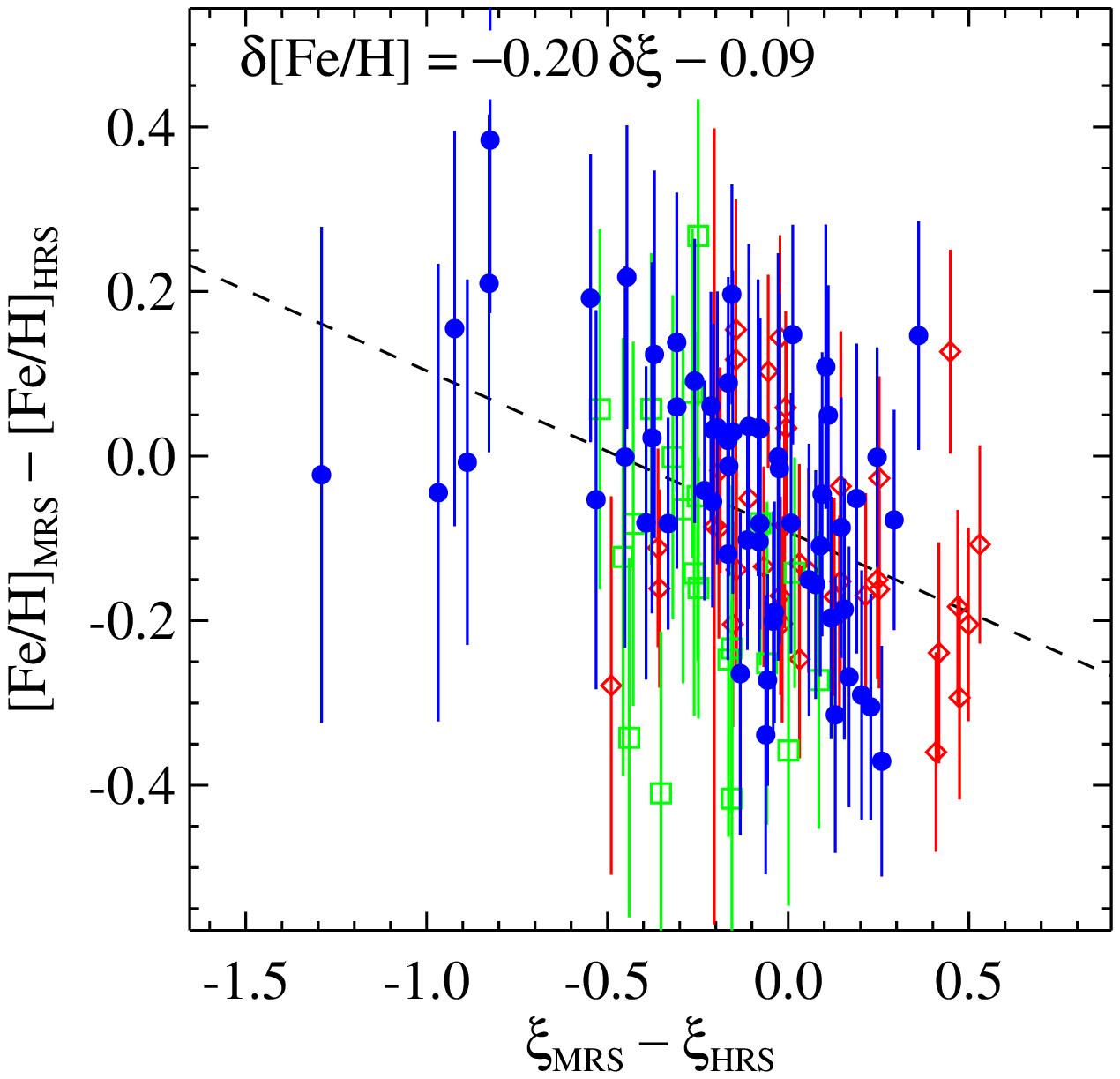}
\caption{Covariance between errors in $\xi$ and \feh.  See
  Fig.~\ref{fig:covartefffeh} for further explanation.  The strong
  anticorrelation illustrates the degeneracy between $\xi$ and
  \feh\ inherent in stellar atmosphere analysis.\notetoeditor{B\&W
    figure caption: Covariance between errors in $\xi$ and \feh.  See
    Fig.~\ref{fig:covartefffeh} for further explanation.  The strong
    anticorrelation illustrates the degeneracy between $\xi$ and
    \feh\ inherent in stellar atmosphere analysis.  (A color version
    of this figure is available in the online
    journal.)}\label{fig:covarvtfeh}}
\end{minipage}
\hfill
\begin{minipage}[t]{0.49\textwidth}
\centering
\includegraphics[width=\textwidth]{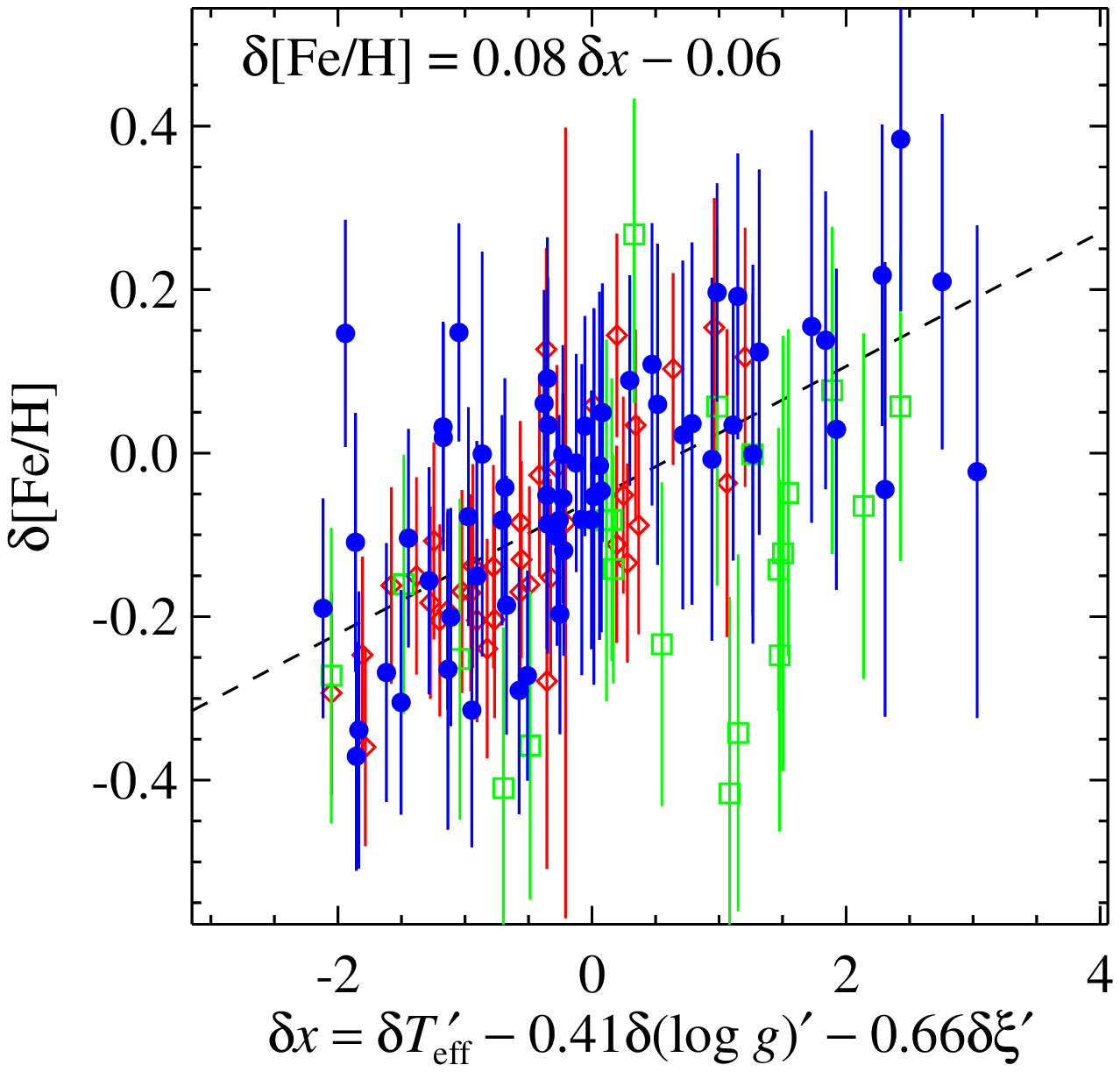}
\caption{Covariance between \feh\ and the linear combination of
  atmospheric parameters that minimizes the $\chi^2$ of the linear fit
  to differences between the MRS and HRS measurements of \feh.  The
  symbol $\delta$ represents the difference between MRS and HRS
  measurements, as in Figs.~\ref{fig:covartefffeh} to
  \ref{fig:covarvtfeh}.  The primes in the $x$-axis label indicate
  that the quantities have been normalized by their standard
  deviations (\allteffdiffsigma~K, \allloggdiffsigma, and
  \allvtdiffsigma~km~s$^{-1}$ for $\delta\mathteff$,
  $\delta(\mathlogg)$, and $\delta\xi$,
  respectively).\notetoeditor{B\&W figure caption: Covariance between
    \feh\ and the linear combination of atmospheric parameters that
    minimizes the $\chi^2$ of the linear fit to differences between
    the MRS and HRS measurements of \feh.  The symbol $\delta$
    represents the difference between MRS and HRS measurements, as in
    Figs.~\ref{fig:covartefffeh} to \ref{fig:covarvtfeh}.  The primes
    in the $x$-axis label indicate that the quantities have been
    normalized by their standard deviations (\allteffdiffsigma~K,
    \allloggdiffsigma, and \allvtdiffsigma~km~s$^{-1}$ for
    $\delta\mathteff$, $\delta(\mathlogg)$, and $\delta\xi$,
    respectively).  (A color version of this figure is available in
    the online journal.)}\label{fig:leastsquares}}
\end{minipage}
\end{figure*}

The most reliable test of the MRS atmospheric parameter and abundance
estimates is to compare with completely independent high-resolution
observations and analyses of the same stars.
Table~\ref{tab:hrscompare} lists the names, references, and
coordinates of stars with previously published HRS measurements that
we have also observed with DEIMOS.  The GC and dSph stars were some of
the targets on multi-slit masks, and the MW halo stars were observed
through a longslit mask.  The one star in M92 was also observed
through a longslit.

Different authors prefer different measurements of the solar
compositions.  We have placed all abundance measurements in this paper
on the same scale by adjusting the published HRS values by the
differences between the solar composition adopted by each HRS study
and the solar composition given in Table~\ref{tab:solar}.  The note at
the bottom of the table explains our choice of solar composition.

All HRS studies share some techniques in common, but some aspects of
the analyses differ.  Table~\ref{tab:hrsmethod} summarizes the
components of the HRS methods that change from study to study.  The
HRS studies we cite here all measure equivalent widths (EWs) of
individual metal lines, and they compute abundances from those EWs,
model atmospheres, and line lists.  The line lists vary from study to
study, causing typically small changes in derived abundances ($\la
0.1$~dex).  Also, different studies use different codes to calculate
model atmospheres, e.g., ATLAS9 \citep{kur93,cas04} and MARCS
\citep{gus75}.  The different codes used to calculate the abundances
are given in footnote b of Table~\ref{tab:hrsmethod}.

Most importantly, different HRS studies determine \teff\ and \logg\ in
different ways.  Sometimes effective temperature is determined through
excitation equilibrium, wherein \teff\ is adjusted to minimize the
trend of abundance with the excitation potential of the Fe transition.
Surface gravity is sometimes determined through ionization balance,
wherein \logg\ is adjusted until the abundance measured from lines of
\ion{Fe}{1} agrees with the abundance measured from lines of
\ion{Fe}{2}.  \citet{gra08} explains these methods in detail.  Some
authors of HRS studies use photometric determinations of \teff\ and
\logg\ only, relying on empirical calibrations or theoretical
isochrones.  For a discussion on the merits and disadvantages of
photometrically determined atmospheric parameters, we refer the reader
to \citet{iva01}.

It is tempting to assume that our HRS comparison set is a flawless
standard, but the heterogeneity of the sources of these abundance
estimates introduces systematic error.  Measurement uncertainties from
HRS are generally smaller than from MRS, but systematic offsets
between studies arise from different choices of methods of determining
atmospheric parameters, line lists, model atmospheres, and spectral
synthesis codes.  Table~\ref{tab:hrsmethod} is meant to illustrate the
diversity of ways to measure elemental abundances spectroscopically.
These differences should be borne in mind when examining the HRS
measurements presented below.  The sample sizes are often too small to
make a meaningful statistical quantification of bias between HRS
studies of the same stars or stellar systems.  Careful attention
should be paid to potential systematic offsets between HRS studies.
Although our analysis is also subject to its own random uncertainties
and systematic errors, one of the principal advantages of our sample
is that the abundances of all of the stars have been measured in a
homogeneous fashion, eliminating most relative systematic offsets from
star to star.

Table~\ref{tab:hrsdiff} lists the differences and standard deviations
between MRS and HRS measurements for all three types of stellar
systems.  The quantities are very similar to comparisons between
different HRS measurements of the same stars \citep[e.g.,][their
  Appendix~B]{coh08}.

Figures~\ref{fig:teff} to \ref{fig:alphaavg} show the comparison
between HRS measurements ($x$-axes) and MRS measurements ($y$-axes).
The points are coded by their membership in GCs, the MW halo field, or
dSphs.  Coding the points by individual system---such as the identity
of the GC or dSph---and also by HRS reference is also instructive, but
the plots contain too many points for this coding to be feasible.
Tables~\ref{tab:hrscompare} and \ref{tab:hrsmethod} contain more
complete descriptions of the comparisons between MRS and HRS
measurements.

The bottom panels of Figs.~\ref{fig:teff}--\ref{fig:alphaavg} show the
differences between MRS and HRS measurements ($y-x$).  Instead of
plotting $y-x$ versus $x$, we have plotted $y-x$ versus $(y+x)/2$, the
average of $x$ and $y$.  This is effectively a scaled orthogonal
distance from the one-to-one line.  (A pure rotation of the upper
panel, $y$ versus $x$, would be $(y-x)/\sqrt{2}$ versus
$(y+x)/\sqrt{2}$, but $y-x$ is easier to interpret than
$(y-x)/\sqrt{2}$.)  We have chosen the orthogonal distance as the
abscissa because a plot of $y-x$ versus $x$ would show trends even if
$x$ and $y$ are drawn from the same distribution with random scatter.
In fact, any random uncertainty in $x$ would cause a downward sloping
trend in $y-x$ versus $x$.  In the orthogonal residual plots, random
scatter along the one-to-one line does not produce a trend as long as
the magnitude of the scatter in $x$ is close to magnitude of the
scatter in $y$.  The uncertainties in the MRS measurements are
generally slightly larger than the uncertainties in the HRS
measurements, but not enough to produce these trends in the orthogonal
residuals.


Figure~\ref{fig:teff} compares $T_{\rm eff,HRS}$ and $T_{\rm
  eff,MRS}$.  The average difference in \teff\ between MRS and HRS
depends on the source of the HRS measurement.  As one example,
consider the MW halo field star sample.  \citet{joh02} calculate
\teff\ spectroscopically, whereas \citet{lai04,lai07} calculate
\teff\ photometrically.  Our measurements of \teff\ are typically
$\sim 100$~K below those of \citet{joh02} and $\sim 100$~K above those
of \citet{lai04,lai07}.  The differences possibly result from the
different methods of measuring \teff.

As another example of a trend seen with the source of the HRS
measurements, consider the study of \citet{fre10b}.  Five of the six
stars in the Ursa Major~II and Coma Berenices data set have $T_{\rm
  eff,MRS}$ at least 175~K higher than $T_{\rm eff,HRS}$.
Furthermore, $\xi_{\rm MRS}$ is between 0.3 and $0.8~{\rm km~s}^{-1}$
less than $\xi_{\rm HRS}$ for all of the stars in
\citeauthor{fre10b}'s sample.  Our measurements of $\mathfeh_{\rm
  MRS}$ for the five stars with the significantly different
temperature measurements exceed $\mathfeh_{\rm HRS}$ by 0.1--0.4~dex.
In contrast to our measurements, \citeauthor{fre10b}\ used a blue
spectral range, higher resolution, and spectroscopically derived
surface gravities and microturbulent velocities.  Our choices are not
better, simply different.  The differences partly explain our
different offsets from different studies.

The standard deviation of $\mathfeh_{\rm MRS} - \mathfeh_{\rm HRS}$ is
$\allfehdiffsigma$.  The minimum uncertainty that we quote on
$\mathfeh_{\rm MRS}$ (in the limit of infinite spectral S/N) is
$\fehsyserr$.  The small difference between our minimum estimate of
uncertainty and the typical difference between MRS and HRS
measurements---which includes the error on $\mathfeh_{\rm HRS}$ in
addition to the error on $\mathfeh_{\rm MRS}$---indicates that we have
not underestimated our measurement uncertainties, even on an absolute
scale.  Furthermore, Fig.~\ref{fig:feh} shows no systematic trend in
$\mathfeh_{\rm MRS} - \mathfeh_{\rm HRS}$ as a function of \feh.
Therefore, our MRS measurements of \feh\ are consistent with HRS
measurements at least over the range $-4.0 < \mathfeh < -0.5$.

The MRS $\alpha$ element abundances also agree with HRS measurements.
(However, we added a constant to $\mathrm{[Si/Fe]}_{\mathrm{MRS}}$ to
force better agreement with $\mathrm{[Si/Fe]}_{\mathrm{HRS}}$.  See
Sec.~\ref{sec:continuum}.)  The standard deviation of the differences
between MRS and HRS $\langle[\alpha/\rm{Fe}]\rangle$ (an average of
the four measured \xfe\ ratios) is $\allalphafediffsigma$, about the
same as the standard deviation for \feh.  The number of detectable Mg,
Si, Ca, and Ti absorption lines in the DEIMOS spectral range is
comparable to the number of Fe absorption lines for red giants.
Therefore, it is reassuring that the precision of the
$\langle\mathafe\rangle_{\rm MRS}$ measurements is comparable to the
precision of the $\mathfeh_{\rm MRS}$ measurements.

We point out in particular an interesting feature of the individual
alpha element ratios concerning GC stars (red points in
Fig.~\ref{fig:alpha}).  Internal variations in Ca and Ti have not been
detected in the GCs presented here.  Furthermore, the [Ca/Fe] and
[Ti/Fe] ratios vary little from cluster to cluster.  Therefore, it is
expected that we see small or zero correlation between MRS and HRS
measurements for these elements.  In fact, the linear Pearson
correlation coefficients between MRS and HRS measurements are
$\gccacorcoeff$ and $\gcticorcoeff$ for [Ca/Fe] and [Ti/Fe],
respectively.  On the other hand, [Mg/Fe] shows a significant spread
within many GCs \citep[e.g.,][]{gra04}.  As a result, we see a
correlation between ${\rm [Mg/Fe]}_{\rm MRS}$ and ${\rm [Mg/Fe]}_{\rm
  HRS}$.  The correlation coefficient is $\gcmgcorcoeff$.  In other
words, the $\alpha$ element ratio measurements from DEIMOS can sort
out which of the four [X/Fe] ratios has a dispersion between GCs.

The large sample of stars observed with at least two independent
measurements provides a unique opportunity to examine the influence of
errors in atmospheric parameters on \feh.
Figures~\ref{fig:covartefffeh}--\ref{fig:covarvtfeh} show how
differences in \teff, \logg, and $\xi$ affect the measurement of \feh.
Each figure shows the least-squares linear fit.  The vertical error
bars are the quadrature sum of the total errors on \feh\ from MRS and
HRS, including the error introduced by uncertainty in atmospheric
parameters.  Horizontal error bars are not shown because most HRS
studies do not include errors on atmospheric parameters.  Not
surprisingly, \teff\ is the atmospheric parameter that most affects
the measurement of \feh.  In general, underestimating \teff\ leads to
an underestimate of \feh.  Reassuringly, the intercept of the
least-squares linear fit to $(\mathfeh_{\rm MRS} - \mathfeh_{\rm
  HRS})$ versus $(T_{\rm eff,MRS} - T_{\rm eff,HRS})$ is close to zero
($\tefffehint \pm \tefffehinterr$).  In other words, if \teff\ were
determined perfectly, $\mathfeh_{\rm MRS}$ and $\mathfeh_{\rm HRS}$
would agree extremely well.  Surface gravity does not have a strong
influence on the measurement of \feh, and microturbulent velocity has
a moderately strong influence.  Underestimating $\xi$ leads to an
overestimate of \feh.

In order to quantify the total effect of errors on atmospheric
parameters, we have identified the linear combination of \teff, \logg,
and $\xi$ differences that minimizes the scatter about the
least-squares linear fit to $(\mathfeh_{\rm MRS} - \mathfeh_{\rm
  HRS})$.  Figure~\ref{fig:leastsquares} shows the result.  To remove
dimensionality, $\delta\mathteff$, $\delta(\mathlogg)$, and
$\delta\xi$ have been normalized by their standard deviations.  As
expected, \teff\ has the most influence on the \feh\ measurement by
far.  Surface gravity and $\xi$ have about one half of the influence
of \teff.  The rms scatter about the line is \fehscatter~dex, compared
to \allfehdiffsigma~dex, which is the rms scatter in $(\mathfeh_{\rm
  MRS} - \mathfeh_{\rm HRS})$ without removing the effect of errors
from atmospheric parameters.  Therefore, uncertainty in \teff\ and
$\xi$ do inflate the error on \feh.  This result possibly indicates
that S/N (for these bright comparison stars) and spectral resolution
are not limiting the precision of $\mathfeh_{\rm MRS}$.  Instead,
improving the determinations of \teff\ and $\xi$ would do much to
improve the measurement of $\mathfeh_{\rm MRS}$.  The problem of
measuring \teff\ and $\xi$ is not unique to MRS.
Table~\ref{tab:hrsmethod} shows that many HRS studies, particularly
for the more distant stars, employ photometry---at least in part---for
determining \teff, \logg, and $\xi$, as we do.



\section{Summary}
\label{sec:summary}

We have presented a catalog of Fe, Mg, Si, Ca, and Ti abundance
measurements for \ndsphstars\ stars in eight dwarf satellite galaxies
of the Milky Way.  Medium-resolution spectroscopy from the Keck/DEIMOS
spectrograph provided the throughput and multiplexing necessary to
perform these measurements for the large number of faint, distant
stars.  The majority of these stars are inaccessible to
high-resolution spectrographs even on the largest telescopes.  The
measurements relied on a spectral synthesis technique discussed
previously (KGS08 and \citeauthor*{kir09}) with some modifications
detailed in this paper.

We have estimated the uncertainty on every measurement of \teff,
\logg, \feh, [Mg/Fe], [Si/Fe], [Ca/Fe], and [Ti/Fe].  The
uncertainties on the abundances include the effect of spectral noise,
spectral modeling uncertainties, and uncertainties in atmospheric
parameters (effective temperature and surface gravity).  The estimated
uncertainties were shown to be accurate based on duplicate
observations of stars in the scientific targets, dwarf galaxies.

Finally, we have quantified the accuracy of our medium-resolution
measurements by observing with DEIMOS a sample of stars with
high-resolution spectroscopic measurements.  We deliberately targeted
stars in dwarf galaxies with previous HRS measurements, and we
observed stars in globular clusters and in the field of the Milky Way
stellar halo.  The mean difference in \feh\ and
$\langle[\alpha/\rm{Fe}]\rangle$ between the HRS and MRS measurements
of these \nhrs\ stars is $\allfehdiffmean$ and $\allalphafediffmean$,
with standard deviations of $\allfehdiffsigma$ and
$\allalphafediffsigma$.

The next paper in this series focuses on the metallicity evolution of
the individual dSphs.  We fit chemical evolution models to the
metallicity distributions.  The shapes of the distributions and
parameters of the fits show trends with dSph luminosity.  The
following paper in the series addresses the star formation timescale
and chemical evolution as revealed by the distribution of the $\alpha$
elements.

\acknowledgments

The authors thank the referee, Piercarlo Bonifacio, for his thoughtful
suggestions, which improved this manuscript.  We gratefully
acknowledge Sandra Faber, Ricardo Schiavon, and Michael Cooper of the
DEEP2 team for acquiring Keck/DEIMOS spectroscopy of bright Milky Way
halo field stars during nights of poor transparency.  We also thank
Peter Stetson for providing additional globular cluster photometry and
Bob Kraft for helpful discussions and for providing some of the
globular cluster spectroscopy for this work.  The generation of
synthetic spectra made use of the University of California Santa Cruz
Pleiades supercomputer and the Yale High Performance Computing cluster
Bulldog.  We thank Joel Primack for sharing his allocation of Pleiades
supercomputer time and Mario Juri{\'c} for a helpful discussion on the
mass of the components of the Milky Way.

Support for this work was provided by NASA through Hubble Fellowship
grant HST-HF-51256.01 awarded to ENK by the Space Telescope Science
Institute, which is operated by the Association of Universities for
Research in Astronomy, Inc., for NASA, under contract NAS 5-26555.  PG
acknowledges NSF grants AST-0307966, AST-0607852, and AST-0507483.  MG
acknowledges support from NSF grant AST-0908752.  CS acknowledges NSF
grant AST-0909978.  JGC acknowledges NSF grant AST-090109.  MHS was
supported at PSU by NASA contract NAS5-00136.  This research used the
facilities of the Canadian Astronomy Data Centre operated by the
National Research Council of Canada with the support of the Canadian
Space Agency.

The authors wish to recognize and acknowledge the very significant
cultural role and reverence that the summit of Mauna Kea has always
had within the indigenous Hawaiian community.  We are most fortunate
to have the opportunity to conduct observations from this mountain.

{\it Facility:} \facility{Keck:II (DEIMOS)}

\clearpage
\LongTables
\begin{deluxetable*}{llcr@{ }c@{ }lccl}
\tablenum{2}
\tablewidth{0pt}
\tablecolumns{9}
\tablecaption{DEIMOS Observations} 
\tablehead{\colhead{Object} & \colhead{Slitmask} & \colhead{\# targets} & \multicolumn{3}{c}{Date} & \colhead{Airmass} & \colhead{Seeing} & \colhead{Exposures}}
\startdata
\cutinhead{Globular Clusters}
NGC~288          & n288     &       119 & 2008 & Nov & 24 & 1.92 & $1\farcs3$\phn & 300~s, 2 $\times$ 420~s \\*
                 &          &           & 2008 & Nov & 25 & 1.86 & $1\farcs23$    & 4 $\times$ 300~s \\
M79       & n1904\tablenotemark{a} & \phn22 & 2006 & Feb & 2  & 1.42 & unknown        & 2 $\times$ 300~s \\*
                 & ng1904   &       104 & 2009 & Feb & 22 & 1.40 & $1\farcs12$    & 2 $\times$ 600~s, 2 $\times$ 1200~s \\
NGC~2419  & n2419\tablenotemark{a} & \phn70 & 2006 & Feb & 2  & 1.21 & unknown        & 4 $\times$ 300~s \\*
                 & n2419c   &    \phn94 & 2009 & Oct & 13 & 1.15 & $0\farcs56$    & 1200~s, 900~s \\*
                 &          &           & 2009 & Oct & 14 & 1.15 & $0\farcs51$    & 200~s, 900~s \\
M5               & ng5904   &       181 & 2009 & Feb & 22 & 1.05 & $0\farcs64$    & 1200~s, 900~s, 600~s, 480~s \\
M13              & n6205    &    \phn93 & 2007 & Oct & 12 & 1.35 & unknown        & 3 $\times$ 300~s \\
M92              & LVMslits & \phn\phn1 & 2008 & May & 6  & 1.09 & unknown        & 500~s, 2 $\times$ 300~s \\
M71              & n6838    &       104 & 2007 & Nov & 13 & 1.09 & $0\farcs6$\phn & 3 $\times$ 300~s \\
NGC~7006         & n7006    &       105 & 2007 & Nov & 15 & 1.01 & $0\farcs57$    & 2 $\times$ 300~s \\
M15              & n7078    &    \phn63 & 2007 & Nov & 14 & 1.01 & $0\farcs77$    & 2 $\times$ 300~s \\*
                 & n7078d   &       164 & 2009 & Oct & 13 & 1.01 & $0\farcs53$    & 3 $\times$ 900~s \\*
                 & n7078e   &       167 & 2009 & Oct & 14 & 1.01 & $0\farcs61$    & 3 $\times$ 900~s \\
M2               & n7089b   &    \phn91 & 2009 & Oct & 13 & 1.09 & $0\farcs57$    & 3 $\times$ 900~s \\
Pal 13           & pal13    &    \phn33 & 2009 & Oct & 13 & 1.48 & $0\farcs60$    & 2 $\times$ 900~s \\*
                 &          &           & 2009 & Oct & 14 & 1.50 & $0\farcs74$    & 900~s, 822~s \\
NGC~7492         & n7492    &    \phn38 & 2007 & Nov & 15 & 1.30 & $0\farcs57$    & 2 $\times$ 210~s \\
\cutinhead{Halo Field Stars}
HE 0012-1441     & LVMslits & \phn\phn1 & 2010 & Aug & 11 & 1.75 & $> 1.5\arcsec$ & 2 $\times$ 900~s \\*
HD 88609         & LVMslits & \phn\phn1 & 2008 & May & 6  & 1.22 & unknown        & 30~s, 2 $\times$ 100~s \\*
HD 115444        & LVMslits & \phn\phn1 & 2008 & May & 6  & 1.04 & unknown        & 200~s, 100~s \\*
BS 16467-062     & LVMslits & \phn\phn1 & 2008 & May & 6  & 1.00 & unknown        & 4 $\times$ 1000~s \\*
HD 122563        & LVMslits & \phn\phn1 & 2008 & May & 6  & 1.20 & unknown        & 3 $\times$ 5~s \\*
BS 16550-087     & LVMslits & \phn\phn1 & 2008 & May & 6  & 1.04 & unknown        & 4 $\times$ 1000~s \\*
CS 30325-028     & LVMslits & \phn\phn1 & 2008 & May & 6  & 1.10 & unknown        & 3 $\times$ 500~s \\*
CS 30329-129     & LongMirr & \phn\phn1 & 2008 & Apr & 11 & 1.52 & unknown        & 2 $\times$ 120~s \\*
BD $+$5 3098     & LVMslits & \phn\phn1 & 2008 & May & 6  & 1.07 & unknown        & 3 $\times$ 50~s \\*
BS 16084-160     & LongMirr & \phn\phn1 & 2008 & Apr & 11 & 1.29 & unknown        & 2 $\times$ 60~s \\*
BD $+$9 3223     & LVMslits & \phn\phn1 & 2008 & May & 6  & 1.02 & unknown        & 3 $\times$ 50~s \\*
BS 16080-054     & LongMirr & \phn\phn1 & 2008 & Apr & 11 & 1.36 & unknown        & 3 $\times$ 120~s \\*
CS 22878-101     & LongMirr & \phn\phn1 & 2008 & Apr & 8  & 1.10 & unknown        & 3 $\times$ 120~s \\*
BS 16080-093     & LongMirr & \phn\phn1 & 2008 & Apr & 11 & 1.33 & unknown        & 3 $\times$ 120~s \\*
BD $+$23 3130    & LongMirr & \phn\phn1 & 2008 & Apr & 11 & 1.00 & unknown        & 2 $\times$ 45~s \\*
HD 165195        & LVMslits & \phn\phn1 & 2008 & May & 6  & 1.05 & unknown        & 3 $\times$ 10~s \\*
HD 186478        & LVMslits & \phn\phn1 & 2008 & May & 6  & 1.47 & unknown        & 2 $\times$ 50~s, 300~s \\*
BD $-$18 5550    & LVMslits & \phn\phn1 & 2008 & May & 6  & 1.30 & unknown        & 3 $\times$ 100~s\\*
BD $-$17 6036    & LVMslits & \phn\phn1 & 2008 & May & 6  & 1.47 & unknown        & 3 $\times$ 200~s \\*
CS 22880-086     & LVMslits & \phn\phn1 & 2008 & May & 6  & 1.40 & unknown        & 4 $\times$ 1000~s \\*
HE 2323-0256     & LVMslits & \phn\phn1 & 2010 & Aug & 11 & 1.18 & $> 1.5\arcsec$ & 2 $\times$ 600~s \\
\cutinhead{dSphs}
Sculptor         & scl1     &    \phn86 & 2008 & Aug & 3  & 1.79 & $0\farcs85$    & 3 $\times$ 1200~s \\*
                 & scl2     &       106 & 2008 & Aug & 3  & 1.68 & $0\farcs85$    & 2 $\times$ 900~s \\*
                 & scl3     &    \phn87 & 2008 & Aug & 4  & 1.67 & $0\farcs94$    & 462~s \\*
                 &          &           & 2008 & Aug & 31 & 1.67 & $0\farcs77$    & 1000~s, 834~s \\*
                 & scl5     &    \phn95 & 2008 & Sep & 1  & 1.73 & $0\farcs84$    & 3 $\times$ 720~s \\*
                 & scl6     &    \phn91 & 2008 & Sep & 1  & 1.88 & $1\farcs23$    & 3 $\times$ 720~s \\
Fornax           & for1B    &       166 & 2008 & Sep & 1  & 2.00 & $1\farcs08$    & 3 $\times$ 500~s \\*
                 &          &           & 2008 & Nov & 25 & 1.86 & $0\farcs7$\phn & 2 $\times$ 1200~s \\*
                 & for3B    &       169 & 2008 & Sep & 1  & 1.71 & $0\farcs85$    & 3 $\times$ 500~s \\*
                 & for4B    &       164 & 2008 & Nov & 26 & 1.74 & $1\farcs1$\phn & 1200~s, 1020~s \\*
                 & for6     &       169 & 2008 & Aug & 31 & 1.73 & $0\farcs68$    & 3 $\times$ 500~s \\*
                 &          &           & 2008 & Nov & 25 & 2.31 & $1\farcs2$\phn & 2 $\times$ 1200~s \\*
                 & for7     &       169 & 2008 & Aug & 31 & 1.71 & $0\farcs76$    & 2 $\times$ 500~s, 460~s \\*
                 &          &           & 2008 & Sep & 30 & 1.71 & unknown        & 3 $\times$ 600~s \\
Leo I            & LeoI\_1\tablenotemark{b}  &    \phn42 & 2003 & Oct & 29 & 1.86 & unknown        & 4800~s \\*
                 & LeoI\_2\tablenotemark{b}  &    \phn83 & 2004 & Oct & 15 & 1.9 \phn & unknown        & 6900~s \\*
                 & LIN1\_1  &       112 & 2006 & Feb & 2  & 1.82 & unknown        & 3600~s (total) \\*
                 & LIN1\_2  &       100 & 2006 & Feb & 2  & 1.23 & unknown        & 3600~s (total) \\*
                 & LIN1\_3  &    \phn92 & 2006 & Feb & 2  & 1.04 & unknown        & 3600~s (total) \\*
                 & LIN1\_4  &       100 & 2006 & Feb & 2  & 1.01 & unknown        & 2520~s (total) \\*
                 & LIN2\_1  &       104 & 2006 & Feb & 3  & 1.83 & unknown        & 4800~s (total) \\*
                 & LIN2\_2  &    \phn98 & 2006 & Feb & 3  & 1.20 & unknown        & 4800~s (total) \\*
                 & LIN2\_3  &    \phn67 & 2006 & Feb & 3  & 1.01 & unknown        & 3600~s (total) \\*
                 & LIN2\_4  &       102 & 2006 & Feb & 3  & 1.07 & unknown        & 3600~s (total) \\*
                 & LIN3\_1  &    \phn86 & 2006 & Feb & 4  & 1.43 & unknown        & 3011~s (total) \\*
                 & LIN3\_2  &    \phn67 & 2006 & Feb & 4  & 1.21 & unknown        & 4800~s (total) \\*
                 & LIN3\_3  &    \phn69 & 2006 & Feb & 4  & 1.02 & unknown        & 4500~s (total) \\*
                 & LIN3\_4  &    \phn88 & 2006 & Feb & 4  & 1.04 & unknown        & 1800~s (total) \\
Sextans          & sex1     &       108 & 2009 & Feb & 22 & 1.53 & $0\farcs96$    & 3 $\times$ 1200~s, 396~s \\*
                 & sex2     &    \phn85 & 2009 & Feb & 22 & 1.21 & $0\farcs70$    & 3 $\times$ 1200~s \\*
                 & sex3     &    \phn88 & 2009 & Feb & 22 & 1.09 & $0\farcs93$    & 4 $\times$ 1200~s \\*
                 & sex4     &       109 & 2009 & Feb & 23 & 1.51 & $1\farcs06$    & 4 $\times$ 1200~s \\*
                 & sex6     &       100 & 2009 & Feb & 23 & 1.19 & $0\farcs86$    & 3 $\times$ 1200~s \\
Leo II           & L2A      &    \phn72 & 2006 & Feb & 2  & 1.00 & unknown        & 3300~s (total) \\*
                 & L2B      &    \phn56 & 2006 & Feb & 3  & 1.29 & unknown        & 1080~s (total) \\*
                 & L2C      &    \phn77 & 2006 & Feb & 4  & 1.01 & unknown        & 3960~s (total) \\*
                 & L2D      &    \phn70 & 2006 & Feb & 2  & 1.09 & unknown        & 3600~s (total) \\*
                 & L2E      &    \phn57 & 2006 & Feb & 3  & 1.07 & unknown        & 3600~s (total) \\*
                 & L2F      &    \phn62 & 2006 & Feb & 4  & 1.13 & unknown        & 3400~s (total) \\
Canes Venatici I & CVn1-1\tablenotemark{a}   &    \phn91 & 2007 & Feb & 14 & 1.14 & unknown        & 4140~s (total) \\*
                 & CVn1-2\tablenotemark{a}   &    \phn94 & 2007 & Feb & 14 & 1.06 & unknown        & 4140~s (total) \\*
                 & CVn1-3\tablenotemark{a}   &    \phn90 & 2007 & Feb & 14 & 1.03 & unknown        & 4860~s (total) \\*
                 & CVn1-dp\tablenotemark{a}  &       115 & 2007 & Feb & 15 & 1.03 & unknown        & 9000~s (total) \\
Ursa Minor       & umi1     &       125 & 2009 & Feb & 22 & 1.99 & $0\farcs71$    & 3 $\times$ 1200~s \\*
                 & umi2     &       134 & 2009 & Feb & 22 & 1.79 & $1\farcs00$    & 2 $\times$ 1200~s, 1400~s \\*
                 & umi3     &       137 & 2009 & Feb & 23 & 1.48 & $0\farcs98$    & 3 $\times$ 1200~s \\*
                 & umi6     &       137 & 2009 & Feb & 23 & 1.57 & $0\farcs93$    & 3 $\times$ 1200~s \\
Draco            & dra1     &       151 & 2009 & May & 23 & 1.27 & $0\farcs67$    & 3 $\times$ 1200~s \\*
                 & dra2     &       167 & 2009 & May & 23 & 1.28 & $0\farcs74$    & 1200~s, 1000~s, 900~s \\*
                 & dra3     &       140 & 2009 & May & 23 & 1.35 & $0\farcs73$    & 3 $\times$ 960~s \\*
                 & dra4     &       140 & 2009 & May & 23 & 1.50 & $0\farcs63$    & 2 $\times$ 960~s, 700~s \\*
                 & dra5     &    \phn78 & 2009 & May & 24 & 1.28 & $0\farcs69$    & 3 $\times$ 1200~s \\*
                 & dra7     &    \phn75 & 2009 & May & 24 & 1.35 & $0\farcs75$    & 3 $\times$ 960~s, 1080~s \\*
                 & dra8     &    \phn60 & 2009 & May & 24 & 1.55 & $0\farcs57$    & 3 $\times$ 840~s \\*
                 & dra9     &       106 & 2009 & May & 24 & 1.28 & $0\farcs66$    & 3 $\times$ 960~s \\
\enddata
\tablenotetext{a}{Observations by \citet{sim07}.}
\tablenotetext{b}{Observations by \citet{soh07}.}
\end{deluxetable*}

\clearpage
\renewcommand{\thetable}{\arabic{table}}
\setcounter{table}{3}
\begin{deluxetable}{llcccccccc}
\tablewidth{0pt}
\tablecolumns{10}
\tablecaption{DEIMOS Multi-Element Abundances Catalog\label{tab:catalog}}
\tablehead{\colhead{dSph} & \colhead{Name} & \colhead{$\mathteff \pm \delta_{\rm spec} \pm \delta_{\rm phot}$} & \colhead{\logg} & \colhead{$\xi$} & \colhead{\feh} & \colhead{[Mg/Fe]} & \colhead{[Si/Fe]} & \colhead{[Ca/Fe]} & \colhead{[Ti/Fe]} \\
\colhead{ } & \colhead{ } & \colhead{(K)} & \colhead{(cm~s$^{-2}$)} & \colhead{(km~s$^{-1}$)} & \colhead{(dex)} & \colhead{(dex)} & \colhead{(dex)} & \colhead{(dex)} & \colhead{(dex)}}
\startdata
Scl & 1002473 & $5085 \pm     101 \pm     258$ & $2.02 \pm 0.08$ & 1.66 & $-2.30 \pm 0.15$ & $+0.65 \pm 0.93$ & $+0.59 \pm 0.17$ & $+0.64 \pm 0.26$ & $+0.26 \pm 0.48$ \\
Scl & 1002447 & $4689 \pm \phn 47 \pm     109$ & $1.27 \pm 0.06$ & 1.84 & $-2.04 \pm 0.12$ & $+0.36 \pm 0.28$ & $+0.47 \pm 0.12$ & $+0.33 \pm 0.16$ & $+0.28 \pm 0.09$ \\
Scl & 1002888 & $4818 \pm     109 \pm     209$ & $1.87 \pm 0.08$ & 1.70 & $-1.97 \pm 0.13$ &     \nodata      & $+0.30 \pm 0.15$ & $+0.47 \pm 0.21$ & $-0.02 \pm 0.61$ \\
Scl & 1003386 & $4660 \pm \phn 41 \pm     125$ & $1.49 \pm 0.06$ & 1.79 & $-1.30 \pm 0.11$ & $-0.50 \pm 0.44$ & $+0.21 \pm 0.23$ & $+0.14 \pm 0.16$ & $+0.03 \pm 0.21$ \\
Scl & 1003505 & $4366 \pm \phn 31 \pm \phn 71$ & $0.88 \pm 0.06$ & 1.93 & $-1.82 \pm 0.11$ & $+0.26 \pm 0.19$ & $+0.20 \pm 0.14$ & $+0.20 \pm 0.14$ & $+0.00 \pm 0.10$ \\
Scl & 1003443 & $4732 \pm \phn 41 \pm     110$ & $1.26 \pm 0.06$ & 1.84 & $-1.62 \pm 0.11$ & $+0.51 \pm 0.34$ & $+0.06 \pm 0.16$ & $+0.20 \pm 0.18$ & $+0.05 \pm 0.11$ \\
Scl & 1003537 & $4259 \pm \phn 23 \pm \phn 58$ & $0.58 \pm 0.07$ & 2.01 & $-2.28 \pm 0.11$ & $+0.42 \pm 0.21$ & $+0.40 \pm 0.12$ & $+0.34 \pm 0.14$ & $+0.20 \pm 0.09$ \\
Scl & 1003694 & $3923 \pm \phn 12 \pm \phn 58$ & $0.32 \pm 0.09$ & 2.07 & $-1.63 \pm 0.11$ & $+0.09 \pm 0.17$ & $+0.10 \pm 0.11$ & $+0.03 \pm 0.14$ & $-0.10 \pm 0.08$ \\
Scl & 1003702 & $4581 \pm \phn 58 \pm     111$ & $1.58 \pm 0.06$ & 1.77 & $-1.89 \pm 0.12$ &     \nodata      & $+0.34 \pm 0.36$ & $+0.10 \pm 0.14$ & $+0.11 \pm 0.17$ \\
Scl & 1003967 & $4525 \pm \phn 58 \pm \phn 86$ & $1.11 \pm 0.05$ & 1.88 & $-2.66 \pm 0.13$ &     \nodata      & $+0.28 \pm 0.82$ & $+0.60 \pm 0.22$ & $+0.40 \pm 0.23$ \\
\enddata
\tablerefs{In the online version of this paper, photometry is given
  from the following sources: Sculptor, \citet{wes06}; Fornax,
  \citet{ste98}; Leo~I, \citet{soh07}; Sextans, \citet{lee03}; Leo~II,
  unpublished, but obtained in the same fashion as for Leo~I; Canes
  Venatici~I, SDSS DR5 \citet{ade07}; Ursa Minor, \citet{bel02}; and
  Draco, \citet{seg07}.}
\tablecomments{Table~\ref{tab:catalog} is published in its entirety in
  the electronic edition of the Astrophysical Journal.  Some columns
  (right ascension, declination, and $B$, $V$, $R$, and $I$
  magnitudes) are suppressed in the printed edition.}
\end{deluxetable}

\setcounter{table}{5}
\begin{deluxetable}{llccccccccccccccccccccc}
\tablewidth{0pt}
\tablecolumns{22}
\tablecaption{Effect of Errors in Atmospheric Parameters on Abundances\label{tab:atmparerr}}
\tablehead{\colhead{ } & \colhead{ } & \multicolumn{4}{c}{$\delta\mathfeh$} & \multicolumn{4}{c}{$\delta$[Mg/Fe]} & \multicolumn{4}{c}{$\delta$[Si/Fe]} & \multicolumn{4}{c}{$\delta$[Ca/Fe]} & \multicolumn{4}{c}{$\delta$[Ti/Fe]} \\
\colhead{ } & \colhead{ } & \multicolumn{2}{c}{$\mathteff \pm$} & \multicolumn{2}{c}{$\mathlogg \pm$} & \multicolumn{2}{c}{$\mathteff \pm$} & \multicolumn{2}{c}{$\mathlogg \pm$} & \multicolumn{2}{c}{$\mathteff \pm$} & \multicolumn{2}{c}{$\mathlogg \pm$} & \multicolumn{2}{c}{$\mathteff \pm$} & \multicolumn{2}{c}{$\mathlogg \pm$} & \multicolumn{2}{c}{$\mathteff \pm$} & \multicolumn{2}{c}{$\mathlogg \pm$} \\
\colhead{dSph} & \colhead{Name} & \colhead{125~K} & \colhead{250~K} & \colhead{0.3} & \colhead{0.6} & \colhead{125~K} & \colhead{250~K} & \colhead{0.3} & \colhead{0.6} & \colhead{125~K} & \colhead{250~K} & \colhead{0.3} & \colhead{0.6} & \colhead{125~K} & \colhead{250~K} & \colhead{0.3} & \colhead{0.6} & \colhead{125~K} & \colhead{250~K} & \colhead{0.3} & \colhead{0.6}}
\startdata
Scl & 1002473 & 0.10 & 0.20 & 0.00 & 0.00    & 0.02 & 0.04 & 0.03 & 0.06 & 0.06 & 0.16 & 0.00 & 0.03 & 0.06 & 0.11 & 0.03 & 0.06 & 0.05 & 0.08 & 0.02 & 0.03 \\
Scl & 1002447 & 0.13 & 0.26 & 0.00 & 0.00    & 0.09 & 0.12 & 0.02 & 0.03 & 0.07 & 0.16 & 0.00 & 0.01 & 0.07 & 0.12 & 0.02 & 0.03 & 0.02 & 0.08 & 0.04 & 0.04 \\
Scl & 1002888 & 0.13 & 0.24 & 0.00 & 0.00    & 0.01 & 0.01 & 0.04 & 0.06 & 0.09 & 0.17 & 0.01 & 0.00 & 0.05 & 0.09 & 0.02 & 0.03 & 0.03 & 0.06 & 0.01 & 0.01 \\
Scl & 1003386 & 0.13 & 0.26 & 0.02 & 0.04    & 0.03 & 0.04 & 0.02 & 0.05 & 0.08 & 0.19 & 0.04 & 0.03 & 0.04 & 0.09 & 0.01 & 0.04 & 0.06 & 0.11 & 0.04 & 0.04 \\
Scl & 1003505 & 0.15 & 0.29 & 0.01 & 0.01    & 0.09 & 0.16 & 0.01 & 0.02 & 0.11 & 0.21 & 0.02 & 0.03 & 0.07 & 0.11 & 0.01 & 0.03 & 0.08 & 0.18 & 0.00 & 0.00 \\
Scl & 1003443 & 0.13 & 0.25 & 0.01 & 0.02    & 0.11 & 0.11 & 0.01 & 0.01 & 0.08 & 0.14 & 0.01 & 0.02 & 0.03 & 0.05 & 0.01 & 0.02 & 0.03 & 0.06 & 0.01 & 0.02 \\
Scl & 1003537 & 0.16 & 0.31 & 0.01 & \nodata & 0.08 & 0.13 & 0.02 & 0.02 & 0.12 & 0.24 & 0.01 & 0.02 & 0.06 & 0.12 & 0.01 & 0.02 & 0.08 & 0.16 & 0.00 & 0.01 \\
Scl & 1003694 & 0.09 & 0.16 & 0.07 & \nodata & 0.02 & 0.01 & 0.05 & 0.06 & 0.12 & 0.23 & 0.02 & 0.01 & 0.09 & 0.12 & 0.06 & 0.08 & 0.20 & 0.29 & 0.02 & 0.06 \\
Scl & 1003702 & 0.14 & 0.27 & 0.01 & 0.02    & 0.00 & 0.00 & 0.00 & 0.00 & 0.08 & 0.16 & 0.00 & 0.01 & 0.05 & 0.06 & 0.00 & 0.00 & 0.04 & 0.10 & 0.05 & 0.03 \\
Scl & 1003967 & 0.14 & 0.27 & 0.01 & 0.02    & 0.00 & 0.00 & 0.00 & 0.00 & 0.08 & 0.15 & 0.01 & 0.02 & 0.04 & 0.08 & 0.01 & 0.01 & 0.01 & 0.08 & 0.01 & 0.02 \\
\enddata
\tablecomments{Table~\ref{tab:atmparerr} is published in its entirety in
  the electronic edition of the Astrophysical Journal.}
\end{deluxetable}

\setcounter{table}{6}
\begin{deluxetable}{lllcccccccccc}
\tablewidth{0pt}
\tablecolumns{13}
\tablecaption{Comparison Between High-Resolution and DEIMOS Abundances\label{tab:hrscompare}}
\tablehead{\colhead{ } & \colhead{ } & \colhead{ } & \multicolumn{5}{c}{HRS} & \multicolumn{5}{c}{MRS} \\
\colhead{System} & \colhead{Name} & \colhead{HRS Reference} & \colhead{\teff} & \colhead{\logg} & \colhead{$\xi$} & \colhead{\feh} & \colhead{[Mg/Fe]} & \colhead{\teff} & \colhead{\logg} & \colhead{$\xi$} & \colhead{\feh} & \colhead{[Mg/Fe]} \\
\colhead{ } & \colhead{ } & \colhead{ } & \colhead{(K)} & \colhead{(cm~s$^{-2}$)} & \colhead{(km~s$^{-1}$)} & \colhead{(dex)} & \colhead{(dex)} & \colhead{(K)} & \colhead{(cm~s$^{-2}$)} & \colhead{(km~s$^{-1}$)} & \colhead{(dex)} & \colhead{(dex)}}
\startdata
M79      & N1904-S80    & \citet{gra89}  & 4250 & 0.75 & 2.50 & $-1.28 \pm 0.20$ & $+0.05 \pm 0.20$ & 4055 & 0.56 & 2.01 & $-1.56 \pm 0.11$ & $+0.07 \pm 0.11$ \\
NGC 2419 & N2419-S1305  & \citet{she01a} & 4275 & 0.70 & 2.10 & $-2.32 \pm 0.11$ & $+0.30 \pm 0.18$ & 4395 & 0.79 & 1.96 & $-2.20 \pm 0.11$ & $+0.58 \pm 0.20$ \\
NGC 2419 & N2419-S1305  & \citet{she01a} & 4275 & 0.70 & 2.10 & $-2.32 \pm 0.11$ & $+0.30 \pm 0.18$ & 4366 & 0.79 & 1.96 & $-2.17 \pm 0.11$ & $+0.17 \pm 0.24$ \\
M5       & III-149      & \citet{iva01}  & 4200 & 0.91 & 1.75 & $-1.15 \pm 0.04$ &     \nodata      & 4218 & 1.05 & 1.89 & $-1.30 \pm 0.11$ & $+0.17 \pm 0.11$ \\
M5       & G18155\_0228 & \citet{ram03}  & 5270 & 3.25 & 1.44 & $-1.31 \pm 0.04$ & $+0.28 \pm 0.06$ & 5286 & 3.25 & 1.37 & $-1.44 \pm 0.11$ & $+0.26 \pm 0.14$ \\
M5       & II-59        & \citet{iva01}  & 4450 & 1.27 & 1.30 & $-1.15 \pm 0.04$ &     \nodata      & 4443 & 1.32 & 1.83 & $-1.26 \pm 0.11$ & $+0.14 \pm 0.19$ \\
M5       & G18447\_0453 & \citet{ram03}  & 5275 & 3.15 & 1.44 & $-1.37 \pm 0.05$ & $+0.13 \pm 0.06$ & 5279 & 3.06 & 1.42 & $-1.23 \pm 0.11$ & $+0.17 \pm 0.12$ \\
M5       & 1-31         & \citet{ram03}  & 4880 & 2.25 & 1.64 & $-1.30 \pm 0.03$ & $+0.24 \pm 0.08$ & 4903 & 2.14 & 1.63 & $-1.27 \pm 0.11$ & $+0.22 \pm 0.11$ \\
M5       & IV-59        & \citet{iva01}  & 4229 & 0.79 & 2.10 & $-1.25 \pm 0.07$ &     \nodata      & 4254 & 0.99 & 1.91 & $-1.34 \pm 0.11$ & $+0.14 \pm 0.12$ \\
M5       & G18484\_0316 & \citet{ram03}  & 4995 & 2.50 & 1.58 & $-1.38 \pm 0.03$ & $+0.22 \pm 0.06$ & 4980 & 2.41 & 1.57 & $-1.32 \pm 0.11$ & $+0.17 \pm 0.11$ \\
\enddata
\tablecomments{Table~\ref{tab:hrscompare} is published in its entirety
  in the electronic edition of the Astrophysical Journal.  Some
  columns ([Si/Fe], [Ca/Fe], and [Ti/Fe] for both MRS and HRS) are
  suppressed in the printed edition.}
\end{deluxetable}


\begin{thebibliography}{}

\bibitem[Adelman-McCarthy et al.(2007)]{ade07} Adelman-McCarthy,
  J.~K., et al. 2007, \apjs, 172, 634


\bibitem[An et al.(2008)]{an08} An, D., et al. 2008, \apjs, 179, 326

\bibitem[Anders \& Grevesse(1989)]{and89} Anders, E., \& Grevesse,
  N.\ 1989, \gca, 53, 197

\bibitem[Aoki et al.(2009)]{aok09} Aoki, W., et al.\ 2009, \aap, 502,
  569

\bibitem[Armandroff \& Da Costa(1991)]{arm91} Armandroff, T.~E., \& Da
  Costa, G.~S.\ 1991, \aj, 101, 1329




\bibitem[Beers et al.(1985)]{bee85} Beers, T.~C., Preston, G.~W., \&
  Shectman, S.~A.\ 1985, \aj, 90, 2089

\bibitem[Beers et al.(1992)]{bee92} ---------.\ 1992, \aj, 103, 1987


\bibitem[Bell et al.(2008)]{bel08} Bell, E.~F., et al.\ 2008, \apj,
  680, 295

\bibitem[Bell et al.(1976)]{bel76} Bell, R.~A., Eriksson, K.,
  Gustafsson, B., \& Nordlund, A.\ 1976, \aaps, 23, 37

\bibitem[Bellazzini et al.(2002)]{bel02} Bellazzini, M., Ferraro,
  F.~R., Origlia, L., Pancino, E., Monaco, L., \& Oliva, E.\ 2002,
  \aj, 124, 3222

\bibitem[Bellazzini et al.(2004)]{bel04} Bellazzini, M., Gennari, N.,
  Ferraro, F.~R., \& Sollima, A.\ 2004, \mnras, 354, 708

\bibitem[Bellazzini et al.(2001)]{bel01} Bellazzini, M., Pecci, F.~F.,
  Ferraro, F.~R., Galleti, S., Catelan, M., \& Landsman, W.~B.\ 2001,
  \aj, 122, 2569


\bibitem[Binney \& Tremaine(2008)]{bin08} Binney, J., \& Tremaine,
  S.\ 2008, Galactic Dynamics (2nd ed.; Princeton UP)

\bibitem[Bond(1980)]{bon80} Bond, H.~E.\ 1980, \apjs, 44, 517


\bibitem[Brocato et al.(1996)]{bro96} Brocato, E., Castellani, V., \&
  Ripepi, V.\ 1996, \aj, 111, 809


\bibitem[Buonanno et al.(1991)]{buo91} Buonanno, R., Pecci, F.~F.,
  Capellaro, E., Ortolani, S., Richtler, T., \& Geyer, E.~H.\ 1991,
  \aj, 102, 1005

\bibitem[Carney et al.(1992)]{car92} Carney, B.~W., Storm, J.,
  Trammell, S.~R., \& Jones, R.~V.\ 1992, \pasp, 104, 44

\bibitem[Carretta et al.(2002)]{car02} Carretta, E., Gratton, R.,
  Cohen, J.~G., Beers, T.~C., \& Christlieb, N.\ 2002, \aj, 124, 481


\bibitem[Castelli(2005)]{cas05} Castelli, F.\ 2005, MSAIS, 8, 34


\bibitem[Castelli \& Kurucz(2004)]{cas04} Castelli, F., \& Kurucz,
  R.~L.\ 2004, arXiv:astro-ph/0405087


\bibitem[Chonis \& Gaskell(2008)]{cho08} Chonis, T.~S., \& Gaskell,
  C.~M.\ 2008, \aj, 135, 264

\bibitem[Clem et al.(2008)]{cle08} Clem, J.~L., Vanden Berg, D.~A., \&
  Stetson, P.~B.\ 2008, \aj, 135, 682


\bibitem[Cohen et al.(2002)]{coh02} Cohen, J.~G., Christlieb, N.,
  Beers, T.~C., Gratton, R., \& Carretta, E.\ 2002, \aj, 124, 470

\bibitem[Cohen et al.(2008)]{coh08} Cohen, J.~G., Christlieb, N.,
  McWilliam, A., Shectman, S., Thompson, I., Melendez, J., Wisotzki,
  L., \& Reimers, D. 2008, \apj, 672, 320

\bibitem[Cohen \& Huang(2009)]{coh09} Cohen, J.~G., \& Huang, W.\
2009, \apj, 701, 1053

\bibitem[Cohen \& Huang(2010)]{coh10} ---------.\ 2010, \apj, 719, 931

\bibitem[Cohen et al.(2010)]{cohetal10} Cohen, J.~G., Kirby, E.~N.,
  Simon, J.~D., \& Geha, M.~C.\ 2010, \apj, in press, arXiv:1010.0031

\bibitem[Cohen \& Mel{\'e}ndez(2005a)]{coh05a} Cohen, J.~G. \&
  Mel{\'e}ndez, J. 2005a, \aj, 129, 303

\bibitem[Cohen \& Mel{\'e}ndez(2005b)]{coh05b} ---------. 2005b, \aj,
  129, 1607

\bibitem[Cohen et al.(2006)]{coh06} Cohen, J.~G., et al.\ 2006, \aj,
  132, 137

\bibitem[Cooper et al.(2010)]{coo10} Cooper, A.~P., et al.\ 2010,
  \mnras, 406, 744

\bibitem[C{\^o}t{\'e} et al.(1991)]{cot91} C{\^o}t{\'e}, P., Richer, H.~B., \&
  Fahlman, G.~G.\ 1991, \aj, 102, 1358


\bibitem[Da Costa et al.(2009)]{dac09} Da Costa, G.~S., Held, 
E.~V., Saviane, I., \& Gullieuszik, M.\ 2009, \apj, 705, 1481 


\bibitem[Demarque et al.(2004)]{dem04} Demarque, P., Woo, J.-H., Kim,
  Y.-C., \& Yi, S.~K.\ 2004, \apjs, 155, 667


\bibitem[Durrell \& Harris(1993)]{dur93} Durrell, P.~R., \& Harris,
  W.~E.\ 1993, \aj, 105, 1420

\bibitem[Eggen, Lynden-Bell, \& Sandage(1962)]{egg62} Eggen, O.~J.,
  Lynden-Bell, D., \& Sandage, A.~R.\ 1962, \apj, 136, 748

\bibitem[Faber et al.(2003)]{fab03} Faber, S.~M., et al.\ 2003,
  \procspie, 4841, 1657


\bibitem[Ferraro et al.(1992)]{fer92} Ferraro, F.~R., Clementini, G.,
  Fusi Pecci, F., Sortino, R., \& Buonanno, R.\ 1992, \mnras, 256, 391

\bibitem[Font et al.(2006)]{fon06} Font, A.~S., Johnston, K.~V.,
  Bullock, J.~S., \& Robertson, B.~E.\ 2006, \apj, 638, 585



\bibitem[Frebel et al.(2010a)Frebel, Kirby, \& Simon]{fre10a} Frebel,
  A., Kirby, E., \& Simon, J.~D.\ 2010, Nature, 464, 72

\bibitem[Frebel et al.(2010b)]{fre10b} Frebel, A., Simon, J.~D., Geha,
  M., \& Willman, B.\ 2010, \apj, 708, 560

\bibitem[Freeman \& Rodgers(1975)]{fre75} Freeman, K.~C., \& Rodgers,
  A.~W.\ 1975, \apjl, 201, L71


\bibitem[Fulbright(2000)]{ful00} Fulbright, J.~P.\ 2000, \aj, 120,
  1841

\bibitem[Fulbright et al.(2004)]{ful04} Fulbright, J.~P., Rich, R.~M.,
  \& Castro, S.\ 2004, \apj, 612, 447

\bibitem[Geffert \& Maintz(2000)]{gef00} Geffert, M., \& Maintz,
  G.\ 2000, \aaps, 144, 227

\bibitem[Geisler et al.(2005)]{gei05} Geisler, D., Smith, V.~V.,
  Wallerstein, G., Gonzalez, G., \& Charbonnel, C.\ 2005, \aj, 129,
  1428



\bibitem[Girardi et al.(2002)]{gir02} Girardi, L., Bertelli, G.,
  Bressan, A., Chiosi, C., Groenewegen, M.~A.~T., Marigo, P.,
  Salasnich, B., \& Weiss, A.\ 2002, \aap, 391, 195

\bibitem[Gratton(1982)]{gra82} Gratton, R.~G. 1982, \apj, 257, 640

\bibitem[Gratton et al.(2003)]{gra03} Gratton, R.~G., Carretta, E.,
  Desidera, S., Lucatello, S., Mazzei, P., \& Barbieri, M.\ 2003,
  \aap, 406, 131

\bibitem[Gratton \& Ortolani(1989)]{gra89} Gratton, R.~G. \& Ortolani,
  S. 1989, \aap, 211, 41

\bibitem[Gratton et al.(2004)]{gra04} Gratton, R., Sneden, C., \&
  Carretta, E.\ 2004, \araa, 42, 385

\bibitem[Gray(2008)]{gra08} Gray, D.~F. 2008, The Observation and
  Analysis of Stellar Photospheres (3rd ed.; Cambridge UP)



\bibitem[Guhathakurta et al.(2006)]{guh06} Guhathakurta, P., et
  al.\ 2006, \aj, 131, 2497

\bibitem[Gustafsson et al.(1975)]{gus75} Gustafsson, B., Bell, R.~A.,
  Eriksson, K., \& Nordlund, A. 1975, \aap, 42, 407

\bibitem[Gustafsson et al.(2003)]{gus03} Gustafsson, B., Edvardsson,
  B., Eriksson, K., J{\o}rgensen, U.~G., Mizuno-Wiedner, M., \& Plez,
  B.\ 2003, in ASP Conf.\ Ser.\ 288, Modelling of Stellar Atmospheres,
  ed.\ I.~Hubeny, D.~Mihalas, \& K.~Werner (San Francisco: ASP), 331

\bibitem[Gustafsson et al.(2008)]{gus08} Gustafsson, B., Edvardsson,
  B., Eriksson, K., J{\o}rgensen, U.~G., Nordlund, {\AA}., \& Plez,
  B.\ 2008, \aap, 486, 951

\bibitem[Harris(1975)]{har75} Harris, W.~E.\ 1975, \apjs, 29, 397

\bibitem[Harris(1996)]{har96} ---------.\ 1996, \aj, 112, 1487

\bibitem[Harris et al.(1997)]{har97} Harris, W.~E., et al.\ 1997, \aj,
  114, 1030










\bibitem[Irwin \& Hatzidimitriou(1995)]{irw95} Irwin, M., \&
  Hatzidimitriou, D.\ 1995, \mnras, 277, 1354


\bibitem[Ivans et al.(2001)]{iva01} Ivans, I.~I., Kraft, R.~P.,
  Sneden, C., Smith, G.~H., Rich, R.~M., \& Shetrone, M.\ 2001, \aj,
  122, 1438

\bibitem[Johnson(2002)]{joh02} Johnson, J.~A.\ 2002, \apjs, 139, 219

\bibitem[Jordi et al.(2006)Jordi, Grebel, \& Ammon]{jor06} Jordi, K.,
  Grebel, E.~K., \& Ammon, K.\ 2006, \aap, 460, 339


\bibitem[Kirby et al.(2010a)Paper~IV]{kir10a} Kirby, E.~N., Cohen,
  J.~G., Smith, G.~H., Majewski, S.~R., Sohn, S.~T., \& Guhathakurta,
  P.\ 2010a, \apj, in press, arXiv:1011.5221 (Paper~IV)

\bibitem[Kirby et al.(2009)Paper~I]{kir09} Kirby, E.~N.,
  Guhathakurta, P., Bolte, M., Sneden, C., \& Geha, M.~C.\ 2009, \apj,
  705, 328 (Paper~I)

\bibitem[Kirby et al.(2008a)Kirby, Guhathakurta, \& Sneden]{kir08a}
  Kirby, E.~N., Guhathakurta, P., \& Sneden, C.\ 2008a, \apj, 682,
  1217 (KGS08)

\bibitem[Kirby et al.(2010b)Paper~III]{kir10b} Kirby, E.~N.,
  Lanfranchi, G.~A., Simon, J.~D., Cohen, J.~G., \& Guhathakurta,
  P.\ 2010b, \apj, in press, arXiv:1011.4937 (Paper~III)

\bibitem[Kirby et al.(2008b)]{kir08b} Kirby, E.~N., Simon, J.~D.,
  Geha, M., Guhathakurta, P., \& Frebel, A.\ 2008b, \apjl, 685, L43

\bibitem[Koch et al.(2008)]{koc08} Koch, A., Grebel, E.~K., Gilmore,
  G.~F., Wyse, R.~F.~G., Kleyna, J.~T., Harbeck, D.~R., Wilkinson,
  M.~I., \& Wyn Evans, N.\ 2008, \aj, 135, 1580





\bibitem[Kravtsov et al.(1997)]{kra97} Kravtsov, V., Ipatov, A.,
  Samus, N., Smirnov, O., Alcaino, G., Liller, W., \& Alvarado,
  F. 1997, \aaps, 125, 1

\bibitem[Kuehn et al.(2008)]{kue08} Kuehn, C., et al.\ 2008, \apjl,
  674, L81

\bibitem[Kupka et al.(1999)]{kup99} Kupka, F., Piskunov, N.,
  Ryabchikova, T.~A., Stempels, H.~C., \& Weiss, W.~W.\ 1999, \aaps,
  138, 119

\bibitem[Kurucz(1993)]{kur93} Kurucz, R.\ 1993, ATLAS9 Stellar
  Atmosphere Programs and 2 km/s grid.~Kurucz CD-ROM No.~13.~
  Cambridge, Mass.: Smithsonian Astrophysical Observatory, 1993, 13

\bibitem[Kurucz(2005)]{kur05} ---------. 2005, MSAIS, 8, 14

\bibitem[Lai et al.(2004)]{lai04} Lai, D.~K., Bolte, M., Johnson,
  J.~A., \& Lucatello, S.\ 2004, \aj, 128, 2402

\bibitem[Lai et al.(2007)]{lai07} Lai, D.~K., Johnson, J.~A., Bolte,
  M., \& Lucatello, S.\ 2007, \apj, 667, 1185


\bibitem[Lee et al.(2003)]{lee03} Lee, M.~G., et al.\ 2003, \aj, 126,
  2840

\bibitem[Letarte et al.(2010)]{let10} Letarte, B., et al.\ 2010, \aap,
  523, A17



\bibitem[Majewski(1993)]{maj93} Majewski, S.~R.\ 1993, \araa, 31, 575

\bibitem[Majewski et al.(1996)]{maj96} Majewski, S.~R., Munn, J.~A.,
  \& Hawley, S.~L.\ 1996, \apjl, 459, L73

\bibitem[Majewski et al.(2000)]{maj00} Majewski, S.~R., Ostheimer,
  J.~C., Kunkel, W.~E., \& Patterson, R.~J.\ 2000, \aj, 120, 2550




\bibitem[Marino et al.(2009)]{mar09} Marino, A.~F., Milone, A.~P.,
  Piotto, G., Villanova, S., Bedin, L.~R., Bellini, A., \& Renzini,
  A.\ 2009, \aap, 505, 1099

\bibitem[Markwardt(2009)]{mark09} Markwardt, C.~B.\ 2009, in ASP
  Conf.\ Ser.\ 411, Astronomical Data Analysis Software and Systems
  XVIII, ed. D.~A. Bohlender, D.~Durand, \& P.~Dowler (San Francisco:
  ASP), 251

\bibitem[Martin et al.(2008)]{mar08} Martin, N.~F., de Jong, J.~T.~A.,
  \& Rix, H.-W.\ 2008, \apj, 684, 1075


\bibitem[Mateo(1998)]{mat98} Mateo, M.~L.\ 1998, \araa, 36, 435 

\bibitem[Mighell \& Burke(1999)]{mig99} Mighell, K.~J., \& Burke,
  C.~J.\ 1999, \aj, 118, 366

\bibitem[Mishenina et al.(2003)Mishenina, Panchuk, \& Samus']{mis03}
  Mishenina, T.~V., Panchuk, V.~E., \& Samus', N.~N. 2003, Astronomy
  Reports, 47, 248





\bibitem[Norris et al.(1999)]{nor99} Norris, J.~E., Ryan, S.~G., \&
  Beers, T.~C.\ 1999, \apjs, 123, 639


\bibitem[Paltrinieri et al.(1998)]{pal98} Paltrinieri, B., Ferraro,
  F.~R., Carretta, E., \& Fusi Pecci, F.\ 1998, \mnras, 293, 434

\bibitem[Pietrzy{\'n}ski et al.(2008)]{pie08} Pietrzy{\'n}ski, G., et
  al.\ 2008, \aj, 135, 1993

\bibitem[Pilachowski et al.(1996)]{pil96} Pilachowski, C.~A., Sneden,
  C., \& Kraft, R.~P.\ 1996, \aj, 111, 1689

\bibitem[Piskunov et al.(2007)]{pis07} Piskunov, A.~E., Schilbach, E.,
  Kharchenko, N.~V., R{\"o}ser, S., \& Scholz, R.-D.\ 2007, \aap, 468,
  151



\bibitem[Pritzl et al.(2005)Pritzl, Venn, \& Irwin]{pri05} Pritzl,
  B.~J., Venn, K.~A., \& Irwin, M. 2005, \aj, 130, 2140


\bibitem[Ram{\'i}rez \& Cohen(2002)]{ram02} Ram{\'{\i}}rez, S.~V. \&
  Cohen, J.~G. 2002, \aj, 123, 3277

\bibitem[Ram{\'i}rez \& Cohen(2003)]{ram03} ---------. 2003, \aj, 125,
  224


\bibitem[Reed et al.(1988)]{ree88} Reed, B.~C., Hesser, J.~E., \&
  Shawl, S.~J.\ 1988, \pasp, 100, 545

\bibitem[Regnault et al.(2009)]{reg09} Regnault, N., et al.\ 2009,
  \aap, 506, 999

\bibitem[Rizzi et al.(2007)]{riz07} Rizzi, L., Held, E.~V., Saviane,
  I., Tully, R.~B., \& Gullieuszik, M.\ 2007, \mnras, 380, 1255

\bibitem[Robertson et al.(2005)]{rob05} Robertson, B., Bullock, J.~S.,
  Font, A.~S., Johnston, K.~V., \& Hernquist, L.\ 2005, \apj, 632, 872

\bibitem[Roederer(2009)]{roe09} Roederer, I.~U.\ 2009, \aj, 137, 272

\bibitem[Rosenberg et al.(2000)]{ros00} Rosenberg, A., Piotto, G.,
  Saviane, I., \& Aparicio, A. 2000, \aaps, 144, 5

\bibitem[Rutledge et al.(1997)]{rut97} Rutledge, G.~A., Hesser, J.~E.,
  \& Stetson, P.~B.\ 1997, \pasp, 109, 907

\bibitem[Sadakane et al.(2004)]{sad04} Sadakane, K., Arimoto, N.,
  Ikuta, C., Aoki, W., Jablonka, P., \& Tajitsu, A. 2004, \pasj, 56,
  1041

\bibitem[Sandquist et al.(1996)]{san96} Sandquist, E.~L., Bolte, M.,
  Stetson, P.~B., \& Hesser, J.~E.\ 1996, \apj, 470, 910




\bibitem[Sbordone(2005)]{sbo05} Sbordone, L.\ 2005, MSAIS, 8, 61

\bibitem[Sbordone et al.(2004)]{sbo04} Sbordone, L., Bonifacio, P.,
  Castelli, F., \& Kurucz, R.~L.\ 2004, MSAIS, 5, 93

\bibitem[Schlegel et al.(1998)]{sch98} Schlegel, D.~J., Finkbeiner,
  D.~P., \& Davis, M.\ 1998, \apj, 500, 525



\bibitem[Searle \& Zinn(1978)]{sea78} Searle, L., \& Zinn, R.\ 1978,
  \apj, 225, 357

\bibitem[S{\'e}gall et al.(2007)]{seg07} S{\'e}gall, M., Ibata, R.~A.,
  Irwin, M.~J., Martin, N.~F., \& Chapman, S.\ 2007, \mnras, 375, 831

\bibitem[Shetrone et al.(1998)Shetrone, Bolte \& Stetson]{she98}
  Shetrone, M.~D., Bolte, M., \& Stetson, P.~B.\ 1998, \aj, 115, 1888

\bibitem[Shetrone et al.(2001)Shetrone, C{\^o}t{\'e} \&
  Sargent]{she01a} Shetrone, M.~D., C{\^o}t{\'e}, P., \& Sargent,
  W.~L.~W.\ 2001, \apj, 548, 592


\bibitem[Shetrone et al.(2009)]{she09} Shetrone, M.~D., Siegel, M.~H.,
  Cook, D.~O., \& Bosler, T.\ 2009, \aj, 137, 62

\bibitem[Shetrone et al.(2003)]{she03} Shetrone, M.~D., Venn, K.~A.,
  Tolstoy, E., Primas, F., Hill, V., \& Kaufer, A.\ 2003, \aj, 125,
  684

\bibitem[Siegel et al.(2001)]{sie01} Siegel, M.~H., Majewski, S.~R.,
  Cudworth, K.~M., \& Takamiya, M.\ 2001, \aj, 121, 935

\bibitem[Siegel et al.(2010)]{sie10} Siegel, M.~H., Majewski, S.~R.,
  Sohn, S.~T., Shetrone, M.~D., Munoz, R.~R., \& Patterson,
  R.~J. 2010, \apj, submitted

\bibitem[Simon et al.(2010)]{sim10} Simon, J.~D., Frebel, A.,
  McWilliam, A., Kirby, E.~N., \& Thompson, I.~B.\ 2010, \apj, 716,
  446

\bibitem[Simon \& Geha(2007)]{sim07} Simon, J.~D., \& Geha, M.\ 2007,
  \apj, 670, 313

\bibitem[Sohn et al.(2007)]{soh07} Sohn, S.~T., et al. 2007, \apj,
  663, 960

\bibitem[Sneden(1973)]{sne73} Sneden, C.~A.\ 1973, Ph.D.~Thesis,
  University of Texas at Austin

\bibitem[Sneden et al.(2004)]{sne04} Sneden, C., Kraft, R.~P.,
  Guhathakurta, P., Peterson, R.~C., \& Fulbright, J.~P. 2004, \aj,
  127, 2162

\bibitem[Sneden et al.(1992)]{sne92} Sneden, C., Kraft, R.~P.,
  Prosser, C.~F., \& Langer, G.~E.\ 1992, \aj, 104, 2121

\bibitem[Sneden et al.(1997)]{sne97} Sneden, C., Kraft, R.~P.,
  Shetrone, M.~D., Smith, G.~H., Langer, G.~E., \& Prosser,
  C.~F. 1997, \aj, 114, 1964

\bibitem[Sneden et~al.(2000)Sneden, Pilachowski, \& Kraft]{sne00}
  Sneden, C., Pilachowski, C.~A., \& Kraft, R.~P. 2000, \aj, 120, 1351

\bibitem[Spite(1967)]{spi67} Spite, M.\ 1967, Annales d'Astrophysique,
  30, 211

\bibitem[Stetson(2000)]{ste00} Stetson, P.~B.\ 2000, \pasp, 112, 925

\bibitem[Stetson et al.(1998)]{ste98} Stetson, P.~B., Hesser, J.~E.,
  \& Smecker-Hane, T.~A. 1998, \pasp, 110, 533


\bibitem[Th{\'e}venin \& Idiart(1999)]{the99} Th{\'e}venin, F., \&
  Idiart, T.~P.\ 1999, \apj, 521, 753


\bibitem[Tolstoy et al.(2001)]{tol01} Tolstoy, E., Irwin, M.~J., Cole,
  A.~A., Pasquini, L., Gilmozzi, R., \& Gallagher, J.~S.\ 2001,
  \mnras, 327, 918



\bibitem[Tucker et al.(2006)]{tuc06} Tucker, D.~L., et al.\ 2006,
  Astronomische Nachrichten, 327, 821





\bibitem[Venn et al.(2004)]{ven04} Venn, K.~A., Irwin, M., Shetrone,
  M.~D., Tout, C.~A., Hill, V., \& Tolstoy, E.\ 2004, \aj, 128, 1177



\bibitem[Webbink(1985)]{web85} Webbink, R.~F.\ 1985, in IAU
  Symp.\ 113, Dynamics of Star Clusters, ed. J.Goodman and P.Hut
  (Dordrecht: Reidel), 541

\bibitem[Westfall et al.(2006)]{wes06} Westfall, K.~B., Majewski,
  S.~R., Ostheimer, J.~C., Frinchaboy, P.~M., Kunkel, W.~E.,
  Patterson, R.~J., \& Link, R.\ 2006, \aj, 131, 375

\bibitem[White \& Rees(1978)]{whi78} White, S.~D.~M., \& Rees,
  M.~J.\ 1978, \mnras, 183, 341



\bibitem[Zacharias et al.(2004)]{zac04} Zacharias, N., Monet, D.~G.,
  Levine, S.~E., Urban, S.~E., Gaume, R., \& Wycoff, G.~L. 2004, in
  Bulletin of the AAS, Vol.~36, AAS Meeting 205 (Washington, DC: AAS),
  1418

\bibitem[Zinn(1985)]{zin85} Zinn, R.\ 1985, \apj, 293, 424

\bibitem[Zucker et al.(2006)]{zuc06} Zucker, D.~B., et al.\ 2006,
  \apjl, 643, L103

\end{thebibliography}
\end{document}